%% file: paper.tex
\documentclass[prd,aps,superscriptaddress,showpacs,preprintnumbers,amsmath,amssymb,floatfix,12pt]{revtex4}

\usepackage{graphicx}
\usepackage{epsf}
\usepackage{dcolumn}
\usepackage{bm}
\input{template}

\begin{document}
\bibliographystyle{apsrev}


\preprint{ BNL-HET-04/11, CU-TP-1115, KANAZAWA-04-14, KEK-TH-992,}
\preprint{ MIT-CTP-3548, RBRC-426, WUB-04-14}

\title{Lattice QCD with two dynamical flavors of domain wall fermions}

\author{Y.~Aoki}
\affiliation{Physics Department, University of Wuppertal, Gaussstr.~20, 42119
  Wuppertal, Germany}

\author{T.~Blum}
\affiliation{RIKEN-BNL Research Center, Brookhaven National Laboratory, Upton, NY 11973}
\affiliation{Physics Department, University of Connecticut, Storrs, CT 06269-3046}

\author{N.~Christ}
\affiliation{Physics Department,Columbia University, New York, NY 10027}

\author{C.~Dawson}
\affiliation{RIKEN-BNL Research Center, Brookhaven National Laboratory, Upton, NY 11973}

\author{K.~Hashimoto}
\affiliation{Institute for Theoretical Physics, Kanazawa University, Kakuma, Kanazawa, 920-1192, Japan}
\affiliation{Radiation Laboratory, RIKEN, Wako 351-0198, Japan}

\author{T.~Izubuchi}
\affiliation{RIKEN-BNL Research Center, Brookhaven National Laboratory, Upton, NY 11973}
\affiliation{Institute for Theoretical Physics, Kanazawa University, Kakuma, Kanazawa, 920-1192, Japan}

\author{J.W.~Laiho\footnote{Present address: Theoretical Physics Department,
    Fermi National Accelerator Lab, Batavia, IL 60510 }}
\affiliation{Physics Department, Princeton University, Princeton, NJ 08544}

\author{L.~Levkova\footnote{Present address: Department of Physics, Indiana University, Bloomington, IN 47405}}
\affiliation{Physics Department, Columbia University, New York, NY 10027}

\author{M.~Lin}
\affiliation{Physics Department, Columbia University, New York, NY 10027}

\author{R.~Mawhinney}
\affiliation{Physics Department, Columbia University, New York, NY 10027}

\author{J.~Noaki\footnote{Present address: 
School of Physics and Astronomy, University of Southampton, Southampton, SO17 1BJ, U.K.}}
\affiliation{RIKEN-BNL Research Center, Brookhaven National Laboratory, Upton, NY 11973}

\author{S.~Ohta}
\affiliation{Institute for Particle and Nuclear Studies, KEK, Tsukuba, Ibaraki, 305-0801, Japan}
\affiliation{RIKEN-BNL Research Center, Brookhaven National Laboratory, Upton, NY 11973}

\author{K.~Orginos\footnote{Present address: Center for Theoretical Physics,
    Massachussetts Institute of Technology, Cambridge MA 02139-4307} }
\affiliation{RIKEN-BNL Research Center, Brookhaven National Laboratory, Upton, NY 11973}

\author{A.~Soni}
\affiliation{Physics Department, Brookhaven National Laboratory, Upton, NY 11973}

\date{\today}

\begin{abstract}
\input{text_sections/abstract.tex}
\end{abstract}

\pacs{11.15.Ha, 
      11.30.Rd, 
      12.38.Aw, 
      12.38.-t  
      12.38.Gc  
}
\maketitle

\newpage


\section{Introduction}
\label{sec:Intro}
\input{text_sections/intro.tex}


\section{Implementation of dynamical DWF }
\label{sec:Implementation}
\input{text_sections/implementation.tex}


\section{Simulation Details}
\label{sec:SimulationDetail}
\input{text_sections/simulation_details.tex}


\section{Thermalization and Auto-correlations}
\label{sec:Thermalization}
\input{text_sections/thermalizations.tex}


\section{Physical Results}
\label{sec:PhysicalResults}
\input{text_sections/physical_spect.tex}

\input{text_sections/physical_static.tex}

\input{text_sections/physical_bk.tex}


\section{Chiral Symmetry}
\label{sec:chiral}
\input{text_sections/chiral.tex}


\section{Conclusions}
\label{sec:Conclusions}
\input{text_sections/conclusions.tex}


\section*{Acknowledgments}

We thank Sinya Aoki, Mike Creutz and Norikazu Yamada for useful discussions.
YA thanks Boris Orth for useful discussions about the effect of finite volume
on the nucleon mass.  TI thanks Takashi Kaneko for useful discussions about
the static quark potential.  We thank RIKEN, Brookhaven National Laboratory
and the U.S.  Department of Energy for providing the facilities essential for
the completion of this work. We also thank the RIKEN Super Combined Cluster at
RIKEN, for the computer resources used for the static quark potential
calculation.  KH thanks RIKEN-BNL Research Center for its hospitality where
this work was performed.  The work of JL was supported in part by the LDRD
funds and (along with AS) by USDOE contract No. DE-AC02-98CH10886. The work of
KO was supported by the RIKEN/BNL Research Center and DOE grant
DF-FC02-94ER40818. JN was partially supported by JSPS Postdoctoral Fellowships
for Research Abroad.


\bibliography{paper}

\clearpage

\section*{Tables}
\input{tables/tables_djk.tex}

\clearpage
\pagebreak


\input{figures/implementation/figures.tex}

\input{figures/thermalization/figures.tex}

\input{figures/physical/figures.tex}

\input{figures/chiral/figures.tex}

\end{document}

%% file: template.tex
\def\mpi2{m_\pi^2}
\def\mK2{m_K^2}

\newcommand{\Dslash}{\rlap{/}\kern-2.0pt D}

\newcommand{\bea}{\begin{eqnarray}}
\newcommand{\eea}{\end{eqnarray}}
\newcommand{\be}{\begin{equation}}
\newcommand{\ee}{\end{equation}}

\def\simge{
    \mathrel{\rlap{\raise 0.511ex
        \hbox{$>$}}{\lower 0.511ex \hbox{$\sim$}}}}
\def\simle{
    \mathrel{\rlap{\raise 0.511ex
        \hbox{$<$}}{\lower 0.511ex \hbox{$\sim$}}}}

\newcommand{\vev}[1]{\left\langle#1\right\rangle}


%% file: text_sections/abstract.tex
We present results from the first large-scale study of two flavor QCD using
domain wall fermions (DWF), a chirally symmetric fermion formulation which has
proven to be very effective in the quenched approximation.  We work on
lattices of size $16^3 \times 32$, with a lattice cut-off of $a^{-1}\approx
1.7$ GeV, and dynamical (or sea) quark masses in the range
$m_{strange}/2\simle m_{sea} \simle m_{strange}$. After discussing the
algorithmic and implementation issues involved in simulating dynamical DWF, we
report on the low-lying hadron spectrum, decay constants, static quark
potential, and the important kaon weak matrix element describing indirect CP
violation in the Standard Model, $B_K$. In the latter case we include the
effect of non-degenerate quark masses ($m_s \neq m_u = m_d$), finding
$B_K^{\overline{MS}}(2\,{\rm GeV}) = 0.495\,(18)$.

%% file: text_sections/intro.tex
%
%

An improved theoretical understanding of the non-perturbative aspects of QCD
is increasingly important due to the continuing advance of experiments
involving hadrons.  Such understanding is an important ingredient in obtaining
precise values of the parameters of the Standard Model and the search for new
physics. It is also key to understanding the fundamental properties of QCD
itself which are under intense investigation at BNL, Jefferson Lab, FNAL,
CERN, and other places.

The fundamental theoretical tool to investigate QCD non-perturbatively is
lattice QCD, the regularized field theory of QCD on a discrete Euclidean
space-time lattice.  Treatment of the fermion field in such a regularization
has been a long-standing difficulty because flavor and chiral symmetry are
badly broken in practical numerical simulations using conventional lattice
fermions.  A revolutionary theoretical framework to realize flavor and chiral
symmetry on the lattice was constructed by Kaplan\cite{Kaplan:1992bt} and
subsequently reformulated and extended in two distinct ways by Narayanan and
Neuberger\cite{Narayanan:1993wx} and Shamir\cite{Shamir:1993zy} who suggested
that the new fermions be used to study vector gauge theories, and especially
QCD.  These new fermions, known as domain wall fermions, turned out to be in a
class of lattice fermions that satisfy the Ginsparg-Wilson
relation\cite{Ginsparg:1982bj}. Later, Neuberger developed still another
closely related lattice fermion called overlap
fermions\cite{Neuberger:1998fp}.

Today, both domain wall and overlap fermions are commonly used in quenched
lattice QCD simulations, those where the fermion determinant is set to one in
all path integrals used to calculate expectation values. In this paper, we
report on the first large scale two flavor dynamical simulations with domain
wall fermions, those where the fermion determinant is included in all path
integrals used to calculate expectation values. From a perturbative point of
view, this is equivalent to keeping all closed fermion loop contributions (at
lowest order, the hadronic vacuum polarization) to all orders in all Feynman
diagrams.

To realize chiral symmetry, domain wall fermions utilize an extra fifth
dimension, demanding increased computational resources that are essentially
linear in this extra dimension. Nevertheless quenched QCD calculations with
domain wall fermions were carried out, showing good chiral symmetry can be
achieved with a practical number of lattice sites in the extra dimension ($L_s
\sim 10 $) \cite{Blum:1997jf,Blum:1997mz,Aoki:2000pc}. 
Subsequently, within the quenched approximation domain wall fermions were used
in many calculations, showing dramatic improvement in scaling toward the
continuum limit over conventional formulations
\cite{AliKhan:2000iv,Blum:2000kn,AliKhan:2001wr,Aoki:2002vt,Sasaki:2001nf}, 
were used in pioneering calculations of weak interaction hadronic matrix
elements\cite{AliKhan:2001wr,Noaki:2001un,Blum:2001xb}, and
proved efficacious in non-perturbative operator renormalization
\cite{Blum:2001sr,Blum:2001xb,Aoki:2003ip}.

As mentioned already, large scale computations using domain wall fermions have
so far been restricted to the quenched approximation.  The problem with such
calculations is that there is no way to systematically reduce the errors due to
quenching without simply performing the unquenched calculation.  Past
experience indicates that the size of the error ranges between 5-10\% for many
observables, but can be much greater, depending on the observable (the
critical temperature of the hadron to quark-gluon plasma phase transition is a
well known example where the quenching error is more than 40\%).  In
\cite{Aoki:2002vt} the ratio of pseudoscalar decay constants in the quenched
approximation was found to be $f_K/f_\pi\approx1.13(3)$, different from the
experimental value, $\approx 1.22$. On the other hand, in this work, we find
$f_K/f_\pi=1.175(11)$, as discussed in Section~\ref{sec:PhysicalResults}.
Moreover, there are important physical objects, the flavor singlet mesons and
scalar mesons, for example, which are believed to be very sensitive to
dynamical, or sea quark effects.  Ultimately, to perform accurate lattice QCD
calculations these effects must be included.

Recently, large-scale dynamical fermion simulations using Wilson fermions
\cite{AliKhan:2001tx,Aoki:2002uc,Kaneko:2003re} and  improved staggered
fermions \cite{Davies:2003ik,Aubin:2004wf} (using two light degenerate quarks
and a heavier quark for the strange quark) have been reported. Both
formulations break chiral symmetry and the former also breaks flavor symmetry
(a severe problem, theoretically and practically).  Nevertheless, calculations
using either type of fermions, done with two or more sufficiently small
lattice spacings, have yielded results which agree accurately with experiment
for some observables.  QCD simulations using staggered fermions are much less
computationally demanding, making them attractive. However, the action
corresponds to four degenerate fermions in the continuum limit, so to simulate
a single flavor requires the fourth root of the determinant. For this to be a
legitimate procedure there must be a local action for a single flavor of
staggered fermions which had the same determinant as this fourth root. While
there is no proof that such a decomposition is impossible, no such action has
been constructed to date.  The flavor symmetry breaking intrinsic to staggered
fermions also leads to larger than expected discretization errors, a problem
that is significantly reduced by improving the naive discretization (see
\cite{Orginos:1998ue,Orginos:1999cr,Lepage:1998vj}).  Even with these
improvements, an extended chiral perturbation theory
\cite{Aubin:2003mg,Aubin:2003uc} 
with many extra low energy constants corresponding to the leading
lattice-spacing errors must be used to do the chiral extrapolations. Without
this extra complexity, the precise agreement with experiment could not be
achieved without costly reductions in the lattice spacing.  While more
computationally demanding than staggered fermions, the Wilson fermion
formulation is still much less demanding than DWF. However, this action
severely breaks chiral symmetry, leading to large discretisation errors
(starting at $O(a)$ rather than $O(a^2)$ for un-improved Wilson fermions) and
a problematic renormalization structure.  In addition to this, the Wilson
Dirac operator for a single flavor cannot be proved to have a positive
determinant for positive quark mass, which is an important prerequisite for
simulating an odd number of dynamical flavors exactly using the currently
available algorithms.

For these reasons, full QCD simulations with DWF are necessary as they avoid
these theoretical uncertainties by providing a fermion formulation with both
exact flavor symmetry, and good chiral properties.  In fact , in the limit of
an infinite extra dimension domain wall fermions posses exact chiral symmetry
at non-zero lattice spacing, that is, away from the continuum limit.  When the
extent of the extra dimension ($L_s$) is finite, explicit chiral symmetry
breaking is induced and is quantified in the form of a small additive quark
mass $m_{res}$.  The domain wall fermion Dirac operator has the nice property
that for positive mass and even $L_S$ its determinant is positive, and so
taking the square-root of the two flavor determinant is a well defined
operation. This, in combination with algorithms such as the polynomial
\cite{Aoki:2001pt} and rational \cite{Clark:2002vz,Clark:2003na} hybrid Monte 
Carlo algorithms, allows an odd number of dynamical flavors to be simulated
exactly.  In addition, if the gauge fields are sufficiently smooth that the
underlying Wilson-Dirac operator is not in the parity broken Aoki phase
\cite{Aoki:1983qi}), then the theory is expected to be local. This condition
has been met in quenched studies. We refer the interested reader to
\cite{Golterman:2003qe} and references therein for a full
explanation.  Note that the same locality condition applies to overlap
fermions.  Thus unquenched lattice QCD with domain wall, or overlap \footnote{
Our preference for domain wall over overlap fermions is largely historical
since they were developed for numerical simulations earlier. However, to our
knowledge domain wall fermions are at present more efficient with respect to
computer resources.}, fermions promises to be the closest non-perturbatively
regularized theory to continuum QCD.  This continuum-like (symmetry) property
is central to the argument that despite Ginsparg-Wilson fermions being naively
more expensive, superior scaling with the lattice spacing $a$ and greatly
simplified renormalization structure may eventually overcome the increased
computational burden since dynamical lattice simulations are known to scale as
$a^{-(7-8)}$ as the continuum, chiral, and infinite lattice volume limits are
approached.

Several years ago some of us performed a study of the finite temperature QCD
phase transition using dynamical domain wall fermions\cite{Chen:2000zu}.
Large explicit chiral symmetry breaking was evident in, for example, the quark
mass dependence of the pion mass, which is now believed to be caused the gauge
field being sufficiently rough that there was a condensation of zero-modes of
the four-dimensional Wilson Dirac operator appearing in the domain wall
fermion operator (the Aoki phase); the lattice scale in these simulations was
below 1 GeV.  Here we have moved to a finer lattice spacing, $a^{-1}\sim 1.7$
GeV, and use an improved gauge action which significantly reduced the explicit
chiral symmetry breaking in quenched calculations by an order(s) of
magnitude\cite{Aoki:2002vt}. As discussed in
Section~\ref{sec:PhysicalResults}, $m_{res}$ is a few MeV in this calculation,
so we are confident that our simulation is not inside the Aoki phase and that,
in fact, explicit chiral symmetry breaking is small and under precise control
for a relatively modest $L_s=12$.  The mass of the two dynamical domain wall
quarks, $m_q = m_f+m_{res}$, is roughly in the range one half to one times the
strange quark mass, meaning the residual quark mass due to the finite size of
the fifth dimension is a small fraction of the input quark mass.

This paper is organized as follows. The lattice action and its numerical
implementation with various improvements are described in
Section~\ref{sec:Implementation}.  The ensemble of gauge field configurations
is summarized in Section~\ref{sec:SimulationDetail}.  Thermalization and
auto-correlations in simulation time are discussed in
Section~\ref{sec:Thermalization}.  Physical results are presented in
Section~\ref{sec:PhysicalResults} with an emphasis on an analysis to
next-to-leading order in chiral perturbation theory. In
section~\ref{sec:chiral} we discuss aspects of chiral symmetry related to
domain wall fermion simulations and compare and contrast dynamical and
quenched simulations. We summarize our results and conclusions in
Section~\ref{sec:Conclusions}.


%% file: text_sections/implementation.tex
In this section we will discuss the algorithmic and implementation details
involved in generating dynamical ensembles with the domain wall fermion
action. As a first step we must precisely define the domain wall fermion
action that we have used.  For the domain wall fermion Dirac operator, $D_{\rm
DWF}$, we use the same definition and conventions as \cite{Blum:2000kn}.
However, were we simply to use the action $S=S_F + S_G$ with
\be
S_F = \sum_x \bar\Psi D_{DWF} \Psi
\ee
and $S_G$ representing the gauge action (in our case the DBW2 action), then we
would face a problem: the fifth direction in the domain wall fermions
framework naturally gives rise to unphysical heavy propagating modes in this
direction, while we are only interested in the physics of the light mode which
is bound to the domain walls.  The number of such modes will diverge in the
$L_s \rightarrow \infty$ limit and so dominate the path integral. To cancel
off this divergence we add a set of Pauli-Villars fields to the action such
that $S=S_F + S_G + S_{PV}$ with
\be
S_{PV} = \sum_x \Phi^{\dagger} D_{DWF}(m_f=1) \Phi \, .
\ee
Note that, while the original formulation of the domain wall fermion action
included Pauli-Villars fields \cite{Furman:1995ky}, here we are using a
slightly different form for these fields which was introduced in
\cite{Vranas:1997da}.

To simulate dynamical fermions on the lattice we have a choice of many
algorithms \cite{Gottlieb:1987mq,Duane:1987de,Luscher:1994xx,
Aoki:2001pt,Clark:2002vz,Clark:2003na}. As in this work we are simulating an
even number of dynamical flavors, it is convenient to use the well-known,
exact, hybrid Monte Carlo (HMC) $\Phi$ algorithm
\cite{Gottlieb:1987mq,Duane:1987de}.  The precise details of the HMC algorithm
can be found in the cited papers, but for convenience we will sketch an
overview here:
\begin{itemize}
\item
The fermionic part of the action is rewritten in terms of a set of bosonic
fields using the relation:
\bea
\int [d\overline{\psi}][d{\psi}] e^{ \overline{\psi} M \psi} &=&
{\rm det} \left(M\right) \\ \nonumber
& =& \int [d\phi] [d\phi^\dagger] e^{ - \phi^\dagger \frac{1}{M} \phi} \, .
\label{eq:phi int}
\eea
For this bosonic integral to converge the matrix, $M$, must have eigenvalues
whose real parts are positive.  Due to this condition this algorithm may not
be applied to QCD-like theories with an odd number of flavors.  However, it
may be applied to theories where quark flavors appear in degenerate mass
pairs. In this situation we may use the fact that
\be
\gamma_5 R D_{DWF} R \gamma_5  = D_{DWF}^\dagger,
\ee
where $R$ is the reflection operator in the fifth dimension, to rewrite ${\rm
det}(D_{DWF})^2$ as ${\rm det}( D_{DWF}^\dagger D_{DWF} )$. The matrix that
will appear in the bosonic integral is therefore
\begin{equation}
\frac{1}{D^{\dagger}_{DWF} D_{DWF}}
\end{equation}
which is the square of a Hermitian matrix and so is positive semi-definite. 

\item
An auxiliary field, $P_\mu$, is also added to the action.  This field plays
the role of a conjugate momenta to the gauge field in a molecular dynamics
evolution.

\item
The evolution of the gauge field is split up into trajectories.  At the
beginning of each trajectory the pseudo-fermion field, denoted $\phi$ in
Equation \ref{eq:phi int}, and the conjugate momenta are chosen from a
heatbath. The gauge field and conjugate momenta are then evolved some distance
in ``molecular dynamics time" using a discretization of Hamilton's equations,
which must be reversible. Several discretization techniques exist in the
literature \cite{Duane:1987de,Takaishi:2002re,Sexton:1992nu}; in this work we
are mainly using that introduced in \cite{Duane:1987de}, although we also
present the results of an exploratory study of the multiple timescale
technique of \cite{Sexton:1992nu}.

\item
If the evolution of the gauge field exactly followed Hamilton's equations then
this algorithm would be complete.  However, to correct for discretization
errors, a Metropolis accept/reject step is performed at the end of every
trajectory.

\end{itemize}

Each step in the molecular dynamics evolution requires an inversion of the
domain wall Dirac operator.  While we use an even-odd preconditioned form for
the operator to speed this up, it is, of course, still the most expensive part
of the gauge field generation. In the following subsections we give details of
our attempts to minimize both the number of inversions needed, and the cost of
each of these inversions.

\subsection{New Force Term}
\label{imp:nforce}

The initial studies of dynamical domain wall fermions \cite{Chen:2000zu} used
separate sets of bosonic fields to simulate the fermionic and Pauli-Villars
parts of the action.  This is potentially wasteful, especially for a large
fifth dimension, as the entire reason for including the Pauli-Villars term was
to cancel much of the contribution from the domain wall fermion Dirac
operator.  By using disparate sets of fields this cancellation is only
apparent after the average over these fields; over a single trajectory the
mismatch between these two pieces of the action may introduce large forces,
and therefore large time-step errors, into the molecular dynamics evolution.

In an attempt to avoid this problem we have implemented a form of the
Hamiltonian for the molecular dynamics evolution, first suggested in
\cite{Vranas:1997da}, that uses a single set of bosonic fields
to estimate both the fermion and Pauli-Villars terms. To do this we simply
note that
\bea
\frac{{\rm det}\left( D^{\dagger}(m_f) D(m_f) \right) }
{ {\rm det}\left( D^{\dagger}(1) D(1) \right) } 
&=&
{\rm det} \left( D(1) \frac{1}{D^{\dagger}(m_f) D(m_f)}
D^{\dagger}(1) \right)^{-1}
\\ \nonumber
&=&
\int [d \Phi][d \Phi^\dagger]
e^{ - S_n ( \Phi ) }
\eea
with 
\be
S_n = \sum_x \Phi^\dagger D(m_f=1)\frac{1}{D^{\dagger}(m_f)D(m_f) } 
D^{\dagger} (m_f=1) \Phi \,.
\ee
This approach needs no more memory space than the previous one and requires
exactly the same number of Dirac operator inversions as before, although they
are performed on different sources.  However, the cancellation between the
fermion and Pauli-Villars contributions to the molecular dynamics force now
happens exactly, step-by-step in the leapfrog evolution, rather than
stochastically.

Table \ref{imp:nforce_small} gives the parameters for, and results of, a small
volume head-to-head comparison of the old and new force terms on an $8^3\times
4\times 8$ ( where the parameters are spatial volume times temporal length
times $L_s$) lattice using the Wilson gauge action with $\beta=5.2$, and
$N_f=2$ domain wall fermions with $m_f=0.02$. As can be seen the new force
term has both a significantly higher acceptance than the old force term when
compared at the same step size, and a (small) reduction in the number of
conjugate gradient iterations needed. The difference of the Hamiltonian used
in the molecular dynamics evolution between the first and the last
configuration in a trajectory, $\Delta H$, is also measured. We find the
theoretical relation to the acceptance\cite{Gupta:1990ka,Takaishi:2002re},
\be
{\rm Prob}_{acc} \cong
{\rm erfc}\left( \sqrt{\vev{\Delta H}}/2\right) \cong
\exp\left( - \frac{\sqrt{\vev{(\Delta H)^2}}}{2\pi} \right) \ ,
\ee
holds to a good accuracy. 
For large space-time volume, $V$, and small size, $\Delta t$, the scaling 
\be
\sqrt{ (\Delta H)^2 } = C_{\Delta H} (\Delta t)^2 \sqrt{V}~~,
\label{eq:Def_C_deltaH}
\ee
is expected, where the coefficient, $C_{\Delta H}$, is independent of volume
and step size at leading order. By switching to the new force term, $C_{\Delta
H}$ is reduced by more than a factor of two, leading to an increased
acceptance, as seen in Table \ref{imp:nforce_small}.

A similar or somewhat better reduction of the discretization error of the new
force term is observed for the larger lattices ($16^3\times32\times 12$) and
smaller couplings on which we base much of this work. A detailed description
of the basic parameters for these ensembles is deferred until Section
\ref{sec:SimulationDetail}; here we give only the details relevant for the HMC
evolution which are summarized in Table~\ref{imp:evols}.  We also present the
results of a short test using the old force term for 45 trajectories, starting
from the thermalized lattice of the lightest sea quark mass. Using the new
force term, $C_{\Delta H}$ is reduced to to $\sim$ 40\% of it's value with the
old force term, and the acceptance is increased from 56\% to 77\%, as shown in
Table~\ref{imp:evols}.  An important observation is that $C_{\Delta H}$ is
almost independent of the sea quark mass for the new force term in our
simulation.  This is in contrast to the empirical assumption
\cite{Christ:2002pb, Ukawa:2002pc}, $C_{\Delta H}\propto m_{\rm
sea}^{-\alpha}$, $\alpha \sim 2$.

In Section \ref{imp:multi} we will discuss a technique that allows us to
exploit this new force term even more succesfully.

\subsection{Chronological Invertor}

For the inversion of the Dirac operator in each molecular dynamics (MD) step
(the most time consuming procedure in the calculation) we use chronological
invertor technique of \cite{Brower:1997vx}. We employ the conjugate gradient
algorithm to find an approximate solution, $\chi$, of the inverse of the
matrix $\tilde{D}^{\dagger}D$ (here $\tilde{D}$ represents the even-odd
preconditioned Dirac operator), acting on the source vector, $\Psi_F$, by
iteratively minimizing
\bea
\label{eq:min}
&\chi^\dagger \tilde D^\dagger \tilde D \chi 
- \chi^\dagger \Psi_F - \Psi^\dagger_F \chi \, ,
\eea
starting from an initial guess.  In the chronological invertor technique this
starting vector is forecast by minimizing Equation \ref{eq:min} in the
subspace spanned by the set of solutions from previous leapfrog steps
\footnote{ We found it is helpful to do Graham-Schmidt orthogonalization twice
in order to find the minimal residual vector from the subspace, while it was
done just once in \cite{Brower:1997vx}.}.  The precise solution is calculated
by the conjugate gradient (CG) algorithm starting from this forecast, so the
number of CG iterations is reduced. In this sub-section we will continue to
discuss the even-odd preconditioned form of the operator.

The first 655 trajectories in the $m_{\rm sea}=0.02$ evolution described in
Table \ref{imp:evols} were generated using a simple linear extrapolation of 
the previous two solution vectors as an initial guess for the conjugate
gradient algorithm \cite{Duane:1987de}, after which we moved to the
chronological invertor using the previous seven vectors (we found little
advantage to using more than this number, as explained below).  Figure
\ref{fig:cgsprof} shows the average and standard deviation of the
conjugate-gradient iteration count versus the leapfrog integration step for
trajectories 500 to 655 and 3000 to 4000. The reduction in the number of CG
iterations needed when using the chronological invertor being readily
apparent. (Note: while the number of inversions is 52, the first and last of
these are half-steps so the total distance moved in molecular dynamics time is
$51/100$). There is, however, the overhead involved in implementing the
chronological forecast. A comparison of the time taken for producing a single
trajectory in our particular implementation shows roughly a factor of two
speed-up whereas the CG iteration count is reduced from that without
forecasting by a factor of 2.6(1), 3.2(2), 3.3(2) for the $m_{\rm sea}=0.02,$
$0.03$ and $0.04$ evolutions, respectively. Table \ref{imp:cg} summarizes the
number of CG iterations for the first 15 steps of the molecular dynamics
trajectory for $m_{\rm sea}=0.02$, $m_{\rm sea}=0.03$, and $m_{\rm sea}=0.04$,
together with the total number of CG iterations per trajectory.  How the CG
converges to the precise solution on a typical configuration of the $m_{\rm
sea}=0.02$ enemble is illustrated in Figure \ref{fig:CG_vs_Nprecog} for
various numbers of previous solutions, $N_p$, used in the forecast.  The
precision of the forecast is improved by increasing $N_p$.  Since the
improvement is saturated for $N_p \geq 7$ for all sea quark masses used, we
carried out our simulation with $N_p=7$.

Using the chronological invertor technique, we must be careful to preserve
reversibility of the MD evolution, which is a condition for the HMC algorithm
to satisfy detailed balance.  To be precise, a trajectory starts with a gauge
configuration and its conjugate momentum, $( U^{(I)}_\mu(x), P^{(I)}_\mu(x)
)$, and evolves to another pair, $( U^{(F)}_\mu(x), P^{(F)}_\mu(x) )$, at the
end of the trajectory.  Starting another evolution with the latter pair with
flipped momentum, $( U^{(F)}_\mu(x), -P^{(F)}_\mu(x) )$, the counterpart of
the first configuration, $( U'^{(I)}_\mu(x), -P'^{(I)}_\mu(x) )$, is produced.
We require that $U'^{(I)}_\mu(x)$ is the same as $U^{(I)}_\mu(x)$ to satisfy
detailed balance.

Exact reversibility is broken in two ways in the numerical simulation.  Due to
round-off errors, the gauge links become non-unitary and are reunitarized at
the end of each trajectory.  This is not a reversible process, but only causes
small changes in the link elements for the parameters
employed in this simulation (This statement quantified below).  As mentioned
previously, a more worrying source of irreversibility is the chronological
invertor. Unless the convergence criteria is stringent enough so that the
solution of the CG is effectively independent from the forecast vector, the
force from the pseudo-fermion action is different from the corresponding
flipped fermion force in the reversed trajectory, and reversibility breaks
down.  Our convergence criteria in the CG is implemented so that the relative
norm of the preconditioned residual vector is equal to or less than a small
number, $R_{conv}$,
\bea
\mbox{\rm res}_{CG}& =& { \frac{ | (\tilde{D}_{DWF}^\dagger \tilde{D}_{DWF})  \chi - \Phi_F | 
 }{ | \Phi_F | }} ~ \leq R_{conv},
\label{eq:resCG}
\eea
where $\chi$ is the solution vector and $\Phi_F$ is the source vector.

We define $U^{(N)}\equiv U'^{(I)}$ to be the configuration obtained by
evolving $U^{(I)}$, with initial momentum $P$, $N/2$ steps followed by $N/2$
steps with the reversed momentum and have calculated the deviation between
$U^{(n)}$ and $U^{(I)}$ after $n$ steps along this path using two different
measures:
\bea
  d(U^{(n)}, U^{(I)}) &=& \sqrt{ \frac{1 }{ 4\cdot 18 N_{vol}} 
\sum_{x,\mu} \left\| U^{(n)}_\mu(x)-U^{(I)}_\mu(x)\right\|_2 }~~,\\
  d_{max}(U^{(n)}_\mu(x), U^{(I)}_\mu(x)) &=&  
\max_{x,\mu,i,j} |\left( U^{(n)}_\mu(x)-U^{(I)}_\mu(x) \right)_{ij}|\, ,
\label{imp:dees}
\eea
where, for a generic matrix $M$, $\left\| M \right\|_2$ represents the $l_2$
norm \cite{cd:Horn}.
These are plotted in Figure~\ref{fig:d_in_evol} for $N=20$, 40, 100, 200, 400,
and 1000 using a starting configuration from the $m_{\rm sea}=0.02$ ensemble
($d(U^{(n)},U^{(I)})$ and $d_{max}(U^{(n)},U^{(I)})$ are relatively
independent of the sea quark mass). The crucial issue is how small $R_{conv}$
must be so that the breaking of reversibility is negligible.  In
Figure~\ref{fig:rev_break_vs_res}, $d(U^{(N)},U^{(I)})$ and
$d_{max}(U^{(N)},U^{(I)})$ are plotted as a function of $R_{conv}$, for a
starting configuration in the $m_{\rm sea}=0.02$ ensemble and $N=102$; the
value we use in our simulations.  For $R_{conv} \geq 10^{-6}$ the MD is not
reversible. $d(U^{(n)},U^{(I)}_\mu)$ reaches the edge of floating point
accuracy at $R_{conv} = 10^{-7}$ but $d_{max}(U^{(n)}_\mu,U^{(I)}_\mu)$ is
still resolved. For $R_{conv} \leq 10^{-8}$, both deviations are below single
precision accuracy and comparable to the deviations due to reunitarization
which are shown by the dotted lines in the graph.  From these checks we choose
$R_{conv} = 10^{-8}$ as the CG convergence criterion in our simulation.

\subsection{Multiple Time-scale Leapfrog evolution}
\label{imp:multi}

In this section we discuss the multiple time-step leapfrog integration scheme
of \cite{Sexton:1992nu}. While this technique was not used in the main part of
this work, a small study was performed to test its usefulness, the results of
which will be included here as they are both encouraging and instructive.

In \cite{Sexton:1992nu} a procedure was outlined for constructing leapfrog
integrators for which a different time-step size could be used for different
parts of the molecular dynamics Hamiltonian. The parts of this Hamiltonian
which produce the dominant contribution to the leap-frog integration
discretization error may then be treated with a finer discretization than the
remainder. In the case where the dominant contribution to the discretization
error comes from the gauge piece of the Hamiltonian, for which the molecular
dynamics force term is relatively inexpensive to compute, a large improvement
in the efficiency of the algorithm is possible.

For the standard actions this does not seem to be the case. However, there is
some evidence that when using the modified force term described in Section
\ref{imp:nforce} the dominant errors are coming from the gauge
part of the action. Figures \ref{fig:hamold} and \ref{fig:hamnew} contrast the
individual contributions of the gauge, momentum, fermion and Pauli-Villars
terms to the total change in the Hamiltonian over a trajectory for the small
volume, large step size  runs tabulated in Table~\ref{imp:nforce_small}.  As
can be seen, for the old force term this change is approximately the same size
for all contributions, while for new force term the gauge and momenta
contributions are much larger than the (combined) fermion contribution. There
is also a noticeable skew in the distributions for the gauge and momenta
pieces of the Hamiltonian, with a positive change in the gauge Hamiltonian
being rejected much more often than a negative change.

To test if the discretization error for the new force term is, indeed,
dominated by the contributions from the gauge and momenta pieces of the
Hamiltonian, we have performed a trial of the multiple gauge-step leapfrog
integrator over 45 trajectories, starting from trajectory 1505 of the
$m_{\rm sea}=0.04$ evolution. Over this range the standard algorithm had an acceptance
of 56\%. Performing two gauge integration steps for every fermion integration
step the acceptance moved up to 91\%, confirming the gauge momenta pieces are
the dominant cause of the discretization error. While we would like to exploit
this fact by using a large fermion step-size combined with a small gauge
step-size, we have found that in the few tests that we have undertaken the
performance of the chronological invertor degrades as the fermion step-size
increases by an amount that almost completely offsets the fewer number of
inversions needed.  However, this is an approach worthy of a more extended
study, and even if it does not allow the move to larger fermion step-sizes,
the gain in acceptance we have observed is appreciable.


%% file: text_sections/simulation_details.tex
%
%

As mentioned previously, our simulations have been made using the standard
domain wall Dirac operator \cite{Furman:1995ky}, combined with the
Paulli-Villars field with the action introduced in
\cite{Vranas:1997da}. For each of our three dynamical ensembles we employ two
dynamical flavors, with $L_s=12$ and $M_5=1.8$ (we use the notation and
conventions introduced in \cite{Blum:2000kn} for the domain wall fermion
action), on lattices of size $16^3 \times 32$. The fermion boundary conditions
are periodic in the spatial directions and anti-periodic in the time direction.
These ensembles have bare quark masses of $0.02$, $0.03$ and $0.04$. The basic
HMC parameters are tabulated in Table~\ref{imp:evols}; all Dirac matrix
inversions were performed using the conjugate gradient algorithm with a
stopping condition of $10^{-8}$.

We use the DBW2 gauge action \cite{Takaishi:1996xj} with $\beta=0.80$, the
same action we have used in previous quenched studies
\cite{Aoki:2002vt,Noaki:2004}. This action approximates the renormalization
group trajectory by using the standard plaquette, $P_{1\times1}$, and a
$1\times 2$ rectangular plaquette, $P_{1\times2}$:
\begin{eqnarray}
S_g &=& - \frac{\beta}{3}\,\left(
\sum_x\, \left(1 - 8c_1\right)\,{\rm Tr}\,P_{1\times1}
+ c_1 \,{\rm Tr\, P_{1\times2}}\right) \, .
\end{eqnarray}
This form was originally suggested by Iwasaki
\cite{Iwasaki:1983ck,Iwasaki:1983cm,Iwasaki:1984cj}, who, using a perturbative
approach, estimated a value of $c_1=-0.331$. In \cite{deForcrand:1999bi} a
non-perturbative estimate, of $c_1=-1.4069$, was made in the quenched
approximation. While we are working with dynamical fermions, this is the value
we use; our intent being to exploit the improvement in the chiral properties
of domain wall fermions that this gauge action provides \cite{Aoki:2002vt},
rather than to be as close as possible to the renormalized trajectory. Of
course, it is possible that the effects of the determinant will negate any such
improvement. This issue is discussed further in section \ref{sec:chiral}.

%% file: text_sections/thermalizations.tex
%
%

In this section we discuss the related issues of thermalization and
auto-correlations for quantities calculated on our ensembles. Each evolution
started from an ordered lattice, running for $O(10)$ trajectories without the
accept/reject step applied, and leaving $O(600)$ trajectories for the lattice
to thermalize. The precise numbers for each evolution are given in Table
\ref{therm:evos}. As a check of the number of trajectories needed to
thermalize the configurations we have calculated chiral condensate
$\langle\overline{q}{q}\rangle(m_{\rm val} = m_{\rm sea})$ on every
trajectory, using a single hit of a random source. Figure \ref{fig:pbptherm}
shows this, together with the average value for trajectory 3000 and above,
which agree well with each other by the 600th trajectory.

For the results presented later in this work we use 94 configurations from
each ensemble, separated by 50 HMC trajectories. For the $m_{\rm sea}=0.02$
and $m_{\rm sea}=0.03$ ensembles these configurations are taken sequentially
from the thermalization point given in Table \ref{therm:evos}. For the $m_{\rm
sea}=0.04$ evolution we leave a gap of 250 trajectories, starting from
trajectory 1775, due to a hardware error on trajectory 1772 that was not
detected until after the entire evolution was finished. The trajectory passed
the accept/reject step, and no effect was seen in any of the physical
observables that we have calculated; nevertheless, to be cautious, we have
allowed this gap of 250 trajectories for the evolution to re-thermalize. Table
\ref{sim:basic} gives the results for the bare lattice values of the chiral
condensate (for $m_{\rm val} = m_{\rm sea}$), and the $r\times t$ on-axis
Wilson loop, $\langle W(r,t)\rangle$, with $(r,t)=\{(1,1),(1,2),(2,2),(3,3)\}$
for these sets of configurations. While we do not use these values in the rest
of this work, we include them here as they may be of utility for others trying
to work at this set of parameters.

To test this spacing of 50 trajectories we have calculated the
auto-correlation function, defined for some observable, ${\cal O}(t)$ (here
$t$ labels the trajectory, and runs from 1 to $N_{data}$), as 
\bea
  \rho(t) &=&  {\frac{1}{ N_{data}-t}}\sum_{t'=1}^{N_{data}-t}
\left({\cal O}(t') - \bar{\cal O}\right)
\left({\cal O}(t'+t) - \bar{\cal O}\right)~~, \\
\bar{\cal O} &=& {\frac{1}{ N_{data}}} \sum_{t'=1}^{N_{data}} {\cal O}(t') \ ,
\label{eq:acorrRho}
\eea
on the $m_{\rm sea}=0.02$ ensemble for the plaquette and the correlation
function of the time component of the local axial vector current at
time slice 12.  The former was measured every trajectory, while the latter was
calculated every 10 trajectories using a Coulomb gauge-fixed box source of
size 10 and a point sink.  Figures \ref{fig:autoplaq} and
\ref{fig:autoax} show both the normalized auto-correlation function, $\rho(t)/\rho(0)$, and the integrated
auto-correlation length,
\be
  \tau_{int}(t_{max}) = {\frac{1}{ 2}} + 
    \frac{1}{\rho(0)}\sum_{t=1}^{t_{max}} \rho(t) \ ,
\label{IntAcorr}
\ee
for these two quantities, versus the separation in trajectories and the
maximum separation over which correlations were calculated, $t_{max}$,
respectively. The quoted error on the integrated auto-correlation length is
calculated using a jackknife procedure: as is standard, each jackknife sample
is constructed by removing a contiguous group of data-points, covering
$N_{block}$ trajectories, from the available data. On the $j^{th}$ jackknife
sample ($j
\in [0,N_{data}/N_{block})$) we construct an estimate of the auto-correlation
function,
\bea
\rho_j(t) &=& \frac{1}{N_{sum}}
\left(
\sum_{t'=1}^{j N_{block}-t} \left({\cal O}(t') - \bar{\cal O}_j\right)
\left({\cal O}(t'+t) - \bar{\cal O}_j\right) \right.
\label{jint}
\\ \nonumber
&+&
\left.
\sum_{t'=(j+1)N_{block}+1}^{N_{data}-t} \left({\cal O}(t') - \bar{\cal O}_j\right)
\left({\cal O}(t'+t) - \bar{\cal O}_j\right)
\right) \ ,
\\ 
\bar{\cal O}_j &=& {\frac{1}{ N_{data}-N_{block}}} 
\left(
\sum_{t'=1}^{jN_{block}} {\cal O}(t')
+\sum_{t'=(j+1)N_{block}+1}^{N_{data}} {\cal O}(t')
\right)
\eea
where $N_{sum}$ is the total number if terms in the two summations; this is
not simply $N_{data}-N_{block}-t$, as if $t$ is greater than or equal to
$jN_{block}$ or $N_{data}-(j+1)N_{block}$ we must drop the first or last
summation, respectively, in Equation \ref{jint}. Estimates of the integrated
auto-correlation length on every jackknife sample are constructed from
Equation \ref{IntAcorr} and used to calculate the error in the standard
fashion. Ideally, we would like to work in the regime such that $N_{block} \gg
t_{max}, \tau_{int}$ for a value of $N_{block}$ which leads to an appreciable
number of jackknife samples. Here we use $N_{block}=100$.  This leads to
acceptable number of jackknife samples ($\sim 50$), but may be too short for a
solid estimation of the error for the axial-axial data.  As can be seen from
Figures \ref{fig:autoplaq} and \ref{fig:autoax}, the integrated
auto-correlation length plateaus at $\approx 3$ for the plaquette and $\approx
35$ for the axial-axial correlator. While these values would suggest that the
auto-correlations may be under control for our configurations, caution should
be taken both because their extraction is insensitive to correlations longer
than $O(100)$ trajectories, and because the auto-correlation length depends on
the quantity.

\newcommand{\plaq}{
   \put(-.5,-.5){\line(1,0){1}}
   \put(.5,-.5){\line(0,1){1}}
   \put(.5,.5){\line(-1,0){1}}
   \put(-.5,.5){\line(0,-1){1}}
}
\newcommand{\recly}{
   \put(-1.,-.5){\line(1,0){2}}
   \put(-1.,.5){\line(1,0){2}}
   \put(-1.,-.5){\line(0,1){1}}
   \put(1.,-.5){\line(0,1){1}}
}

\newcommand{\recst}{
   \put(-.5,-1.){\line(1,0){1}}
   \put(-.5,1.){\line(1,0){1}}
   \put(-.5,-1.){\line(0,1){2}}
   \put(.5,-1.){\line(0,1){2}}
}
\newcommand{\clover}{\setlength{\unitlength}{.4cm}\raisebox{-.4cm}{
   \begin{picture}(2.7,2.3)(-1.15,-1.15)
   \multiput(-1.15,-1.15)(1.15,1.15){2}{\begin{picture}(1.15,1.15)(-.575,-.575)
   \plaq\end{picture}}
   \multiput(-1.15,0)(1.15,-1.15){2}{\begin{picture}(1.15,1.15)(-.575,-.575)
   \plaq\end{picture}}
   \put(0,0){\circle*{.3}}
   \end{picture}}}
\newcommand{\clrecly}{\setlength{\unitlength}{.4cm}\raisebox{-.4cm}
{  \begin{picture}(2.6,2.6)(-1.55,-1.15)
   \multiput(-1.15,-0)(2.15,0.){2}{\begin{picture}(1.15,1.15)(-.575,-.575)
   \recly\end{picture}}
   \multiput(-1.15,-1.15)(2.15,0.){2}{\begin{picture}(1.15,1.15)(-.575,-.575)
   \recly\end{picture}}
   \put(.5,0){\circle*{.3}}
   \end{picture}}}
\newcommand{\clrecst}{\setlength{\unitlength}{.4cm}\raisebox{-.4cm}
{   \begin{picture}(2.6,2.6)(-1.55,-1.55)
   \multiput(-2.15,0)(1.15,0){2}{\begin{picture}(1.15,1.15)(-.575,-.575)
   \recst\end{picture}}  
   \multiput(-2.15,-2.15)(1.15,0){2}{\begin{picture}(1.15,1.15)(-.575,-.575)
   \recst\end{picture}}  
   \put(-1.0,-.5){\circle*{.3}}
   \end{picture}}}

An important example of a quantity which displays correlations over the scale of
many hundreds of trajectories is the topological charge. To calculate the
topological charge the lattices are first smoothed by applying 
20 steps of APE smearing with a coefficient of 0.45; then the
topological charge is measured using a discretization of the operator
\begin{equation}
Q_{\rm top} = \frac{1}{32\pi^2} \epsilon_{i j k l}
\mathrm{Tr}\left[
F_{ij}F_{kl}
\right] \,,
\label{therm:qtop}
\end{equation}
which is based upon two
definitions of the lattice field strength tensor,  $F_{\mu \nu}$: that using the 
clover leaf pattern for the simple ($1\times1$) plaquette,
\be
F^C_{\mu\nu} = -\frac{i}{4} \; {\rm An} \left ( \clover \right ) \, ,
\label{therm:fc}
\ee
and the analogous quantity for the $2 \times1$ rectangle,
\be
F^R_{\mu\nu} = -\frac{i}{16} \; {\rm An} \left ( \clrecly \hspace{0.8cm} 
+ \ \ \clrecst \right ) \, ,
\label{therm:fr}
\ee
where
\be
{\rm An} \left( M \right) = \frac{1}{2}
\left(
M - M^{\dagger} 
\right)\, .
\ee
While equations \ref{therm:fc} and \ref{therm:fr} may simply be substituted into
equation \ref{therm:qtop} to obtain discretized expressions for the topological
charge (which we denote $Q^{C}$ and $Q^{R}$, respectively),
they may also be combined such that the $O(a^2)$ errors cancel
between the two definitions. This gives the 
improved topological charge operator \cite{AliKhan:2001ym}:
\be
Q^{\rm imp}_2
= \frac{5}{3} Q^{C} - \frac{2}{3} Q^{R} \, .
\label{therm:a2}
\ee
We may also construct a definition built up from a classically $O(a^2)$
improved field strength tensor \cite{Bilson-Thompson:2002jk}:
\be
Q^{\rm imp}_4
= \frac{25}{9} Q^{C} - 
\frac{20}{9}\frac{1}{32\pi^2} \epsilon_{i j k l}
\mathrm{Tr}\left[
F^C_{ij}F^R_{kl}
\right]
+
\frac{4}{9} Q^{R}\, ,
\ee
which is also $O(a^2)$ improved, but which has different $O(a^4)$ errors.  We
use this last operator to quote our values of the topological charge, although
there would be little difference if we had decided to use the operator of
Equation \ref{therm:a2}: for the 2565 topological charge measurements made on
our dynamical ensembles, we observed only 3 cases in which these two
discretizations differed when rounded to the the nearest integer.  We have
also compared our values to those calculated using the topological charge
operator of \cite{DeGrand:1998ss}: for a test over 418 measurements from the
$m_{sea}=0.03$ evolution we found $90\%$ agreement on the nearest integer,
with the largest absolute difference being 0.76. Figure \ref{fig:topo} shows
these topological charge values versus trajectory for all of the
ensembles. Note the correlations over many hundreds of HMC
trajectories. Although the DBW2 action suppresses topology changing
configurations (this is the reason for its improved chiral properties in
conjunction with domain wall fermions)\cite{Aoki:2002vt}, considering the
relatively large level of explicit chiral symmetry breaking observed in our
simulations (see sections~\ref{sec:PhysicalResults} and \ref{sec:chiral}), we
believe the long correlation length observed here is due mainly to the small
trajectory length HMC algorithm which is known to move slowly between topological
sectors\cite{Ilgenfritz:2000nj}.


%% file: text_sections/physical_spect.tex
\subsection{hadron spectrum and decay constants}

In this work hadron masses are extracted using standard covariant fits (see,
{\it e.g.} \cite{Toussaint:1989fk}) of two-point correlation functions at
relatively large Euclidean time interval from a single meson or baryon source
located at $t=0$, so that only a single exponential corresponding to the
ground state is fit in each case. The source is either a Coulomb gauge-fixed
wall source or a plain wall source. The latter corresponds to averaging over a
point source on a time-slice after averaging over the gauge field ensemble. We
use only point sinks.  The wall sources generally have better overlap with the
ground states in which we are most interested, and therefore lead to more
accurate determination of the particle masses. However, the decay constants
are more easily obtained from the point source matrix elements.  In the case
of the point source, we also computed the correlation functions using a source
located at $t=16$, {\it i.e.} the maximum distance from the original source on
our periodic lattice, in order to improve our statistics.  We consider zero
momentum states only. All quoted statistical errors are estimated by fitting
the correlation functions under a standard single-elimination jackknife
procedure.

We begin with the calculation of the residual mass, $m_{\rm res}$, in order to
define the chiral limit, $m_f=-m_{\rm res}$. $m_{\rm res}$ is the additive
renormalization of the quark mass caused by mixing between the left and right
handed fermions localized on opposite boundaries of the fifth dimension and is
defined from the explicit chiral symmetry breaking term in the axial
Ward-Takahashi identity for domain wall fermions\cite{Blum:2000kn}.  We refer
the reader to \cite{Blum:2000kn,Aoki:2002vt} for details of the method to
calculate $m_{\rm res}$.  In Figure~\ref{fig:mres plateau} the effective residual
mass, $R(t)$, for $m_{\rm sea}=m_{\rm val}=0.02$ is shown for each time slice (because
the correlator is periodic, we fold the correlation functions about the
mid-point of the lattice in the time direction which is why our plots range
from $t=0$ to 16). It falls by about a factor of 2 over the first couple of
times slices and then remains constant. This non-local effect of the extra
dimension of domain wall fermions is well known
\cite{Aoki:2002vt,Golterman:2003qe}. Taking the error-weighted average over
time slices 4 through 16 and linearly extrapolating the $m_{\rm sea}=m_{\rm val}$
points to $m_f=0$, we find $am_{\rm res}=0.001372(49)$ (see Figure~\ref{fig:mres
extr}). We also use the axial Ward-Takahashi identity and the partially
conserved axial-vector current to extract the renormalization factor $Z_A$
appearing in Eq.~\ref{eq:decay ax}\cite{Blum:2000kn}. We find
$Z_A=0.75734(55)$, defined in the chiral limit, and extracted from a linear
fit to the mass dependence of the fully dynamical points (see
Figure~\ref{fig:za}).

Next, we turn to the vector meson mass. In principle, the vector particle can
decay hadronically into two pseudo-scalars since vacuum sea quark effects are
included in this two flavor simulation. However, in the regime we are working
the sea quarks are still relatively heavy, and, taking into account that two
pions must have relative orbital angular momentum $L=1$ for the decay to be
allowed, it is easy to note that our vector mass is always below the threshold
for this decay. Thus the two pion states, while present in this channel,
represent excited states. 

Figure~\ref{fig:mvec-eff} displays the effective mass in the vector channel,
$m_{\rm sea}=m_{\rm val}$, averaged over all three spatial
polarizations. Tables~\ref{tab:vector mass 0.02}~-\ref{tab:vector mass 0.04}
give the fitted masses from the wall-point correlation functions for all
combinations of sea and valence quark masses for a set of time slice ranges
chosen based on the effective mass plateaus shown in
Figure~\ref{fig:mvec-eff}.  The masses are extracted from a fit to
\bea
\label{eq:corr func}
\lim_{t\to\infty} G(t) &=& A\left(e^{-m\,t} + e^{-m\,(N_t -t)}\right)\\
&=& {2A}e^{-m(N_t/2)} \cosh{(m\,(N_t/2-t))}.
\eea
The mass plateaus are rather poor, especially for $m_{\rm sea}=0.02$, which is
reflected in the values of $\chi^2$ for the fits. Additional statistics are
desirable, though we note our trajectory lengths are already quite long for
dynamical fermion simulations ($\sim 5000$).  Table~\ref{tab:vector mass
chiral limit} shows the results of performing two fits to the data in
Tables~\ref{tab:vector mass 0.02}-\ref{tab:vector mass 0.04}: a linear fit to
the fully dynamical, $m_{\rm sea}=m_{\rm val}$, points,
\be
M_{vector} = a + b ( m_{sea} + m_{res} ) \, ,
\ee
and a partially quenched fit to all the data, 
\be
M_{vector} = a +  b ( m_{sea} + m_{res})  + c ( m_{val} + m_{res} ) \, .
\ee
As can be seen from the consistancy of these two fits, the linear ansatz works
well for this data. Considering the small mass range available, we do not
attempt more sophisticated fits.  The linear extrapolation of the three
(uncorrelated) points $m_{\rm sea}=m_{\rm val}$ to $\bar m=(m_u+m_d)/2$ (later, when we discuss
the pseudo-scalar meson, we will see how $\bar m$ is determined) yields the
value of the vector meson mass corresponding to the physical $\rho$
meson. From $M_\rho = 770 {\rm MeV}$ we find
\be
a^{-1}_{\rho}=1.691(53) \, {\rm GeV}.
\ee

A similar analysis has been carried through for the nucleon mass. This is
calculated from Coulomb gauge-fixed wall-point two point functions of the
interpolating field, $J_N=\epsilon^{ijk}(u^{i T}C\gamma_5d^j)u^k$, using a
positive parity projection.  The results of a fit to a single exponential
($Ae^{-mt}$) are shown in Table \ref{tab:nucleon mass}. Taking the negative
parity projection we may also extract the mass of the $N^*$, the parity
partner of the nucleon, the results for which are tabulated in Table
\ref{tab:N* mass}. Figure \ref{fig:mnuc} shows both these quantities for the
dynamical, $m_{\rm sea}=m_{\rm val}$, points. In Figure \ref{fig:ape} we show
the APE plot ($M_I/M_\rho$ vs.\ $(M_\pi/M_\rho)^2$ where $M_I$ is represents
the mass of the $N$ or $N^{*}$), together with the results from quenched DWF
with DBW2 gauge action \cite{Aoki:2002vt} for the nucleon. We note that the
dynamical and the quenched values are in good agreement. Of course, in a
comparison between the nucleon and rho masses, we must bear in mind that the
rho is stable for all values of the quark masses studied here, which
introduces a systematic error that could easily be 10\% or more in the ratio
$m_N/m_\rho$.  However, we are encouraged to think that our error in $m_\rho$
may be smaller than this, since the value of the lattice spacing determined
from the rho mass agrees quite closely with that implied by our calculation of
$f_\pi$ and of the Sommer scale, $r_0$, as discussed later in this section and
in Section~\ref{subsec:staticquarkpotential}, respectively.

In Figures \ref{fig:mnuc} and \ref{fig:ape}, we show an extrapolation to the
physical point, $m_f=\bar{m}$. To perform this extrapolation we have used a
linear ansatz for the quark mass dependence of the nucleon. The extrapolated
values shown in the figures are taken from a fit to just the dynamical points,
the results of which are given in Table\ref{tab:chiral baryon masses}.  As can
be seen, the nucleon mass is two standard deviations (9\%) larger than
experiment. Note that our spatial lattice size, $L\simeq 1.9$ fm is not small
enough that we would expect to see significant finite volume effects in our
data for the quark masses we are using; $LM_\pi\simeq 4.7$ for our lightest
mass. Systematic numerical studies
\cite{Orth:2003nb,Orth:2004:thesis,AliKhan:2003cu}, as well as theoretical
calculations \cite{Beane:2004tw}, on the finite volume effects in the nucleon
mass with two flavor Wilson fermions indicate a few percent finite volume mass
shift with similar parameters to our lightest point. However, as the $\rho$
mass receives similar finite volume effect\cite{Orth:2003nb,Orth:2004:thesis},
the ratio at the physical limit can shift as little as $\simle 1 \%$.
Assuming that holds also for DWF, finite volume effect may be responsible for
a minor part of the discrepancy. In addition, the systematic error associated with
the linear fit is at least of the scale of this discrepancy. This can be seen
by comparing the diagonal extrapolation and two stage linear extrapolation,
where the valence limit $m_{\rm val}\to \bar{m}$ is taken first with fixed
$m_{\rm sea}$, then dynamical extrapolation $m_{\rm sea}\to \bar{m}$ is performed.
The result is shown in Table
\ref{tab:chiral baryon masses}, Table \ref{tab:mass ratio}, 
and in Figure~\ref{fig:ape} (triangle down). The difference between the two
extrapolations indicates inapplicability of the linear ansatz.  This ansatz
clearly does not properly account for observed mass dependence of the
nucleon. However, the statistical accuracy on the nucleon mass needs to be
improved before more appropriate fits can be used
\cite{Jenkins:1991ts,Leinweber:1999ig,Bernard:2003rp,Procura:2003ig,Beane:2004ks}.

The physical $N^*$ mass is studied similarly. The diagonal and two stage
extrapolations yield different results, but both are consistent with
experiment to within one standard deviation (10\%) below and above.  We note that
the $M_{N^*}$ at the lightest simulated point might suffer from the threshold
effect, since $M_N+M_\pi=1.045 (12)$ and $M_{N^*}=1.021 (71)$. However the
large statistical uncertainty prevents us from discussing further on this
point.

The analysis of the pseudo-scalar mesons will be somewhat more involved since
partially-quenched chiral perturbation theory is at our disposal, and there
are two channels, pseudo-scalar and the axial vector, which couple to the
physical pion and kaon states.
Figures~\ref{fig:mpi-GAM-5-eff}-\ref{fig:mpi-dmes11-eff} show the
pseudo-scalar meson effective mass computed from the point-point and
wall-point correlation functions. In each case we show $m_{\rm sea}=m_{\rm val}$;
however the cases where $m_{\rm sea}\neq m_{\rm val}$ are similar. As expected, the
point-point correlators exhibit plateaus that emerge at later times compared
to the wall-point ones.  As the quark mass increases, even the latter plateau
emerges (from below) at rather late times. The statistical errors for the
point-point axial-vector case are somewhat larger than for the others,
especially at smaller valence quark mass.  Based on the above plateaus we fit
the correlation functions from $t_{min} =9$ to $t_{max}=16$. The results do
not change by more than one standard deviation when $t_{min}$ is changed by
two units in either direction, provided $\chi^2$/dof remains acceptable which
was almost always the case. Results for all quark mass combinations are
summarized in Tables~\ref{tab:pseudo-scalar mass}-\ref{tab:axial-vector mass
point-point}.  Fitted meson masses among these four methods are mostly
consistent within one (statistical) standard deviation, except for the
lightest point ($m_{\rm sea}=0.02$, $m_{\rm val}=0.01$) where the mass in the
point-point pseudo-scalar channel is almost two standard deviations higher
than the rest. This may indicate that our statistical errors are
underestimated due to the limited length in simulation time of our evolution
($\sim 5000$ trajectories).

The pseudo-scalar decay constant $f_{PS}$ is obtained from the point-point
correlation function, either directly (axial-vector), or through the
Ward-Takahashi identity (pseudo-scalar)~\cite{Blum:2000kn,Aoki:2002vt}. When
using a point source, $A$ in Eq.~\ref{eq:corr func} is proportional to the square of
the decay constant.  Specializing to the pseudo-scalar,
\bea
A &=& \frac{|\langle 0 | \chi | PS\rangle|^2}{2 M_{PS}},
\eea
where $\chi=\bar\psi \gamma_\mu\gamma_5\psi$ for the axial vector and 
$\bar\psi \gamma_5\psi$ for the pseudo scalar, and $M_{PS}$ is the pseudo-scalar mass. 
We have for the {\it lattice} matrix elements
\bea
\label{eq:decay ax}
\langle 0 |\bar \psi \gamma_4\gamma_5\psi|PS\rangle &=& f^{lat}_{PS}\,M_{PS} = \frac{f_{PS}}{Z_A}\, M_{PS},\\
\label{eq:decay ps}
\langle 0 |\bar \psi \gamma_5\psi|PS\rangle &=& f_{PS}\, \frac{M_{PS}^2}{2 (m_{val}+m_{res})},
\eea
where the first equation defines the decay constant and the second results
from it and the use of the Ward-Takahashi identity. The finite renormalization
factor $Z_A$ appears in the first equation as we use the local definition
of the flavor non-singlet axial vector current which is not
(partially-)conserved. On the other hand, no such factor appears in
Eq.~\ref{eq:decay ps} because the combination $(m_{\rm val}+m_{\rm res})\langle 0
|\bar \psi \gamma_5\psi|PS\rangle$ is a renormalization group invariant,
protected from renormalization by the Ward-Takahashi identity for all $a$. The
bare quark mass associated with the field $\psi$ is $m_{\rm val}+m_{\rm res}$.

Tables~\ref{tab:pseudo-scalar mass point-point}~and~\ref{tab:axial-vector mass
point-point} show the results for $f_{PS}$, which agree well between the two
methods except for the heaviest valence mass point at $m_{\rm sea}=0.02$ where
there is a $\sim1.5$ standard deviation discrepancy between central
values. Note that the errors from the axial-vector correlator are 2-3 times
larger than from the pseudo-scalar, so we quote central values and statistical
errors from the latter.

Next, we wish to extrapolate our results to the physical light quark masses,
$\bar{m}=(m_u+m_d)/2$ and $m_s$ (our simulation is not at a level of precision
sufficient to account for isospin breaking effects arising from $m_u\neq m_d$;
thus we work with degenerate up and down valence quarks). Since we have
simulated a theory where the strange quark is quenched and $m_{\rm sea}\neq
m_{\rm val}$, the proper framework for such an extrapolation is partially-quenched
chiral perturbation theory\cite{Bernard:1993sv}. The next-to-leading order
(NLO) partially quenched chiral perturbation theory formulae for the
pseudo-scalar mass squared and decay constant made from degenerate valence
quarks are
\cite{Golterman:1997st,Laiho:2002jq}.
\begin{eqnarray}
M^{2}_{PS(1-loop)} &=& M^{2}\left(1+\frac{\Delta M^2}{M^2}\right)
\label{eq:nlo mpisq}
\end{eqnarray}
\begin{eqnarray}
f_{PS(1-loop)}&=& f \left(1+\frac{\Delta f}{f}\right)\\
\label{eq:nlo decay}
\frac{\Delta M^2}{M^2} & = &
\frac{2}{N}\left[\frac{M^2-M^2_{SS}}{16\pi^2f^2}+\frac{2M^2-M^2_{SS}}{M^2}
A_0(M^2)\right] - \frac{16}{f^2} [(L_5 - 2L_8)M^2 \nonumber \\ &&
+ (L_4-2L_6) NM^2_{SS}],
\label{eq:nlo fit 1}\\
\frac{\Delta f}{f} & = & -NA_0(M^2_{vS}) + \frac{8}{f^2} (L_5 M^2
+ L_4 NM^2_{SS}), 
\label{eq:nlo fit 2}
\end{eqnarray}
with 
\begin{eqnarray}
M^2 &=& {2\,B_0}\,(m_{val}+m_{\rm res}),\\
M^2_{SS} &=& {2\,B_0}\,(m_{sea}+m_{\rm res}),\\
M^2_{vS} &=& \frac{(M^2+M^2_{SS})}{2},\\
A_0(M^2)&= &\frac{1}{16\pi^2f^2} M^2 \ln \frac{M^2}{\Lambda_\chi^2}.
\end{eqnarray}
The subscript ``$S$" implies a sea quark inside the meson, $\Lambda_\chi$ is
the chiral perturbation theory cut-off scale, $f$ is the decay constant in the
chiral limit, and $L_i$ are Gasser-Leutwyler low energy constants that appear
in the ${\cal O}(p^4)$ chiral lagrangian of QCD.

We begin by fitting the pseudo-scalar meson mass to Eqs.~\ref{eq:nlo
mpisq}~and~\ref{eq:nlo fit 1}. To be fully consistent we should perform a
simultaneous fit of $M_{PS}$ and $f_{PS}$ as both fit functions depend on
$B_0$ and $f$. However, as we explain in the following, the $f_{PS}$ fit is
problematic. As such, for the the $M_{PS}$ fit we use a fixed value of
$af=0.078$, which is the result of a linear fit to the decay constant mass
dependence. The results are summarized in Table~\ref{tab:nlo mpisq fit} and,
for the wall-point axial vector, shown in Figure~\ref{fig:nlo mpisq fit}. We
have taken $\Lambda_\chi=1$ GeV; a change in this scale is absorbed into the
scale-dependent $L_i$, leaving the value of $B_0$ unchanged. The NLO formula
works well for quark masses up to about 0.04, where the fit quality begins to
degrade. From Table~\ref{tab:lin mpisq fit}, we note that a simple linear fit
(lowest order in chiral perturbation theory) to the $m = m_{\rm sea}=m_{\rm val}$
points,
\be
M_{PS}^2  = c + b (m+m_{\rm res}) \, ,
\ee
is consistent with the NLO fit. This fit and data are shown in
Figure~\ref{fig:lin mpisq fit}. In fact, the deviations from this simple linear
form are quite small. However, so is the contribution of the logarithm
predicted by NLO chiral perturbation theory. We conclude that our data are
consistent with NLO chiral perturbation theory for $m_f\simle 0.04$.

The NLO fit is constrained to vanish in the chiral limit, $m_f = -m_{\rm res}$, as
it must due to the universality of the low energy domain wall fermion
theory~\cite{Blum:2000kn}. Remarkably, the simple linear fit, which is not
constrained, extrapolates to $m_{PS}^2\approx 0$ (within 1 standard deviation
of the statistical error) at $m_f = -m_{\rm res}$ (see Figure~\ref{fig:lin mpisq
fit}).  This is in stark contrast to the situation in the quenched
approximation~\cite{Blum:2000kn,Aoki:2002vt}, where the vanishing point from a
simple linear extrapolation occurred when $m_f\approx -(2-3)\times m_{\rm res}$.
This we attributed to the presence of quenched chiral logarithms, $\sim
m_{PS}^2\,\ln{m_{PS}^2}$. Here Eq.~\ref{eq:nlo fit 1} shows the chiral
logarithms are much weaker, $\sim m_{PS}^4\,\ln{m_{PS}^2}$. Thus this
comparison of the quenched and $N_f=2$ theories nicely confirms the
predictions of chiral perturbation theory and the low-energy effective theory
of domain wall fermions\cite{Blum:2000kn,Blum:2001sr}.

Another interesting feature of Figure~\ref{fig:nlo mpisq fit} is that our
simulations happen to coincide with the region where sea quark mass effects on
$m_{PS}^2$ appear to be the greatest (in an absolute sense).  Near the chiral
limit, $m_{\rm sea}=$ constant curves approach each other since they must vanish
at the same place, and as $m_{\rm val}$ gets heavy, they also merge since the
heavy quarks are insensitive to the light sea quarks.

Now we discuss the determination of $\bar m$. $\bar m$ is found from the
intersection of the NLO fit to $(am_{PS})^2$ with the line $(am_{PS})^2 =
(am_{\pi})^2$ where $m_\pi$ is the mass of the neutral pion, 134.9766 MeV, and
the lattice spacing $a$ is set from the vector meson mass evaluated at $\bar
m$.  This procedure is performed iteratively until convergence (in practice
the number of iterations $\simle 5$).  We find $\bar m = 0.00017(11)$.
Similarly, we find $m_{DK} = 0.0225(15)$
, where $m_{DK}$ is the quark mass for which a pseudo-scalar meson made
of degenerate quarks has the same mass as the neutral kaon,
$m_{K^0}=497.672$ MeV.  Since the NLO formulas for $M_K^2$ and $f_K$ for
non-degenerate valence quarks depend on the same parameters as in
Eqs.~\ref{eq:nlo fit 1}~and~\ref{eq:nlo fit 2}, we make a 
determination of $m_s$ (and $f_K$) by extrapolating to this non-degenerate
limit\cite{Golterman:1997st,Laiho:2002jq}.
These equations read:
\begin{eqnarray}
\label{eq:nlo fit MK}
\frac{\Delta M^2_K}{M^2_K}  & = &
\frac{-1}{N(M^2_K-M^2_{\pi})}\left[(M^2_{\pi}-M^2_{SS})
A_0(M^2_\pi)+(-2M^2_K+M^2_\pi+M^2_{SS}) A_0(M^2_{33})\right]
\nonumber \\ &&
 -\frac{16}{f^2} [(L_5-2L_8) M^2_K +
(L_4-2L_6) N M^2_{SS}]
\label{eq:nlo fK}
\end{eqnarray}
\begin{eqnarray}
\frac{\Delta f_K}{f} & = &
\frac{1}{N16\pi^2f^2}(m^2_K-m^2_{SS})-\frac{m^4_{\pi}+m^2_K(m^2_{SS}-2m^2_{\pi})}{2N(m^2_K-m^2_{\pi})}
\left(\frac{1}{m^2_{\pi}}A_0(m^2_\pi) -
\frac{1}{m^2_{33}}A_0(m^2_{33})\right) \nonumber \\ &&
-\frac{N}{2}( A_0(m^2_{uS})+A_0(m^2_{sS})) +\frac{8}{f^2} (L_5
m^2_K + L_4 Nm^2_{SS}), 
\end{eqnarray}
\begin{eqnarray}
M_{\pi}^2 &= & 2 B_0 (\bar m + m_{\rm res})\\
M_{K}^2 &= & B_0 (\bar m+ m_s+ 2m_{\rm res})\\
M_{33}^2 &= & 2 M_K^2 - M_\pi^2,
\end{eqnarray}
Note: Equations \ref{eq:nlo fit 1} and \ref{eq:nlo fit MK} have different
logarithmic terms.  The mass of the non-degenerate meson with one valence and
both dynamical quark masses fixed to $\bar m$, is also shown in
Figure~\ref{fig:nlo mpisq fit}, plotted versus the mass of the remaining
valence quark. Thus, using Eq.~\ref{eq:nlo fit MK} and the parameters
determined from the degenerate formula, we find $m_s=0.0446(29)$, our final
value. As a consistency check, we may take this value, together with the value
for $\bar{m}$, and use the results of the partially quenched fit to the vector
meson mass (Table \ref{tab:vector mass chiral limit}) to estimate the mass of
the $\phi$ meson. This gives a value of $M_{\phi} = 978(12) {\rm MeV}$;
comfortably close to the experimental value of 1019 MeV.  For comparison we
have also extracted the strange quark mass from the linear fit. Using the
three $m_{\rm sea}=m_{\rm val}$ points, one finds a value for the strange quark
mass of $m_{s} = 0.04177(64)$, $7\%$ smaller than the NLO value. The
difference is easily appreciated from inspection of Figures~\ref{fig:nlo mpisq
fit} and \ref{fig:lin mpisq fit} and demonstrates the significance of the NLO
analysis in this case.  Finally, it is important to note that $\bar m$ and
$m_s$ as quoted above are bare quark masses that correspond to the physical
$\pi$- and $K$- meson states; the renormalized quark mass is defined as
$Z_m(m+m_{res})$, where $Z_m$ is a scheme and scale dependent renormalization
factor.

We now move to the extraction of the decay constants from the point-point
correlators. Recall that the errors on the decay constant from the
axial-vector case are significantly larger than in the pseudo-scalar case, so
we focus on the latter; our conclusions do not depend on this choice.  In
contrast to the above analysis for $m_{PS}^2$, the NLO chiral perturbation
theory formula for the decay constant, Eqs.~\ref{eq:nlo decay} and \ref{eq:nlo
fit 2}, does not fit the data. The results of this fit are presented in
Table~\ref{tab:nlo decay fit}. Restricting $m_f \le 0.03$, the resulting fit
is shown in Figure~\ref{fig:nlo decay fit}.  While it reproduces the data used
in the fit reasonably well, it misses the larger mass points badly. Including
these points in the fit does not change the fit results significantly except to
increase the value of $\chi^2$. Note that the $m_{\rm sea}=\bar m$ line bends
down steeply as $m_{\rm val}\to 0$ which yields a value for $f_\pi= 100(10)$ MeV,
$\sim 30$\% smaller than the physical value.

There is just enough data to attempt a restricted NNLO fit, in which we
include all of the ${\cal O}(p^6)$ analytic terms: $C_1 m^2_{\rm sea}$, $C_2
m^2_{\rm val}$, and $C_{12} m_{\rm sea} m_{\rm val}$. This fit is not a systematic
application of chiral perturbation theory, as we do not include the
(un-calculated) logarithmic terms which also appear at this order. However,
performing this fit allows us to investigate the utility of moving to the next
order, and estimate the size of the terms needed.  The results of the NNLO fit
are summarized in Table~\ref{tab:nnlo decay fit}.  While the value of $\chi^2$
is acceptable, the errors on the fitted parameters are extremely large,
especially in the case of $C_1$.

The basic problem is that the coefficient of the log term, which has been
calculated analytically in the continuum (Eq.~\ref{eq:nlo fit 2}), is
inconsistent with our data, which appears to be mostly linear.  The additional
NNLO terms in the fit do act to reduce the effect of this log, but this
approach is both badly constrained and non-systematic. One interpretation of
our data is that the quark masses we are using are sufficiently heavy that the
NNLO terms are important relative to the NLO terms. In this case, the only way
to do better is to simulate at even lighter values of the quark mass where
(presumably) chiral perturbation theory works well.

There are several other interpretations: both lattice artifacts (${\cal
O}(a^2)$ in this study) and finite volume effects modify the coefficient of
the chiral log, possibly making it smaller. In the former case we expect this
effect to be of the order of a few percent, while including finite size
effects such as the ones in \cite{Bernard:2001yj} should not change the fits
significantly since our smallest mass still corresponds to $m_{PS} L\simge 3$.
Neither of these effects therefore seem large enough to explain the
discrepancy. In \cite{Aubin:2004fs}, it is also suggested that using a
physical parameter, such as $f_K$, as the chiral coupling rather than
$f$, leads to a better behaved chiral expansion.  This approach will have
the effect of significantly reducing the coefficient of the chiral logarithm
when applied to our data. If we allow the coefficient of the continuum chiral
log to be free parameter in the NLO equation, then we are able to make a good
fit ($\chi^2$/dof = $0.48(38)$).  However, we find that this coefficient is
$0.2(4)$, instead of the value 1 predicted by the continuum theory, which is a
large deviation.

A remaining alternative is to forsake most of the higher order terms and do a
simple linear fit. Again, this is not systematic, as this leaves out the
(known) NLO logarithmic term.  We fit the data with three independent terms,
$f$, and terms proportional to $m_{\rm sea}$ and $m_{\rm val}$ (in other words, we
simply set the coefficient of the log term in the NLO formula to zero):
\be
f_{PS} = af + c_1 \frac{m_1 + m_2 + 2m_{\rm res}}{2} + c_2 ( m_{\rm sea} + m_{\rm res} ) 
\ee
As mentioned before, such a fit -- the results of which we summarize in
Table~\ref{tab:lo decay fit} and Figure~\ref{fig:nlo decay fit} -- actually
describes the data quite well.  For comparison, $f$ is also extracted from a
simple two parameter linear fit to the $m_{sea}=m_{val}$ data points (see
Table~\ref{tab:lo decay fit} and Figure~\ref{fig:lin decay fit}); these two
linear fits agree well. Using the three parameter linear fit, restricted to
$m_{val}\le 0.04$, we find $f_\pi =134.0(42)$ MeV, $f_K=157.4$(38) MeV, and
$f_K/f_\pi =1.175(11)$, while the Particle Data Book gives $f_\pi=130.7$
MeV, $f_K=160$ MeV, and $f_K/f_\pi =1.224$. 

While the inability to fit our data to the predicted NLO chiral perturbation
theory form is discouraging, it is not unprecedented: this problem also exists
for the current Wilson fermion \cite{Aoki:2002uc,AliKhan:2001tx}, and
staggered fermion \cite{Aubin:2004fs} simulations. The former simulations, for
dynamical masses in the range such that $M_\pi/M_\rho = 0.6 - 0.8$, saw little
evidence for the existence of the chiral log, extracting final numbers using a
polynomial ansatz for the mass dependence (\cite{Hashimoto:2002vi} discusses
the matter further). The latter simulations, which include much lighter
dynamical masses, also cannot fit the NLO formula for comparable quark masses.
They quote final results from a restricted NNLO fit which, as above, leaves
out the logarithmic terms that should appear at that order.  The advantage of
this work, however, is that we have a much cleaner extraction of the decay
constants due to both flavour and chiral symmetry being respected at finite
lattice spacing, allowing us to be confident that this discrepancy from
continuum NLO chiral perturbation theory is a physical effect.

%% file: text_sections/physical_static.tex
\subsection{Static quark potential}
\label{subsec:staticquarkpotential}

In this section we discuss the extraction of the static quark potential,
$V(r)$, from the Wilson loop, $W(\vec{r}, t)$. To improve the signal we smear
the operator in the spatial coordinates using $n_{smear}$ applications
of APE smearing\cite{Albanese:1987ds}, 
\be
U'_i(x) \leftarrow {\rm Proj_{SU(3)}} \left[ 
U_i(x) + c_{smear} U^{(staple)}_i(x) \right] \ \ \ (i=1,2,3) \ , 
\ee
where $U^{(staple)}_i(x)$ is the sum of four spatial staples. Using these
smeared links we construct the spatial path between the infinitely heavy quark
and anti-quark by employing the Bresenham algorithm \cite{Bolder:2000un}.  

To cut down on short-term noise, we measure the static quark potential more
frequently than every 50 trajectories, although our final results are obtained
from a block-average over 50 trajectories.  Fig.~\ref{fig:Wloop-tauint} shows
the integrated autocorrelation times of a selection of Wilson loops, which are
$\le 25$ trajectories for $r\le8$.  In total we use 941, 559 and 473
configurations for $m_{\rm sea}=0.02, 0.03$ and $ 0.04$ respectively; these
configurations are separated by five trajectories for $m_{\rm sea}=0.02$,
while one in every ten trajectories is measured for $m_{\rm sea}=0.03$ and
0.04. Trajectories from 1775th to 2025th of $m_{\rm sea}=0.04$ are abandoned
due to the hardware error described in Section~\ref{sec:Thermalization}.

We use the theoretical formula 
\be
\frac{\vev{W(\vec{r},t)}}{\vev{W(\vec{r},0)}}
 = C(\vec{r}) \exp[-V(\vec{r}) t]
\ee
which should hold for large $t$, and (following \cite{Aoki:2002vt})
extract  $V(\vec{r},t)$ from 
\bea
V(\vec{r}) = \log\left[\frac{ \vev{W(\vec{r}, t)}}{ \vev{W(\vec{r}, t+1)}}\right]~~,
\label{eq:potextract}
\eea
together with $C(\vec{r})$ from
\be
C(\vec{r}) = {\frac{\vev{W(\vec{r}, t)}^{t+1}}{ 
              \vev{W(\vec{r},0)} \vev{W(\vec{r},t+1)}^{t}}} \ .
\label{eq:wloopoverlap}
\ee
To maximize the overlap factor, $C(\vec{r})$, we have explored the two
dimensional parameter space for the smearing, $(c_{smear}, n_{smear})$, in the
range $ 0 \le c_{smear}\le 1 $, $ 0 \le n_{smear} \le 60$.  We conclude
$(c_{smear},n_{smear})=(0.5, 20)$ is a reasonable choice for all of $r$ we
use.

The dependence of the potential to the temporal length of the Wilson loop is
carefully examined to control the contamination from excited states.  We found
$V(\vec{r})$ and $C(\vec{r})$ extracted from $t \geq 5$ and $r\leq 8$ is
relatively independent of $t$, and we therefore extract our results from this
range.  This can be clearly seen in Figure ~\ref{fig:pot_vs_t}, which shows
$V(\vec{r})$ for $m_{\rm sea}=0.02$.  The effect of the positivity violation in
improved gauge actions\cite{Luscher:1984is} for small $t$ is evident in the
graph: $V(r)$ extracted from $t < 3$ approaches its asymptotic value from
below for $r=\sqrt{2}$.  This was also observed in quenched simulations
\cite{Necco:2003vh}, and the effect on $C(\vec{r})$ is discussed in
\cite{RBC:KoichiLattice2004,Iwasaki:1996sn}; our selection of temporal length
is set as large as possible to exclude this lattice artifact.  The potential
extracted from $t=4,5$ and 6 is shown in Fig.~\ref{fig:pot} for the $m_{\rm
sea}=0.02$ ensemble versus $r$. We see no evidence of string-breaking for
large $r$.

For ease of comparison with other work, we also
fit the potential to the form
\bea
&&V(\vec{r}) = V_{cont}(r) - l \delta V(\vec{r}),~~ r = |\vec{r}| \ , \\
&&\ \ \ V_{cont}(r) = V_0 - { \frac{\alpha}{r} } + \sigma r  \ ,\\
&&\ \ \ \delta V(\vec{r}) = \left[{\frac{1}{\vec{r}}}\right] - {\frac{1}{ r}} \ ,
\label{eq:potfit}
\eea
where $[{1/ \vec{r}}]$ is the lattice Coulomb potential
\cite{Michael:1992nj, Lang:1982tj, Weisz:1982zw, Iwasaki:1984cj},  
\be
\left[\frac{1 }{\vec{r}} \right] =
\int_{-\pi}^{\pi} \frac{dk^3 }{ 8\pi^2}
\frac{\exp( i \vec{k}\cdot\vec{r}) 
}{
\sum_i \sin^2(k_i/2)- 4 c_1  \sum_i \sin^4(k_i/2) } \, . 
\label{eq:LatCoulomb}
\ee
This form is often used in an attempt to correct the breaking of the
rotational symmetry of the lattice.  The  correction
term, $\delta V(\vec{r})$, describes the corresponding data, $[ V^{\rm
(data)}(\vec{r})- V_{cont}(r) ] / l$, qualitatively well as seen in
Figure~\ref{fig:cmp_lat_coulomb}, in which $V_{cont}(r)$ and $l$ are from the
fit result using $t=5, r \in [\sqrt{3}, 8]$ on $m_{sea}=0.02$ ensemble.
We also note that the fit parameters, especially $\alpha$, 
become less sensitive to the selection of $r_{min}$ 
by adding the lattice Coulomb term, although the errors become larger.

The fit results (both with and without the lattice Coulomb term), 
together with Sommer scale\cite{Sommer:1994ce},
\be
r_0 = \sqrt{\frac{ 1.65 - \alpha }{\sigma }} \ ,
\ee
are presented in Table.~\ref{tab:pot}.  As can be seen in
Figure~\ref{fig:r0_vs_tmin}, a major source of systematic error is the
selection of $t$, with the $r_{max}$ dependence being almost negligible.  To
be precise, one can see a mild but continuous increase of $r_0$ as $t$ becomes
large.  At $t=7$, the signal to noise ratio becomes poor, and the statistical
error of the data at $t=6$ is less controlled as seen in
Figure~\ref{fig:pot_vs_t}.  We therefore choose to take our central values
from $t=5$ and $r\in [r_{min},r_{max}]=[\sqrt{3},8]$, quoting a systematic
error due to the selection of $t$, as well as fit range by the shift of
central values of these parameters. The selection is varied in either
direction in $t, r_{min}, r_{max}$ at once.  The fit ranges $r_{min}\in
[\sqrt{2},\sqrt{6}]$, $r_{max}\in [7,9]$, and $t=5,6$ are swept. More detailed
results including the direct evaluation of the force, $\nabla V(\vec{r})$, and
comparison with quenched simulation will be presented elsewhere
\cite{RBC:KoichiLattice2004,RBC:KoichiInPrep}.

Although in this paper we use the hadronic observables to set the lattice
scale, we may also set the scale from $r_0$.  The sea quark mass dependency of
$r_0$ is shown in Figure~\ref{fig:r0_vs_msea}. This mass dependence is
consistent with that observed using other fermion formulations
\cite{Allton:1998gi, Bali:2000vr, Bernard:2000gd, AliKhan:2001tx, Aoki:2002uc}. 
Linearly extrapolating to the chiral limit, we obtain a value of $r_0$ 
in lattice unit of
\be
r_0|_{m_{sea}\to-m_{res}} =  4.278(54) \left(^{+174}_{-011}\right)~~.
\ee
This value is obtained from the fit without the lattice Coulomb term and the
systematic error in the second parenthesis is estimated by the various choice
of fit ranges.  We note that the large positive shift of the central value
(+174) is due to the rise of the $r_0$ at $r_{min}=\sqrt{3}$ from $t=5$ to
$t=6$. This could be a sign of the remaining excited states contamination but
it is less conclusive with the comparably large statistical error at $t=6$ as
seen in Figure~\ref{fig:r0_vs_tmin}.  We note that $r_0$ depends on $m_{\rm
sea}$ so mildly that linear fit of $1/r_0$ yields a very close value,
$r_0|_{m_{\rm sea}\to-m_{\rm res}} = 4.287(58)$, at the chiral limit.  The fit
with the lattice Coulomb term yields similar value, $r_0 = 4.235(56)$.  All of
these three numbers are within the quoted statistical error.

For our final results we use the continuum fitting form, and a linear fit to
the mass dependence; taking $r_0=0.5$ fm we get 
\be
a^{-1}_{r_0}= 1.688(21)\left(^{+69}_{-04}\right)
\mbox{\rm GeV}~~,
\ee
which, in contrast to the situation in the quenched
approximation\cite{Aoki:2002vt}, is consistent with the value, $a_{\rho}^{-1}
= 1.691(53) {\rm GeV}$, extracted from the rho meson mass.

%% file: text_sections/physical_bk.tex
\subsection{Kaon B parameter}
\label{sec:bk}
When studying the mixing of neutral kaon in the standard model,
it is necessary to calculate the low energy matrix element, 
in QCD, of the effective weak interaction operator 
\be
O_{LL} = \bar{s}\gamma_\mu(1-\gamma_5)d\,\bar{s}\gamma_\mu(1-\gamma_5)d \, ,
\ee
between kaon states. This is usually quantified in terms of the kaon B-parameter,
\begin{eqnarray}
B_K &=& 
\frac{\langle \bar{K^0}|
\bar{s}\gamma_\mu(1-\gamma_5)d\,\bar{s}\gamma_\mu(1-\gamma_5)d
|K^0\rangle}
{\frac{8}{3}\,\langle \bar{K^0}|A_4|0\rangle\langle 0|A_4|K^0\rangle},
\label{eq:bk}
\\
&=&
\frac{\langle \bar{K^0}|
\bar{s}\gamma_\mu(1-\gamma_5)d\,\bar{s}\gamma_\mu(1-\gamma_5)d
|K^0\rangle}
{\frac{8}{3}\,f_K^2\,M_K^2}.
\end{eqnarray}

Before describing the lattice calculation of $B_K$, we address the
(unphysical) mixing of $O_{LL}$ with wrong chirality dimension six operators.
Such mixings are allowed in the chiral limit due to the explicit breaking of
chiral symmetry of domain wall fermions with finite $L_s$.  However, since
such explicit breaking is small, it is useful to understand the order of
magnitude of these mixings in terms of $a m_{\rm res}$. This allows us to
argue they may be neglected in our calculation.  The framework for
understanding this mixing within a low energy effective theory of domain wall
fermions was laid out in \cite{Blum:2000kn,Blum:2001sr}, but there the
explicit example of $B_K$ was not discussed. We begin by adding a spurion
field $\Omega$ to the low energy effective action for domain wall fermion QCD,
whose only effect is to modify the symmetry breaking terms in the action so
they transform in the same way as a conventional mass term. Each spurion field
in the modified action or effective weak operator carries a factor of $a
m_{\rm res}$.  After determining the dependence of the correlation function on
$\Omega$, and therefore $am_{\rm res}$, we set $\Omega\to1$ to recover the low
energy theory of domain wall fermions.  In the exact theory, the original
four-quark operator transforms as a (27,1) dimensional representation of
$SU_L(3)\times SU_R(3)$ chiral symmetry which must also be true for the
modified wrong chirality operators if they are to mix. All such
operators\cite{Allton:1998sm} have at least two right-handed quark fields, so
at least two spurion fields are needed to formally rotate these into
left-handed fields.  Thus, to lowest order the mixing is $\sim (a m_{\rm
res})^2$, or ${\cal O}(10^{-6})$; this is small enough that it may be
neglected in the present calculation. We note that the leading explicit chiral
symmetry breaking in the matrix element is still ${\cal O}(a m_{\rm res})$,
however. This is just the statement that the chiral limit for all low energy
observables is $m_f=-m_{\rm res}$\cite{Blum:2000kn}.  Once this trivial shift
is taken into account, the error on $B_K$ is again ${\cal O}((a m_{\rm
res})^2)$.

Details of the method used to calculate $B_K$ can be found
in~\cite{Blum:2001xb}. We employ the so-called conventional method
(Eq.~\ref{eq:bk}), since difficulties of the quenched approximation, such as
significant contamination by topological near-zero modes, do not
apply \cite{Blum:2001xb}.  The lattice B parameter for arbitrary quark mass,
denoted $B_{PS}$, is tabulated in Tables~\ref{tab:bp} and
\ref{tab:bp-nd2}-\ref{tab:bp-nd4}, for
degenerate and non-degenerate valence quark masses respectively.  The values
shown are averages over the time slice range for the operator insertion,
$t_{op}=14-17$.  The source and sink pseudo-scalar meson time slices were
fixed to $t=4$ and 28, respectively. These values were chosen based on the
quenched calculation in~\cite{Blum:2001xb} ($a^{-1}\approx 2$ GeV) where they
lead to reasonable plateaus, which is also the case here (we also averaged the
value of $B_{PS}$ over a larger time slice range, $t_{op} = 10-22$, and found
good agreement in all cases). This can be seen in Figures~\ref{fig:bp plateaus
0.02}-\ref{fig:bp plateaus 0.04}. Note however, that there are large
fluctuations in the plateaus for the smallest valence quark masses.

To extract $B_K$, we fit our data for $B_{PS}$ to the predictions of NLO chiral
perturbation theory, and extrapolate/interpolate to the physical quark masses.
Our preferred way to do this is, of course, to take the (non-degenerate )
limit
\be
m_{\rm sea}  \rightarrow  \overline{m} \ ; \
m_{\rm val,1}  \rightarrow   m_s  \ ; \
m_{\rm val,2}  \rightarrow   \overline{m} \, ,
\ee
corresponding to dynamical, degenerate up and down quarks, and a quenched
strange quark.  However, as previous work has only used degenerate
valence quark masses, extrapolating to the point
\be
m_{\rm val,1} \equiv m_{\rm val,2} = m_{DK} \, ,
\ee
we also present results from this method, so we may compare the two techniques.

The degenerate valence quark mass data is plotted in
Figure~\ref{fig:bp-sea}. As can be seen, $B_{PS}$ displays a rather weak, but
noticeable, dependence on the sea quark mass. To be precise: as $m_{sea}$
decreases so does $B_{PS}$.  The predicted dependence of 
$B_{PS}$ on  $m_{sea}$ and $m_{val}$, when $m_{sea}\neq m_{val}$, is
\begin{eqnarray}
B_{PS} &=& b_{0}\left(1 - 
    \frac{1}{(4\pi f)^2}\left( 6\, M^2\,\log{\frac{M^2}{\Lambda_\chi^2}}
\right)\right)
+ 
(b_1-b_3)M^2 + b_2 M^2_{SS} \, .
\label{eq:bp-sea}
\end{eqnarray}
Fits using Eq.~\ref{eq:bp-sea} are summarized in
Table~\ref{tab:bp-sea}. Eq.~\ref{eq:bp-sea} does not simultaneously fit all of
the data very well. This is evident from Figure~\ref{fig:bp-sea}: the results
of the fit for $0.02 \le m_{\rm val} \le 0.04 $ are shown, and under-predicts both
the lightest and heaviest points. However, as noted previously, we see large
fluctuations in the plateau for the lightest valence quark masses, and the
systematic error on these points probably outweighs the quoted statistical
error.  Therefore, for our final results we exclude both the $m_{\rm val}=0.01$
and $m_{\rm val}=0.05$ points; the latter in an attempt to stay in the region for
which NLO chiral perturbation theory is valid.

We now move to the extraction of $B_K$ including the effect of non-degenerate
valence quarks.  In this case sea quark dependent log terms appear, as well
as many new non-degenerate valence quark terms:
\begin{eqnarray}
\label{eq:bk-nd}
B_{PS} &=& 
b_0\left(  
1 + \frac{M^2_{12}} {(4\pi f)^2}\left(
-2(3 + \epsilon^2) \log{\frac{M^2_{12}}{\Lambda_\chi^2}}
-(2 + \epsilon^2) \log{(1 - \epsilon^2)}
 - 3\epsilon \log{\frac{1 + \epsilon}{ 1 -\epsilon}}\right)\right.
\\\nonumber
&&\left. +\frac{2}{3}\frac{1}{(4\pi f)^2}\left(
 \frac{3}{N}M_{SS}^2\left(
 \frac{2- \epsilon^2}{2 \epsilon} 
 \log{\frac{1 + \epsilon}{ 1 - \epsilon}}-2\right.\right.\right) 
\\\nonumber
 &&\left.\left.+ \frac{3}{N}\,M^2_{12}\left(2+\epsilon^2 
 - \frac{1 - 2\epsilon^2 - \epsilon^3}{ \epsilon} \log{\frac{1 + \epsilon }{1
 - \epsilon}} 
+ 2\epsilon^2 \log{\frac{M^2_{12} (1 -
\epsilon)}{\Lambda_\chi^2}}\right)\right)\right)
\\\nonumber
 && + b_1\, M_{12}^2
+ b_2\, M_{SS}^2
+ b_3\,M_{PS}^2(-2+\frac{M_{PS}^2}{M^2_{12}}) \, .
\end{eqnarray}
In the above
\begin{eqnarray} 
\epsilon&\equiv& \frac{m_2 - m_1}{m_1+m_2+2m_{\rm res}},\\ 
M_{PS}^2&\equiv& 2 B_0\, (m_1+m_{\rm res}),\\ 
M_{12}^2&\equiv& 2 B_0 \frac{m_1+m_2+2m_{\rm res}}{2},
\label{eq:dummy}
\end{eqnarray}
where $m_1$, $m_2$ ($m_1 \le m_2$) are the valence quark masses. The log terms
were computed in \cite{Golterman:1998st}.  $B_{PS}$ is shown in
Figure~\ref{fig:bp nondegenerate} for both the degenerate and non-degenerate
mass points, plotted as a function of $(m_1+m_2)/2$. As can be seen, the
non-degenerate points appear to lie on a smooth line connecting the degenerate
mass points, a clear sign that the non-degenerate mass effects are small.
Fits to Eq.~\ref{eq:bk-nd} are summarized in Table~\ref{tab:bp-nd-sea}. Note
that the value of $B_K$ is determined from the fit to Eq.~\ref{eq:bk-nd}
evaluated at $m_{s}$ and $\bar m$. As in the degenerate case, we use valence
quark masses in the range $ 0.02
\le m_{\rm val} \le 0.04$ for the extraction of our final result.

It is customary to quote $B_K$ in the NDR-$\overline{MS}$ scheme at $\mu=2$
GeV. We accomplish this in two steps. First we renormalize non-perturbatively
in the RI scheme at a scale $\mu\sim 1/a$ \cite{Izubuchi:2003rp} (Note that
this extraction makes use of the pertubative two-loop, continuum, results for
the running ), and then match to $\overline{MS}$ in the continuum using
one-loop perturbation theory~\cite{Ciuchini:1995cd}. Details of our method are
given in \cite{Noaki:2004,Izubuchi:2003rp}. With $Z^{\overline{MS}}_{B_K}=0.93(2)$ 
\cite{Izubuchi:2003rp} we find (adding errors in quadrature)
\begin{eqnarray}
B_K^{\overline{MS}}(2\,{\rm GeV})&=&0.509\,(18).
\end{eqnarray}
 for the degenerate case, and
\begin{eqnarray}
B_K^{\overline{MS}}(2\,{\rm GeV})&=&0.495\,(18),
\end{eqnarray}
for the non-degenerate case. We take this as our final value for $B_K$.  While
these two numbers agree within the quoted statistical error, taking the
jackknife difference shows that the 3\% difference between the non-degenerate
and degenerate extraction is statistically well resolved.

This two flavor, non-degenerate valence quark result is roughly 10\% smaller
than recent quenched values obtained with domain wall or overlap
fermions \cite{AliKhan:2001wr,Blum:2001xb,DeGrand:2003in,Noaki:2004} and
roughly 20\% smaller than the quenched values reported in
\cite{Aoki:1998nr,Garron:2003cb} (Kogut-Susskind and overlap fermions,
respectively).  We caution that there are as yet unquantified systematic
errors in our determination of $B_K$, most notably non-zero lattice spacing
and finite volume effects. Recent quenched studies
\cite{AliKhan:2001wr,Noaki:2004} find these to be on the order of 5\% each,
though they could differ in the dynamical case.

Of course, the quantity of direct relevance to experiment is $B_K$ at the kaon
mass, as quoted. However, in passing, we also mention our value for this
parameter in the chiral limit, $B_K^{\chi}( \overline{MS}, {\rm 2 GeV})$, as
it provides a useful comparison with phenomenological
models~\cite{Bijnens:1995br,Peris:2000sw,Cata:2003mn}, where calculations in the chiral
limit are often under better control. We find, $B_K^{\chi}( \overline{MS},{\rm
2 GeV} )=0.241(10)$. This should be compared with a value of 0.267(14), obtained in
our previous quenched study~\cite{Blum:2001xb}.

In summary: our study shows that the inclusion of sea quarks and
non-degenerate valence quarks tends to lower the value of $B_K$ by roughly
10\% and 3\%, respectively, which represents a very important step in
estimating systematic uncertainties in $B_K$.  We note that some time ago
$N_f=2$ staggered fermion calculations found no sea quark effects on $B_K$
outside of quoted errors\cite{Kilcup:1993pa,Ishizuka:1993ya}. These studies
used unimproved staggered fermions which have large lattice spacing errors;
thus we consider our new determination to be more reliable. It should also be
noted that early Wilson fermion results \cite{Soni:1995qq}, as well as the
more recent \cite{Flynn:2004au} suggest a small decrease in $B_K$ when
including the effects of dynamical quarks.  In \cite{Sharpe:1996ih} the effect
of sea quarks on $B_K$ was estimated in chiral perturbation theory from the
difference of the chiral logs between the quenched and dynamical
theories. This comparison suggested that the sea quark effects increase the
value of $B_K$, however it assumed that the analytic terms remain the same
between the quenched and dynamical theories and so, again, we consider our
determination to be more reliable.

%% file: text_sections/chiral.tex
In this section we will discuss our understanding of the (relatively large)
breaking of chiral symmetry observed in this work, starting with our choice of
gauge action.  As mentioned previously, the DBW2 gauge action has been used
successfully in the quenched approximation to improve the chiral properties of
domain wall fermions. To give an example: when comparing the Wilson, Iwasaki,
and DBW2 actions for inverse lattice spacings of approximately 2GeV, the
residual masses for $L_s=16$ are around 3MeV, 0.3MeV and 0.03MeV respectively
\cite{Aoki:2002vt}.  This dramatic improvement can be ascribed to the
interplay of the positively weighted plaquette term and the negatively
weighted rectangle term leading to a strong suppression of dislocations of the
lattice.  While we would like to exploit this mechanism in the dynamical case,
it is not a priori obvious what form of bare gauge action we should take such
that, when combined with the effects of the fermion determinant, we have an
effective short-distance gauge action with a similar form to that used in the
quenched approximation.  Over large (physical) distance scales it is
well-known that the effect of adding the determinant is to smooth out the
gauge field, and so, in order to perform a dynamical simulation with the same
lattice spacing as a quenched simulation, we must increase the gauge coupling.
The question we are interested in here, however, is how the inclusion of the
fermion determinant modifies the short-distance properties of the gauge
field. One particular worry, for example, is that the addition of the fermion
determinant may effectively induce a rectangle term of the opposite sign to
the one in the gauge action, leading to reduced suppression of dislocations.

To study this we solve for the short distance effective gauge action using the
Schwinger-Dyson equations, following the approach of
\cite{deForcrand:1999bi}. Briefly summarizing: for an ansatz of the effective
gauge action of 
\be 
S = \beta^{\alpha} S^{\alpha} \, ,
\ee 
a set of operators, $O_i$, and some variation of the gauge fields, $\delta$,
the Schwinger-Dyson equations read
\be 
\langle \delta O_i \rangle = 
-\langle O_i \left[ \delta S^{\alpha} \right] \rangle  
\beta^{\alpha}
\, .
\label{eq}
\ee
Calculating $\langle \delta O_i \rangle$ and $\langle O_i \left[ \delta
S^{\alpha} \right] \rangle$ for the same number of independent operators as
there are independent terms in the ansatz for the gauge action, this equation
may be solved to give $\beta^{\alpha}$. In the following we use
\be
S^{\alpha} = 1 - \frac{1}{3}\sum_l {\rm Re} {\rm Tr} 
\left[ U_l G_l^{\alpha}
\right]
\ee
where $U_l$ denotes a specific link and $G_l^\alpha$ 
the sum of the (forward) staples of type
$\alpha$ for link $l$, and the variation
\be
\delta^a_l U_l = -i \epsilon^{a} \lambda^{a} U_l
\ee
for observables
\be
O^{a, i}_l = {\rm Im} {\rm Tr} \left[ \lambda^a  U_l G_l^{i} \right] \, ,
\ee
and then sum Eq.~\ref{eq} over $a$ and $l$.

Some idea of the utility of this method can be gained by studying quenched
configurations.  Obviously, should the ansatz of the gauge action include the
terms that constitute the action used to generate the ensemble, then this is
precisely the result this method will give.  However, in Table
\ref{tab:dbw2sch} we show the results of applying this method using a quenched
ensemble of 404 configurations of size $16^3\times 32$ generated with the DBW2
action with $\beta=1.04$ -- corresponding to an inverse lattice spacing of
$\approx 2 {\rm GeV}$ -- and solving for the effective plaquette gauge
action. Using an observable based on the simple plaquette staple we get an
answer of $\beta_P \approx 9 $.  An ensemble generated with this plaquette
gauge action would have a much finer lattice spacing than the one we are
studying; our interpretation of this result is that, at the short distances
probed by this observable, lattices generated with the DBW2 action are much
smoother than those of the same lattice spacing generated with the plaquette
action.  As can also be seen, as the size of the observable used to calculate
the effective $\beta_P$ is increased to the $2\times1$ planar rectangle;
$2\times 2$ square and then $3\times3$ square, the calculated value of
$\beta_P$ decreases towards, presumably, the value that would generate one
ensemble with the same lattice spacing as the one we are studying
($\beta\approx 6$).

Table \ref{tab:dyn_sch} shows the results of applying this method to each of
our dynamical ensembles, using relatively localized observables.  In each case
we see that the short distance properties of the gauge field are dominated by
the bare form of the gauge action: the effective gauge action, up to some very
small deviations, is identical to this bare gauge action. We may therefore
qualitatively understand the larger observed chiral symmetry breaking for the
dynamical lattices versus quenched lattices as being due to the smaller value
of the (bare) $\beta$ used in the gauge action leading to reduced suppression
of dislocations; the effects of the determinant seem to only become
significant for larger distance observables.

While the chiral symmetry breaking observed for the ensemble studied here is
small enough to be a negligible correction to the physical quantities we are
calculating, it is still interesting to know how the residual mass varies both
with the size of the fifth dimension and the gauge coupling. A reasonable
estimate of the former may be made by calculating the residual mass for
different valence values of $L_s$.  Table \ref{tab:valmres} shows this for
fifth dimension of extent 8 to 32 for the $m_{\rm sea}=0.04$ ensemble at the
dynamical point. For $L_s \neq 12$ the residual mass is calculated on 45
configurations, leaving 100 trajectories between configurations starting from
trajectory 1005 (we skip trajectories 1805 and 1905 due to the hardware error
mentioned in Section \ref{sec:Thermalization}).  Figure~\ref{lsmres} shows
these dynamical results, and compares them with the residual mass on three
quenched ensembles: Wilson $\beta=6.0$ \cite{Blum:2000kn}, Iwasaki $\beta=2.6$
\cite{AliKhan:2000iv} and DBW2 $\beta=1.04$ \cite{Aoki:2002vt}; all of which
have inverse lattice spacings $\approx  2 {\rm GeV}$.  As can be seen,
while the residual mass in the dynamical simulation is much larger than that
in the comparable quenched DBW2 calculation, it is still smaller than the
value when using the Wilson gauge action in the quenched approximation.

We have also made a few exploratory studies using stronger couplings for the
DBW2 gauge action in dynamical simulations \cite{Levkova:2003dq}, namely
$\beta=0.75$ and $0.70$. For both these simulations we used $M_5=1.8$,
$L_s=12$ on $16^3\times 32$ lattices, as with the rest of this work.  Table
\ref{tab:betmres} gives values of the input quark mass, number of
configurations collected, the rho meson mass, and the value of the residual
mass.  As for each value of the coupling we used only one value for the
dynamical quark mass, the latter two quantities are quoted in the valence
chiral limit.  While the residual mass at $\beta=0.70$ is not prohibitively
large, caution should be taken; we may also look at at the spectral flow of
the Hermitian Wilson Dirac operator.  A transfer matrix in the fifth dimension
for domain wall fermions may be written in terms of this operator, with the
success of the domain wall fermion mechanism being dependent on the existence
of a gap in the spectral flow for negative Wilson masses.  Figure
\ref{spectral} shows a typical spectral flow for the $\beta=0.8$ ensembles, while
Figure \ref{spectralbeta} shows typical spectral flows for all the gauge
couplings we have studied.  As can be seen, the gap in the spectral flow
rapidly closes as we move to stronger gauge coupling. This is, perhaps, the
main challenge for the domain wall fermion/overlap approach: the computational
cost of lattice generation falls so quickly as the lattice spacing increases
that the extra cost of domain wall fermions versus other approaches could
easily be amortized by working on slightly coarser lattices. However, for the
current formulation, the domain wall mechanism begins to fail as we move to
lattice spacings significantly coarser then $a^{-1} \approx 2 {\rm
GeV}$. There have been some preliminary attempts to modify the gauge and/or
fermion action in such a way that the domain wall mechanism persists at
stronger couplings \cite{Levkova:2003dq}. These have so far met with limited
success, but this is clearly a direction of research which has not been
exhausted.

%% file: text_sections/conclusions.tex
We have presented a large scale lattice QCD calculation using two flavors of
dynamical domain wall fermions with small quark masses on lattices with large
volumes. Domain wall fermions possess exact chiral symmetry in the limit $L_s
\rightarrow \infty$ even at finite lattice spacing -- a symmetry fundamental
to much of the physics of QCD.  The $\sim 3 \times 5000$ Monte Carlo
trajectories in this work obtained with light dynamical quarks and small
residual quark mass, $m_{\rm strange}/2 \simle m_{\rm sea}+m_{\rm res} \simle
m_{\rm strange}
$, represent a substantial computational undertaking that took almost two years to
complete.

The first physics results to come from this endeavor are both interesting
and encouraging. The decay constants $f_\pi = 134.0(42)$ and $f_K = 157.4(38)$
agree with experiment well within $\simle$ 5\% statistical errors. Their
ratio, determined to 1\% statistical accuracy, $f_K/f_\pi = 1.175(11)$,
slightly under-predicts the experimental value [1.223] obtained from the
Particle Data Book.

The Kaon B parameter is the hadronic matrix element of the $\Delta S=2$ weak
interaction operator that governs neutral kaon mixing.  It is required to
determine the level of indirect CP violation in the kaon system that is
predicted in the Standard Model. We find that inclusion of sea quarks tends to
decrease the value of $B_K$ relative to our comparable quenched
calculations\cite{Blum:2001xb,Noaki:2004} by about two (statistical) standard
deviations, or roughly 10\%. Non-degenerate valence quark effects further lower
the value by 3\%.  At a renormalization scale of 2~GeV in the continuum
$\overline{MS}$ scheme, we find $B_K^{\overline{MS}} = 0.495(18)$ (statistical
error only) which is significantly lower than previously reported quenched
values and could impact the extraction of the phase of Standard Model CKM
quark mixing matrix if this value accurately describes the chiral, infinite
volume and continuum limits.

Besides physical results, two important and closely related theories, the
low-energy effective theory of domain wall fermions and (partially-) quenched
chiral perturbation theory, are strongly supported by the pattern of explicit
chiral symmetry breaking in our calculations.  In quenched calculations the
pion mass squared, when fit to a simple linear function of the quark mass,
significantly over-shot the chiral limit, $m_f = -m_{\rm res}$. Through careful
study, it was shown that this effect could be explained by a prediction of
quenched chiral perturbation theory, that is to a logarithmic singularity
unique to the chiral approximation.  When accounting for this term, $m_\pi^2$
was shown to vanish at the correct chiral limit.  In the two-flavor case this
offending logarithm does not appear.  Such a logarithm appears only at higher
order (see Eq.~\ref{eq:nlo fit 1}) and does not effect the chiral limit.  The
net result is that a simple linear extrapolation of $m_\pi^2$ should come
closer to the true chiral limit. That this indeed happens was demonstrated in
Figure~\ref{fig:nlo mpisq fit} and Table~\ref{tab:nlo mpisq fit}.

These initial results, obtained from two flavor calculations with three
relatively heavy sea quark masses, on a single volume and lattice spacing, and
fifth dimension $L_s=12$, while encouraging, may still suffer significant
systematic uncertainties. We stress that the use of next-to-leading order
partially-quenched chiral perturbation theory was crucial in our analysis of
$m_\pi^2$ and $B_K$, where it worked reasonably well. The analysis for the
decay constant was more problematic. This lead us to quote results for $f_\pi$
and $f_K$ from linear fits, {\it i.e.} the NLO analytic terms were included
but not the logarithms.  In addition, while the number of trajectories in our
study is large from a historical perspective, large scale auto-correlations
which can only be detected in longer runs may still be present.
In fact, here as in all dynamical simulations, the number of trajectories
studied is determined at least as much by the amount of available computer
resources as by established scientic criteria. All of these issues must be
further addressed by future calculations. The proven scaling
behavior of domain wall fermions gives us confidence that the results
presented here provide a solid foundation on which to build.

%% file: tables/tables_djk.tex
\begin{table}[ht]
\caption{Small lattice comparison of HMC evolutions. All these evolutions use
the Wilson gauge action with $\beta=5.2$ and two flavours of Domain Wall
Fermions with a bare mass of $m_{\rm sea} = 0.02$.
}
\begin{tabular}{ccccccc}\hline \hline
Force Term & $\Delta t$ & Steps/Trajectory & Trajectories & Acceptance &
CG-iterations/Trajectory  & $C_{\Delta H}$\\ \hline
Old        & $1/64$    & 33               & 1000-1880    & 87\%       &   8336
&  26.5
\\ 
Old        & $1/32$    & 17               & 1000-1929    & 59\%       &   4310
&  30.0
\\
New        & $1/32$    & 17               & 1000-1936    & 79\%       &   4179
&  12.9
\\ \hline \hline
\end{tabular}
\label{imp:nforce_small}
\end{table}

\begin{table}[ht]
\caption{Parameters for the large lattice HMC evolutions. The averaged elapse
time per one trajectory of our paricular implementation on 32 mother boards
(32MB $\sim$ 100GFLOPS theoretical peak speed) or 64 mother boards (64MB
$\sim$ 200GFLOPS) QCDSP, and the observed acceptance in the Metropolis test
are also quoted. The CPU time includes the computation time for the chiral
condensation, $\langle \bar q q \rangle^L_{m_v = m_{sea}}$, and the $r\times
t$ on-axis Wilson loop, $\langle W(r,t)\rangle$ with
$(r,t)=\{(1,1),(1,2),(2,1)\}$, but does not include the time for I/O.  The
scaled squared energy difference between the first and the last configuration
in a trajectory, $C_{\Delta H}=\sqrt{ \vev{(\Delta H)^2}/V }/(\Delta t)^2 $,
is quoted with the standard deviation error.  ``$0.02$ (OLD)'' are the results
of a small number of trajectories using the old force term.  }
\begin{tabular}{lcccccc}\hline \hline
 $m_{\rm sea}$    &  $\Delta t$ & Steps/Trajectory& Trajectories & Acceptance &
$C_{\Delta H}$ & Time/Trajectory(machine) \\ \hline
$0.02$ (OLD) &  $1/100$   &  51          &  45         & 56\% 
& 39(4) &  \\
 $0.02$   &  $1/100$   &  51             &  656-5361        & 77\% 
& 16.2(2) & 0.8784(6) hours (64MB)\\
 $0.03$   &  $1/100$   &  51             &  615-6195        & 78\% 
& 15.8(1) & 0.8324(4) hours (32MB)\\
 $0.04$   &  $1/80$    &  41             &  625-5605        & 68\% 
& 16.4(2) & 0.7116(2) hours (32MB)\\ \hline \hline
\end{tabular}
\label{imp:evols}
\end{table}

\begin{table}[htbp] 
\caption{Average cg-count and standard deviation (shown in square brackets)
versus molecular dynamics step for trajectories 3000 to 4000 of each ensemble,
together with the average for the total number of cg-iterations per
trajectory.}
\begin{tabular}{lccc}\hline \hline
Leapfrog step & $m_{\rm sea}=0.02$ & $m_{\rm sea}=0.03$ & $m_{\rm sea}=0.04$ \\ \hline
 $1/2$ step &    715[11] &  513.9[49] &  401.5[32] \\
    step  1 &    626[11] &  435.0[49] &  340.4[32] \\
    step  2 &    543[11] &  363.3[47] &  284.7[29] \\
    step  3 &    477[13] &  297.6[50] &  231.7[32] \\
    step  4 &    411[19] &  227.3[90] &  172.3[48] \\
    step  5 &    346[15] &  181.1[61] &  137.2[46] \\
    step  6 &    280[12] &  175.0[54] &  128.2[41] \\
    step  7 &    277[13] &  157.9[58] &  120.5[29] \\
    step  8 &    269[10] &  149.7[39] &  117.8[21] \\
    step  9 &    274[14] &  146.0[60] &  116.4[48] \\
    step 10 &    275[15] &  154.5[74] &  115.6[51] \\
    step 11 &    275[14] &  158.3[47] &  127.3[28] \\
    step 12 &    269[17] &  153.6[78] &  122.0[65] \\
    step 13 &    282[15] &  152.6[76] &  120.5[52] \\
    step 14 &    279[17] &  158.8[76] &  122.1[51] \\
\hline
 Total &   16014[396] &    9214[130] &  5964[71] \\
\hline \hline
\end{tabular}
\label{imp:cg}
\end{table}

\begin{table}[ht]
\caption{Evolution details and number of configuration used for physical results.}
\begin{tabular}{lcccc}\hline \hline
 $m_{\rm sea}$    & Until Accept/Reject & Thermalisation & Trajectories & Configurations
 \\ \hline
 $0.02$  &  28      &    656          &    5361   & 94\\
 $0.03$  &  25      &    615          &    6195   & 94 \\
 $0.04$  &  5       &    625          &    5605   & 94 \\ \hline \hline
\end{tabular}
\label{therm:evos}
\end{table}

\begin{table}
\caption{The bare chiral condensate, $\langle \bar q q \rangle^L_{m_{\rm val}
    = m_{\rm sea}}$,
and the $r\times t$ on-axis Wilson loop, $\langle W(r,t)\rangle$.}
\begin{tabular}{llccccc}\hline \hline
$m_{sea}$ &
$N_{configs}$ &
$\langle \bar q q \rangle^L_{m_{val} = m_{sea}}$ &
$\langle W(1,1)\rangle$ &
$\langle W(2,1)\rangle$ &
$\langle W(2,2)\rangle$ &
$\langle W(3,3)\rangle$ \\
\hline 
0.02 &
94 &
0.002542(11) &
0.646706(50) &
0.407777(80) &
0.17601(10) &
0.033633(80) \\
0.03 &
94 &
0.0034419(98) &
0.646594(47) &
0.407629(82) &
0.17572(10) &
0.033287(81) \\
0.04 &
94 &
0.0043255(81) &
0.646561(48) &
0.407562(71) &
0.175615(97) &
0.033330(81)\\
\hline \hline
\end{tabular}
\label{sim:basic}
\end{table}

\begin{table}[htdp]
\caption{The fitted vector meson mass from the wall-point correlation
  function for $m_{\rm sea}=0.02$.}
\begin{center}
\begin{tabular}{lllcc}\hline \hline
 $m_{sea}$ & $m_{val}$ & fit range & $\chi^2/$dof & mass \\\hline
0.02 & 0.01 & 5-16 & $1.9(10)$ & $0.511(10)$ \\
 0.02 & 0.015 & 5-16 & $2.0(11)$ & $0.5278(77)$ \\
 0.02 & 0.02 & 5-16 & $1.8(11)$ & $0.5425(64)$ \\
 0.02 & 0.025 & 5-16 & $1.7(11)$ & $0.5567(56)$ \\
 0.02 & 0.03 & 5-16 & $1.6(11)$ & $0.5712(50)$ \\
 0.02 & 0.035 & 5-16 & $1.7(11)$ & $0.5860(47)$ \\
 0.02 & 0.04 & 5-16 & $1.9(11)$ & $0.6010(44)$ \\
 0.02 & 0.045 & 5-16 & $2.1(11)$ & $0.6160(41)$ \\
 0.02 & 0.05 & 5-16 & $2.3(12)$ & $0.6309(39)$ \\
\hline\hline
\end{tabular}
\end{center}
\label{tab:vector mass 0.02}
\end{table}
                                                                                                                   
\begin{table}[htdp]
\caption{The fitted vector meson mass from the wall-point correlation
  function for  $m_{\rm sea}=0.03$.}
\begin{center}
\begin{tabular}{lllcc}\hline \hline
 $m_{\rm sea}$ & $m_{\rm val}$ & fit range & $\chi^2/$dof & mass \\\hline
 0.03 & 0.01 & 6-16 & $0.21(32)$ & $0.537(13)$ \\
 0.03 & 0.015 & 6-16 & $0.24(34)$ & $0.5522(96)$ \\
 0.03 & 0.02 & 6-16 & $0.37(42)$ & $0.5669(79)$ \\
 0.03 & 0.025 & 6-16 & $0.55(52)$ & $0.5809(67)$ \\
 0.03 & 0.03 & 6-16 & $0.75(62)$ & $0.5946(58)$ \\
 0.03 & 0.035 & 6-16 & $0.93(70)$ & $0.6079(52)$ \\
 0.03 & 0.04 & 6-16 & $1.07(76)$ & $0.6213(47)$ \\
 0.03 & 0.045 & 6-16 & $1.17(81)$ & $0.6347(43)$ \\
 0.03 & 0.05 & 6-16 & $1.25(85)$ & $0.6481(40)$ \\
\hline \hline
\end{tabular}
\end{center}
\label{tab:vector mass 0.03}
\end{table}

\begin{table}[htdp]
\caption{The fitted vector meson mass from the wall-point correlation
  function for $m_{\rm sea}=0.04$.}
\begin{center}
\begin{tabular}{lllcc}\hline \hline
 $m_{\rm sea}$ & $m_{\rm val}$ & fit range & $\chi^2/$dof & mass \\\hline
 0.04 & 0.01 & 7-16 & $1.72(99)$ & $0.580(25)$ \\
 0.04 & 0.015 & 7-16 & $1.8(10)$ & $0.586(19)$ \\
 0.04 & 0.02 & 7-16 & $1.7(10)$ & $0.591(15)$ \\
 0.04 & 0.025 & 7-16 & $1.58(99)$ & $0.598(11)$ \\
 0.04 & 0.03 & 7-16 & $1.51(95)$ & $0.6084(92)$ \\
 0.04 & 0.035 & 7-16 & $1.47(95)$ & $0.6198(79)$ \\
 0.04 & 0.04 & 7-16 & $1.44(97)$ & $0.6323(70)$ \\
 0.04 & 0.045 & 7-16 & $1.41(99)$ & $0.6455(64)$ \\
 0.04 & 0.05 & 7-16 & $1.37(100)$ & $0.6590(59)$ \\
\hline \hline
\end{tabular}
\end{center}
\label{tab:vector mass 0.04}
\end{table}

\begin{table}[htdp]
\caption{The results of a fit to the quark mass dependence of vector meson
  mass. }
\begin{center}
\begin{tabular}{lcccc} \hline \hline
fit                 & $\chi^2$/dof & $a$       & $b$     & $c$      \\
\hline
dynamical           & 0.9 (19)    & 0.448(15) & 4.52(47)&          \\
partially quenched  & 0.34(40)     & 0.448(13) & 1.78(43)&  2.78(13)\\
\hline \hline
\end{tabular}
\end{center}
\label{tab:vector mass chiral limit}
\end{table}%

\begin{table}[ht]
\caption{The fitted nucleon mass from the wall-point correlation function.}
\label{tab:nucleon mass}
\begin{tabular}{ccccc}
\hline
\hline
$m_{\rm sea}$ & $m_{\rm val}$ &  fit range & $\chi^2/$dof & mass \\
\hline
0.020 & 0.010 & 6-16 & 0.76 (69) & 0.679 (19) \\
0.020 & 0.015 & 6-16 & 0.58 (62) & 0.719 (12) \\
0.020 & 0.020 & 7-16 & 0.90 (80) & 0.755 (12) \\
0.020 & 0.025 & 7-16 & 1.28 (96) & 0.789 (10) \\
0.020 & 0.030 & 7-16 & 1.6 (1.1) & 0.8209 (94) \\
0.020 & 0.035 & 7-16 & 1.8 (1.1) & 0.8510 (89) \\
0.020 & 0.040 & 7-16 & 2.0 (1.1) & 0.8796 (85) \\
0.020 & 0.045 & 7-16 & 2.1 (1.1) & 0.9071 (83) \\
0.020 & 0.050 & 7-16 & 2.2 (1.1) & 0.9338 (80) \\
\hline
0.030 & 0.010 & 7-16 & 1.43 (88) & 0.747 (30) \\
0.030 & 0.015 & 7-16 & 1.29 (93) & 0.766 (19) \\
0.030 & 0.020 & 7-16 & 1.13 (84) & 0.788 (14) \\
0.030 & 0.025 & 8-16 & 1.14 (86) & 0.816 (16) \\
0.030 & 0.030 & 8-16 & 1.05 (80) & 0.844 (14) \\
0.030 & 0.035 & 8-16 & 1.01 (77) & 0.870 (12) \\
0.030 & 0.040 & 8-16 & 1.02 (77) & 0.896 (11) \\
0.030 & 0.045 & 8-16 & 1.06 (79) & 0.922 (10) \\
0.030 & 0.050 & 8-16 & 1.12 (81) & 0.9481 (95) \\
\hline
0.040 & 0.010 & 8-15 & 1.5 (1.1) & 0.748 (44) \\
0.040 & 0.015 & 8-16 & 0.75 (70) & 0.778 (35) \\
0.040 & 0.020 & 8-16 & 0.53 (56) & 0.808 (25) \\
0.040 & 0.025 & 8-16 & 0.53 (59) & 0.835 (19) \\
0.040 & 0.030 & 8-16 & 0.51 (60) & 0.860 (15) \\
0.040 & 0.035 & 9-16 & 0.33 (54) & 0.875 (15) \\
0.040 & 0.040 & 8-16 & 0.40 (52) & 0.910 (11) \\
0.040 & 0.045 & 8-16 & 0.40 (53) & 0.935 (10) \\
0.040 & 0.050 & 8-16 & 0.47 (59) & 0.9602 (92) \\
\hline
\hline
\end{tabular}
\end{table}

\begin{table}[htbp]
\caption{The $N^*$ mass from the wall-point correlation function.}
\label{tab:N* mass}
\begin{tabular}{ccccc}
\hline
\hline
$m_{\rm sea}$ & $m_{\rm val}$ &  fit range & $\chi^2/$dof & mass \\
\hline
0.020 & 0.010 & 3-5 & 0.10 (67) & 1.006 (46) \\
0.020 & 0.015 & 5-9 & 1.6 (1.4) & 1.021 (92) \\
0.020 & 0.020 & 5-9 & 1.5 (1.4) & 1.021 (71) \\
0.020 & 0.025 & 5-9 & 1.4 (1.3) & 1.038 (59) \\
0.020 & 0.030 & 5-9 & 1.3 (1.3) & 1.061 (50) \\
0.020 & 0.035 & 5-9 & 1.1 (1.2) & 1.086 (42) \\
0.020 & 0.040 & 5-9 & 0.9 (1.1) & 1.111 (37) \\
0.020 & 0.045 & 5-9 & 0.72 (99) & 1.136 (32) \\
0.020 & 0.050 & 5-9 & 0.54 (87) & 1.161 (28) \\
\hline
0.030 & 0.010 & 5-8 & 0.6 (1.2) & 1.21 (75) \\
0.030 & 0.015 & 5-8 & 0.39 (90) & 0.93 (12) \\
0.030 & 0.020 & 5-9 & 1.1 (1.3) & 0.950 (65) \\
0.030 & 0.025 & 5-9 & 0.8 (1.1) & 0.990 (48) \\
0.030 & 0.030 & 5-9 & 0.64 (98) & 1.026 (38) \\
0.030 & 0.035 & 5-9 & 0.59 (93) & 1.059 (32) \\
0.030 & 0.040 & 5-9 & 0.60 (93) & 1.089 (28) \\
0.030 & 0.045 & 5-9 & 0.62 (94) & 1.118 (25) \\
0.030 & 0.050 & 5-9 & 0.64 (95) & 1.145 (23) \\
\hline
0.040 & 0.010 & 5-7 & 0.21 (97) & 0.69 (17) \\
0.040 & 0.015 & 5-8 & 0.16 (63) & 0.92 (10) \\
0.040 & 0.020 & 4-8 & 0.19 (52) & 1.020 (47) \\
0.040 & 0.025 & 5-9 & 0.50 (84) & 1.057 (49) \\
0.040 & 0.030 & 5-9 & 0.55 (88) & 1.091 (39) \\
0.040 & 0.035 & 5-9 & 0.56 (88) & 1.119 (33) \\
0.040 & 0.040 & 5-9 & 0.54 (87) & 1.144 (28) \\
0.040 & 0.045 & 5-9 & 0.52 (85) & 1.168 (25) \\
0.040 & 0.050 & 5-9 & 0.50 (84) & 1.192 (23) \\
\hline
\hline
\end{tabular}
\end{table}

\begin{table}[htbp]
\caption{Physical baryon mass at the light quark mass ($\bar{m}$) 
 by diagonal ($m_{\rm val}=m_{\rm sea}\to\bar{m}$) or two stage
($m_{\rm val}\to\bar{m}, m_{\rm sea}\to\bar{m}$) extrapolation.
The valence extrapolation ($m_{\rm val}\to\bar{m}$) results
are obtained with linear fits using all valence masses in
Tables \ref{tab:nucleon mass} and \ref{tab:N* mass}.}
\label{tab:chiral baryon masses}
\begin{tabular}{lcccc}
\hline
\hline
$m_{\rm sea}$ & $M_N$ & $\chi^2/$dof & $M_{N^*}$ & $\chi^2/$dof \\
\hline
 $\bar{m}$ (diagonal)  & 0.605 (26) & 0.5 (1.5) & 0.81 (12) & 1.1 (2.1)\\
 $\bar{m}$ (two stage) & 0.556 (44) & 0.4 (1.3) & 1.00 (13) & 0.7 (1.7)\\
\hline
 0.02 & 0.633 (15) & 0.27 (32) &  0.951 (57) & 0.04 (18) \\
 0.03 & 0.687 (21) & 0.024 (67) & 0.841 (84) & 0.035 (97) \\
 0.04 & 0.704 (35) & 0.05 (11) &  0.892 (73) & 0.50 (48) \\
\hline
\hline
\end{tabular}
\end{table}

\begin{table}[htbp]
\caption{Mass ratio of physical $N$ ($N^*$) and $\rho$.
 $M_N$ and $M_{N^*}$ are calculated with both diagonal and two
 stage extrapolations (Table \ref{tab:chiral baryon masses}), 
while $M_\rho$ is always from diagonal extrapolation.}
\label{tab:mass ratio}
\begin{tabular}{ccc}
\hline
\hline
Fit  & $M_N/M_\rho$ & $M_{N^*}/M_\rho$\\
\hline
 diagonal  & 1.329 (59) & 1.77 (26) \\
 two stage & 1.221 (98) & 2.20 (30) \\
\hline
\hline
\end{tabular}
\end{table}

\clearpage

\begin{table}[htdp]
\caption{The fitted pseudo-scalar meson mass from the wall-point correlation function.}
\begin{center}
\begin{tabular}{llccc}
\hline \hline 
$m_{\rm sea}$ & $m_{\rm val}$ &  fit range & $\chi^2/$dof & mass \cr
\hline
0.02 & 0.01 & 9-16 & $0.79(85)$ & $0.2160(37)$  \\
0.02 & 0.015 & 9-16 & $1.05(95)$ & $0.2556(31)$  \\
0.02 & 0.02 & 9-16 & $1.3(11)$ & $0.2902(28)$  \\
0.02 & 0.025 & 9-16 & $1.6(11)$ & $0.3213(26)$  \\
0.02 & 0.03 & 9-16 & $1.8(12)$ & $0.3501(25)$  \\
0.02 & 0.035 & 9-16 & $2.0(13)$ & $0.3771(24)$  \\
0.02 & 0.04 & 9-16 & $2.2(13)$ & $0.4026(23)$  \\
0.02 & 0.045 & 9-16 & $2.3(14)$ & $0.4269(22)$  \\
0.02 & 0.05 & 9-16 & $2.5(14)$ & $0.4502(21)$  \\
0.03 & 0.01 & 9-16 & $1.6(12)$ & $0.2240(28)$  \\
0.03 & 0.015 & 9-16 & $1.3(11)$ & $0.2631(24)$  \\
0.03 & 0.02 & 9-16 & $1.04(94)$ & $0.2975(22)$  \\
0.03 & 0.025 & 9-16 & $0.90(85)$ & $0.3287(20)$  \\
0.03 & 0.03 & 9-16 & $0.81(80)$ & $0.3575(19)$  \\
0.03 & 0.035 & 9-16 & $0.77(77)$ & $0.3844(18)$  \\
0.03 & 0.04 & 9-16 & $0.76(76)$ & $0.4098(17)$  \\
0.03 & 0.045 & 9-16 & $0.78(77)$ & $0.4339(16)$  \\
0.03 & 0.05 & 9-16 & $0.82(79)$ & $0.4570(15)$  \\
0.04 & 0.01 & 9-16 & $1.06(91)$ & $0.2254(38)$  \\
0.04 & 0.015 & 9-16 & $1.08(91)$ & $0.2639(35)$  \\
0.04 & 0.02 & 9-16 & $1.11(91)$ & $0.2978(33)$  \\
0.04 & 0.025 & 9-16 & $1.14(91)$ & $0.3287(31)$  \\
0.04 & 0.03 & 9-16 & $1.15(91)$ & $0.3573(29)$  \\
0.04 & 0.035 & 9-16 & $1.13(91)$ & $0.3840(27)$  \\
0.04 & 0.04 & 9-16 & $1.09(89)$ & $0.4094(25)$  \\
0.04 & 0.045 & 9-16 & $1.03(87)$ & $0.4336(24)$  \\
0.04 & 0.05 & 9-16 & $0.96(84)$ & $0.4568(23)$  \\
\hline \hline
\end{tabular}
\end{center}
\label{tab:pseudo-scalar mass}
\end{table}%

\begin{table}[htdp]
\caption{The fitted pseudo-scalar meson mass from the axial-vector wall-point
correlation function.}
\begin{center}
\begin{tabular}{llccc} \hline \hline
$m_{\rm sea}$ & $m_{\rm val}$ &  fit range & $\chi^2/$dof & mass \cr
\hline
 0.02 & 0.01 & 9-16 & $0.62(79)$ & $0.2154(35)$ \\
 0.02 & 0.015 & 9-16 & $0.30(56)$ & $0.2558(28)$ \\
 0.02 & 0.02 & 9-16 & $0.22(45)$ & $0.2910(24)$ \\
 0.02 & 0.025 & 9-16 & $0.27(44)$ & $0.3227(22)$ \\
 0.02 & 0.03 & 9-16 & $0.40(50)$ & $0.3518(21)$ \\
 0.02 & 0.035 & 9-16 & $0.55(59)$ & $0.3789(20)$ \\
 0.02 & 0.04 & 9-16 & $0.72(68)$ & $0.4045(20)$ \\
 0.02 & 0.045 & 9-16 & $0.90(76)$ & $0.4288(19)$ \\
 0.02 & 0.05 & 9-16 & $1.06(84)$ & $0.4520(19)$ \\
 0.03 & 0.01 & 9-16 & $0.56(94)$ & $0.2221(38)$ \\
 0.03 & 0.015 & 9-16 & $0.50(77)$ & $0.2620(32)$ \\
 0.03 & 0.02 & 9-16 & $0.48(70)$ & $0.2967(29)$ \\
 0.03 & 0.025 & 9-16 & $0.48(66)$ & $0.3280(27)$ \\
 0.03 & 0.03 & 9-16 & $0.51(65)$ & $0.3568(25)$ \\
 0.03 & 0.035 & 9-16 & $0.55(66)$ & $0.3837(24)$ \\
 0.03 & 0.04 & 9-16 & $0.61(70)$ & $0.4091(23)$ \\
 0.03 & 0.045 & 9-16 & $0.67(74)$ & $0.4333(22)$ \\
 0.03 & 0.05 & 9-16 & $0.73(79)$ & $0.4564(21)$ \\
 0.04 & 0.01 & 9-16 & $0.40(55)$ & $0.2241(35)$ \\
 0.04 & 0.015 & 9-16 & $0.56(64)$ & $0.2621(29)$ \\
 0.04 & 0.02 & 9-16 & $0.64(70)$ & $0.2962(26)$ \\
 0.04 & 0.025 & 9-16 & $0.62(70)$ & $0.3273(25)$ \\
 0.04 & 0.03 & 9-16 & $0.55(68)$ & $0.3562(23)$ \\
 0.04 & 0.035 & 9-16 & $0.47(64)$ & $0.3831(22)$ \\
 0.04 & 0.04 & 9-16 & $0.40(61)$ & $0.4086(21)$ \\
 0.04 & 0.045 & 9-16 & $0.34(60)$ & $0.4329(20)$ \\
 0.04 & 0.05 & 9-16 & $0.31(61)$ & $0.4561(20)$ \\
\hline \hline
\end{tabular}
\end{center}
\label{tab:axial-vector mass}
\end{table}%

\begin{table}[htdp]
\caption{The pseudo-scalar meson mass and decay constant 
computed from the pseudo-scalar point-point
correlation function.}
\begin{center}
\begin{tabular}{llcccc}\hline \hline
$m_{\rm sea}$ & $m_{\rm val}$ &  fit range & $\chi^2/$dof & mass  & decay constant \cr
\hline
 0.02 & 0.01 & 9-16 & $0.95(81)$ & $0.2211(28)$ & $8.80(11) \times 10^{-2}$ \\
 0.02 & 0.02 & 9-16 & $0.39(52)$ & $0.2938(18)$ & $9.494(62) \times 10^{-2}$ \\
 0.02 & 0.03 & 9-16 & $0.57(63)$ & $0.3528(19)$ & $0.10161(78)$ \\
 0.02 & 0.04 & 9-16 & $0.54(62)$ & $0.4051(17)$ & $0.10720(77)$ \\
 0.02 & 0.05 & 9-16 & $0.52(62)$ & $0.4525(16)$ & $0.11244(77)$ \\
 0.03 & 0.02 & 9-16 & $0.53(66)$ & $0.3050(26)$ & $9.740(86) \times 10^{-2}$ \\
 0.03 & 0.03 & 9-16 & $0.68(68)$ & $0.3610(18)$ & $0.10253(56)$ \\
 0.03 & 0.04 & 9-16 & $0.61(69)$ & $0.4123(20)$ & $0.10926(76)$ \\
 0.04 & 0.04 & 9-16 & $1.4(11)$ & $0.4087(16)$ & $0.11059(57)$ \\
\hline \hline
\end{tabular}
\end{center}
\label{tab:pseudo-scalar mass point-point}
\end{table}%

\begin{table}[htdp]
\caption{The pseudo-scalar meson mass and decay constant 
computed from the axial-vector point-point
correlation function.}
\begin{center}
\begin{tabular}{llcccc}\hline \hline
$m_{\rm sea}$ & $m_{\rm val}$ &  fit range & $\chi^2/$dof & mass  & decay constant \cr
\hline
 0.02 & 0.01 &  9-16 & $0.43(54)$ & $0.2110(76)$ & $8.97(32) \times 10^{-2}$ \\
 0.02 & 0.02 &  9-16 & $0.47(63)$ & $0.2891(39)$ & $9.55(18) \times 10^{-2}$ \\
 0.02 & 0.03 &  9-16 & $0.58(76)$ & $0.3487(40)$ & $0.1011(20)$ \\
 0.02 & 0.04 &  9-16 & $0.73(88)$ & $0.4014(34)$ & $0.1058(18)$ \\
 0.02 & 0.05 &  9-16 & $0.78(92)$ & $0.4492(30)$ & $0.1100(17)$ \\
 0.03 & 0.02 &  9-16 & $0.65(69)$ & $0.3065(59)$ & $9.75(29) \times 10^{-2}$ \\
 0.03 & 0.03 &  9-16 & $1.5(10)$ & $0.3610(31)$ & $0.1032(14)$ \\
 0.03 & 0.04 &  9-16 & $0.59(71)$ & $0.4134(40)$ & $0.1086(24)$ \\
 0.04 & 0.04 &  9-16 & $0.52(64)$ & $0.4124(25)$ & $0.1100(16)$ \\
\hline \hline
\end{tabular}
\end{center}
\label{tab:axial-vector mass point-point}
\end{table}%

\begin{table}[htdp]
\caption{Results from a fit of the pseudo-scalar meson mass squared to a
  linear form. Only the dynamical, $m_{sea}=m_{val}$, points are included. In
  contrast to the case in the quenched approximation, when extrapolated to
  zero quark mass the result is consistent with zero. This may be explained by
  the absence of a contribution from the quenched chiral logarithm. }
\begin{center}
\begin{tabular}{cccc}\hline\hline
fit range & $\chi^2/$dof &  c & b \\
\hline
\multicolumn{4}{c}{Pseudo-scalar} \\
\hline
9-16 & 1.0(20) & $-0.0047(39)\times 10^{-3}$ & 4.19(13) \\
\hline
\multicolumn{4}{c}{Axial} \\
\hline
9-16 & 0.5(14) & $-3.1(34)\times 10^{-3}$ & 4.12(11) \\
\hline
\hline
\end{tabular}
\end{center}
\label{tab:lin mpisq fit}
\end{table}

\begin{table}[htdp]
\caption{Parameters from chiral perturbation theory fits to the values of
$m_{ps}^2$ computed from the pseudo-scalar wall-point,
and axial-vector wall point. (Tables~\protect\ref{tab:pseudo-scalar mass}
and \protect\ref{tab:axial-vector mass} respectively).  
$\chi^2$ is from uncorrelated fits in $m_f(=m_{\rm sea,val})$.  }
\begin{center}
\begin{tabular}{cccccc}\hline \hline
$m_f$ range & fit range & $\chi^2/$dof & $2\,B_0$ &  $L_5-2L_8$  &
$L_4-2L_6$ \\\hline
\multicolumn{6}{c}{Pseudo-scalar}\\\hline
 0.01-0.03 & 9-16 & $0.12(13)$ & $3.94(27)$ & $-1.51(74) \times 10^{-4}$ & $-1.9(12) \times 10^{-4}$  \\
 0.01-0.04 & 9-16 & $1.7(10)$ & $4.18(16)$ & $-1.4(44) \times 10^{-5}$ &
 $-1.17(43) \times 10^{-4}$  \\ \hline
\multicolumn{6}{c}{Axial}\\
\hline
  0.01-0.03 & 9-16 & $0.22(17)$ & $4.04(28)$ & $-1.87(90) \times 10^{-4}$ & $-1.2(11) \times 10^{-4}$  \\
  0.01-0.04 & 9-16 & $1.66(80)$ & $4.23(14)$ & $-4.0(48) \times 10^{-5}$ & $-7.7(33) \times 10^{-5}$  \\
\hline \hline
\end{tabular}
\end{center}
\label{tab:nlo mpisq fit}
\end{table}%

\begin{table}[htdp]
\caption{Parameters from next-to-leading order chiral perturbation theory fits
 to values of $f_{ps}$ listed in Table~\protect\ref{tab:pseudo-scalar mass
 point-point}. $L_{i}$ refer to Gasser-Leutwyler low energy constants
 evaluated at $\mu=1$ GeV.  $\chi^2$ is from uncorrelated in $m_f(=m_{\rm sea,}$,
 $m_{\rm val})$ fits.}
\begin{center}
\begin{tabular}{cccccc}\hline \hline
$m_f$ range & fit range & $\chi^2/$dof & $a\,f$  & $L_5$ & $L_4$ \cr
\hline
 0.01-0.03 & 9-16 & $0.14(32)$ & $5.36(48) \times 10^{-2}$ & $7.92(96) \times 10^{-4}$ & $7.2(63) \times 10^{-5}$ \\
 0.01-0.04 & 9-16 & $3.2(18)$ & $4.54(33) \times 10^{-2}$ & $7.14(80) \times
 10^{-4}$ & $1.29(23) \times 10^{-4}$ \\
\hline \hline
\end{tabular}
\end{center}
\label{tab:nlo decay fit}
\end{table}%

\begin{table}[htdp]
\caption{Parameters from next-to-next-to-leading order chiral perturbation
theory fits to values of $f_{ps}$ listed in
Table~\protect\ref{tab:pseudo-scalar mass point-point}. All quark mass points
were used in the fit. $L_{i}$ refer to Gasser-Leutwyler low energy constants
evaluated at $\mu=1$ GeV. $C_i$ are ${\cal O}(p^6)$ counter-terms.  $\chi^2$
is from uncorrelated in $m_f(=m_{\rm sea,}$, $m_{\rm val})$ fits.}
\begin{center}
\begin{tabular}{ccccccc}\hline \hline
 $\chi^2/$dof & $a\,f$  & $L_5$ & $L_4$ & $C_1$ & $C_2$ & $C_{12}$ \cr
\hline
 $0.56(58)$ & $6.43(91) \times 10^{-2}$ & $5.6(27) \times 10^{-4}$ & $-2.2(44) \times 10^{-4}$ & $3.2(76)$ & $1.62(100)$ &$6.8(28)$ \\
\hline \hline
\end{tabular}
\end{center}
\label{tab:nnlo decay fit}
\end{table}%

\begin{table}[htdp]
\caption{Parameters from linear fits to values of $f_{ps}$ listed in
Table~\protect\ref{tab:pseudo-scalar mass point-point}. $\chi^2$ is from
uncorrelated in $m_f(=m_{\rm sea,}$, $m_{\rm val})$ fits. For comparison,
results from a two parameter linear fit, $f_P = af + c_1 ( m + m_{res} )$, to
the $m_f = m_{\rm sea}=m_{\rm val}$ data points are included in the last
line.}
\begin{center}
\begin{tabular}{ccccccc}\hline \hline
$m_f$ range & fit range & $\chi^2/$dof & $a\,f$  & $c_1$ & $c_2$ \cr
\hline
0.01-0.04 & 9-16 & $0.41(43)$ & $7.81(16) \times 10^{-2}$ & $0.622(20)$ &
 $0.164(51)$ \\
0.02-0.04 & 9-16 & $0.12(72)$ & $7.81(14) \times 10^{-2}$ & 
$0.783(42)$ & \\
\hline \hline
\end{tabular} 
\end{center}
\label{tab:lo decay fit}
\end{table}%

\begin{table}[htdp]
\caption{
The results of fit Eq.~\ref{eq:potfit} without ($l=0$) and with ($l\neq0$) 
the lattice Coulomb correction, $l \delta V(\vec{r})$,
to  data of the static quark potential extracted at 
$t=5, r\in [r_{min},r_{max}]=[\sqrt{3},8]$ for
$m_{sea}=0.02,0.03,0.04$. 
The first and the second errors are the statistical error and
the estimation of the systematic error due to the selection of
the fit ranges. $t, [r_{min},r_{max}]$ are swept in the region of
$t=5,6$, $\sqrt{2}\leq r_{min}\leq \sqrt{6}$, and $7 \leq r_{max}\leq 9$. 
}
\begin{center}
\begin{tabular}{lccccc}\hline \hline
$m_{sea}$ & $r_0$ & $\alpha$ & $\sigma$ & $V_0$ & $l$  \cr
\hline
0.02   & 4.177(22)(99)  & 0.398(7)(47)  & 0.0718(11)(49) & 0.753(6)(25) & 0 (fixed) \cr
0.02   & 4.126(23)(106) & 0.518(18)(57) & 0.0665(14)(65) & 0.805(10)(38) & 0.35(4)(15)  \cr
\hline
0.03   & 4.066(25)(32) & 0.368(8)(29)  & 0.0776(14)(20) & 0.728(7)(16)  & 0 (fixed) \cr
0.03   & 4.026(26)(62) & 0.457(21)(55) & 0.0736(17)(29) & 0.766(11)(26) & 0.26(5)(17) \cr
\hline
0.04   & 4.076(27)(29)  & 0.399(9)(57)  & 0.0753(15)(33) & 0.749(7)(29)  & 0 (fixed) \cr
0.04   & 4.020(29)(34)  & 0.520(23)(28) & 0.0699(18)(23) & 0.801(12)(13)& 0.35(5)(14) \cr
\hline \hline
\end{tabular}
\end{center}
\label{tab:pot}
\end{table}%


\begin{table}[htdp]
\caption{The pseudo-scalar B parameter using degenerate valence quarks.}
\begin{center}
\begin{tabular}{cccc}
\hline \hline
$m_{sea}$ & $m_{val}$ & $t_{op} $ range & $B^{\rm (lat)}_{ps}$  \cr
\hline
0.02 & 0.01 & 14-17 & $0.488(14)$ \\
0.02 & 0.02 & 14-17 & $0.5524(92)$ \\
0.02 & 0.03 & 14-17 & $0.5923(72)$ \\
0.02 & 0.04 & 14-17 & $0.6229(61)$ \\
0.02 & 0.05 & 14-17 & $0.6485(55)$ \\
0.03 & 0.01 & 14-17 & $0.525(14)$ \\
0.03 & 0.02 & 14-17 & $0.5771(83)$ \\
0.03 & 0.03 & 14-17 & $0.6104(64)$ \\
0.03 & 0.04 & 14-17 & $0.6368(55)$ \\
0.03 & 0.05 & 14-17 & $0.6596(50)$ \\
0.04 & 0.01 & 14-17 & $0.512(12)$ \\
0.04 & 0.02 & 14-17 & $0.5747(85)$ \\
0.04 & 0.03 & 14-17 & $0.6133(69)$ \\
0.04 & 0.04 & 14-17 & $0.6425(58)$ \\
0.04 & 0.05 & 14-17 & $0.6662(51)$ \\
\hline \hline
\end{tabular}
\end{center}
\label{tab:bp}
\end{table}%

\begin{table}[htdp]
\caption{The bare pseudo-scalar B parameter using non-degenerate valence quarks,
  $m_1$ and $m_2$, for a dynamical mass of 0.02.}
\begin{center}
\begin{tabular}{ccccc}
\hline \hline
$m_{sea}$ & $m_{1}$ & $m_2$ & $t_{op} $ range & $B^{\rm (lat)}_{ps}$  \cr
\hline
0.02 & 0.02 & 0.01 & 14-17 & $0.526(12)$ \\
0.02 & 0.03 & 0.01 & 14-17 & $0.555(11)$ \\
0.02 & 0.03 & 0.02 & 14-17 & $0.5742(83)$ \\
0.02 & 0.04 & 0.01 & 14-17 & $0.577(10)$ \\
0.02 & 0.04 & 0.02 & 14-17 & $0.5929(77)$ \\
0.02 & 0.04 & 0.03 & 14-17 & $0.6085(67)$ \\
0.02 & 0.05 & 0.01 & 14-17 & $0.596(11)$ \\
0.02 & 0.05 & 0.02 & 14-17 & $0.6093(74)$ \\
0.02 & 0.05 & 0.03 & 14-17 & $0.6231(64)$ \\
0.02 & 0.05 & 0.04 & 14-17 & $0.6362(58)$ \\
\hline \hline
\end{tabular}
\end{center}
\label{tab:bp-nd2}

\end{table}
\begin{table}[htdp]
\caption{The bare pseudo-scalar B parameter using non-degenerate valence
  quarks, $m_1$ and $m_2$, for a dynamical mass of 0.03.}
\begin{center}
\begin{tabular}{ccccc}
\hline \hline
$m_{sea}$ & $m_{1}$ & $m_2$ & $t_{op} $ range & $B^{\rm (lat)}_{ps}$  \cr
\hline
0.03 & 0.02 & 0.01 & 14-17 & $0.556(11)$ \\
0.03 & 0.03 & 0.01 & 14-17 & $0.5799(95)$ \\
0.03 & 0.03 & 0.02 & 14-17 & $0.5954(73)$ \\
0.03 & 0.04 & 0.01 & 14-17 & $0.5999(91)$ \\
0.03 & 0.04 & 0.02 & 14-17 & $0.6116(67)$ \\
0.03 & 0.04 & 0.03 & 14-17 & $0.6244(59)$ \\
0.03 & 0.05 & 0.01 & 14-17 & $0.6171(90)$ \\
0.03 & 0.05 & 0.02 & 14-17 & $0.6262(64)$ \\
0.03 & 0.05 & 0.03 & 14-17 & $0.6375(56)$ \\
0.03 & 0.05 & 0.04 & 14-17 & $0.6487(52)$ \\
\hline \hline
\end{tabular}
\end{center}
\label{tab:bp-nd3}
\end{table}

\begin{table}[htdp]
\caption{The bare pseudo-scalar B parameter using non-degenerate valence
  quarks, $m_1$ and $m_2$, for a dynamical mass of 0.04.}
\begin{center}
\begin{tabular}{ccccc}
\hline \hline
$m_{sea}$ & $m_{1}$ & $m_2$ & $t_{op} $ range & $B^{\rm (lat)}_{ps}$  \cr
\hline
0.04 & 0.02 & 0.01 & 14-17 & $0.551(10)$ \\
0.04 & 0.03 & 0.01 & 14-17 & $0.5806(97)$ \\
0.04 & 0.03 & 0.02 & 14-17 & $0.5963(77)$ \\
0.04 & 0.04 & 0.01 & 14-17 & $0.6045(96)$ \\
0.04 & 0.04 & 0.02 & 14-17 & $0.6152(74)$ \\
0.04 & 0.04 & 0.03 & 14-17 & $0.6291(64)$ \\
0.04 & 0.05 & 0.01 & 14-17 & $0.6248(97)$ \\
0.04 & 0.05 & 0.02 & 14-17 & $0.6321(72)$ \\
0.04 & 0.05 & 0.03 & 14-17 & $0.6435(61)$ \\
0.04 & 0.05 & 0.04 & 14-17 & $0.6551(55)$ \\
\hline \hline
\end{tabular}
\end{center}
\label{tab:bp-nd4}
\end{table}%

\begin{table}[htdp]
\caption{The kaon B parameter from NLO fit, including extrapolation of the sea
quark mass to the physical point, $m_{\rm sea}=m_{light}$. $B_0$ is the value of
the B parameter in the chiral limit.}
\begin{center}
\begin{tabular}{cccccccc}
\hline \hline
 $m_{\rm sea}$ range & $m_{\rm val}$ range & $t_{op} $ range & $\chi^2$/dof & $b_0$  
&  $b_1-b_3$ & $b_2$ 
& $B^{\rm (lat)}_{K}$ \cr
\hline
0.02-0.03 & 0.01-0.03 & 14-17 &   $0.39(39)$ &  $0.260(21)$ &  $0.527(99)$ &   $0.54(26)$ &  $0.521(30)$  \\
0.02-0.04 & 0.01-0.04 & 14-17 &   $1.27(89)$ &  $0.265(11)$ &  $0.744(44)$ &   $0.26(12)$ &  $0.550(16)$  \\
0.02-0.04 & 0.01-0.05 & 14-17 &   $2.34(84)$ & $0.2522(86)$ &  $0.927(29)$ &   $0.27(10)$ &  $0.545(14)$  \\
0.02-0.03 & 0.02-0.03 & 14-17 &   $0.18(25)$ &  $0.258(18)$ &  $0.616(64)$ &   $0.49(25)$ &  $0.525(28)$  \\
0.02-0.04 & 0.02-0.04 & 14-17 &   $0.80(93)$ & $0.2594(96)$ &  $0.826(33)$ &   $0.24(12)$ &  $0.547(15)$  \\
0.02-0.04 & 0.02-0.05 & 14-17 &   $1.45(72)$ & $0.2476(79)$ &  $0.987(25)$ &   $0.24(10)$ &  $0.543(14)$  \\
\hline \hline
\end{tabular}
\end{center}
\label{tab:bp-sea}
\end{table}%

\begin{table}[htdp]
\caption{Same as Table~\ref{tab:bp-sea}, but including
non-degenerate valence quarks. }
\begin{center}
\begin{tabular}{ccccccccc}
\hline \hline
 $m_{\rm sea}$ range & $m_{\rm val}$ range & $t_{op}$ range & $\chi^2$/dof & $b_0$  & $b_1$  & $b_2$  & $b_3$& $B^{\rm (lat)}_{K}$ \cr
\hline
0.02-0.03 & 0.01-0.03 & 14-17 & $0.20(20)$ &  $0.260(22)$ &   $0.51(15)$ &  $0.56(28)$ & $-1(96) \times 10^{-3}$ &  $0.524(30)$ \\
0.02-0.04 & 0.01-0.04 & 14-17 & $0.84(74)$ &  $0.266(12)$ &  $0.713(70)$ &  $0.27(13)$ & $-1(59) \times 10^{-3}$ &  $0.554(18)$ \\
0.02-0.04 & 0.01-0.05 & 14-17 & $1.36(62)$ & $0.2536(94)$ &  $0.853(49)$ &  $0.28(11)$ & $-3.9(47) \times 10^{-2}$ &  $0.546(16)$ \\
0.02-0.03 & 0.02-0.03 & 14-17 & $9(12) \times 10^{-2}$ &  $0.258(18)$ &   $0.25(13)$ &  $0.49(25)$ & $-0.370(86)$ &  $0.498(24)$ \\
0.02-0.04 & 0.02-0.04 & 14-17 & $0.55(79)$ & $0.2591(98)$ &  $0.601(54)$ &  $0.25(12)$ & $-0.220(40)$ &  $0.533(15)$ \\
0.02-0.04 & 0.02-0.05 & 14-17 & $0.77(59)$ & $0.2475(83)$ &  $0.784(36)$ &  $0.25(11)$ & $-0.193(31)$ &  $0.530(14)$ \\
\hline \hline
\end{tabular}
\end{center}
\label{tab:bp-nd-sea}
\end{table}%
\newpage

\begin{table}[ht]
\caption{The effective $\beta$ for the plaquette action for an ensemble of
  404 quenched DBW2 configurations versus the observable used in 
Equation~\protect\ref{eq}. In each case the operator chosen is planar.
}
\begin{center}
\begin{tabular}{cc}
\hline \hline
Observable  & Effective $\beta_P$ \\
\hline
 $1\times 1$ & $9.10039(34)$ \\
 $2\times 1$ & $7.87358(35)$ \\
 $2\times 2$ & $6.92080(50)$ \\
 $3\times 3$ & $6.6815(13)$ \\
\hline \hline
\end{tabular}
\end{center}
\label{tab:dbw2sch}
\end{table}

\begin{table}[ht]
\caption{ The effective, short-distance, gauge action, solved for using the
schwinger dyson equation with an anzatz for the form of: plaquette ($\beta_{1
\time 1}$), rectangle ($\beta_{1\times 2}$) and the two hypercubic, five-link,
loops ($\beta_{5,1}$ and $\beta_{5,2}$). Up to small corrections the results
are equal to the bare input parameters.  }
\begin{center}
\begin{tabular}{ccccc}
\hline \hline
dynamical mass & $\beta_{1\times 1}$ & $\beta_{1\times 2}$ & 
$\beta_{5,1}$ & $\beta_{5,2}$ \\
\hline
 0.02 & $9.7384(36)$ & $-1.11392(53)$ & $2.99(56) \times 10^{-3}$ & $6.35(77)
 \times 10^{-3}$ \\
 0.03 & $9.7388(43)$ & $-1.11523(50)$ & $3.35(67) \times 10^{-3}$ & $6.18(93)
 \times 10^{-3}$ \\
 0.04 & $9.7471(44)$ & $-1.11531(54)$ & $2.15(73) \times 10^{-3}$ & $7.5(10)
 \times 10^{-3}$ \\
\hline \hline
\end{tabular}
\end{center}
\label{tab:dyn_sch}
\end{table}

\begin{table}[ht]
\caption{ Parameters and results for the stronger coupling
dynamical studies}
\begin{center}
\begin{tabular}{ccccccc}
\hline \hline
$\beta$ & $m_f$  & $m_\rho$ & $\chi^2/Dof$  & $m_{\rm res}$ & \# conf.\\
\hline 
0.70 	& 0.026 & 0.830(5) & 1.4 & $0.0094(1)$&   32          \\

0.75   & 0.022 &  0.667(7) & 0.8    & $0.00405(7)$ &  41     \\
\hline \hline
\end{tabular}
\end{center}
\label{tab:betmres}
\end{table}

\begin{table}[ht]
\caption{Residual mass versus valence $L_s$ for the $m_{sea}=0.04$
ensemble}
\begin{center}
\begin{tabular}{ccccccc}\hline \hline
$m_{val}$ & $L_s=8$& $L_s=12$& $L_s=16$& $L_s=24$& $L_s=32$\\\hline
0.04 & $4.787(37) \times 10^{-3}$& $1.347(20) \times 10^{-3}$& $5.75(21)
\times 10^{-4}$& $2.13(17) \times 10^{-4}$& $1.30(15) \times 10^{-4}$\\\hline \hline
\end{tabular}
\end{center}
\label{tab:valmres}
\end{table}

%% file: figures/implementation/figures.tex
\begin{figure}[htbp]
\centerline {
\includegraphics[width=6in]{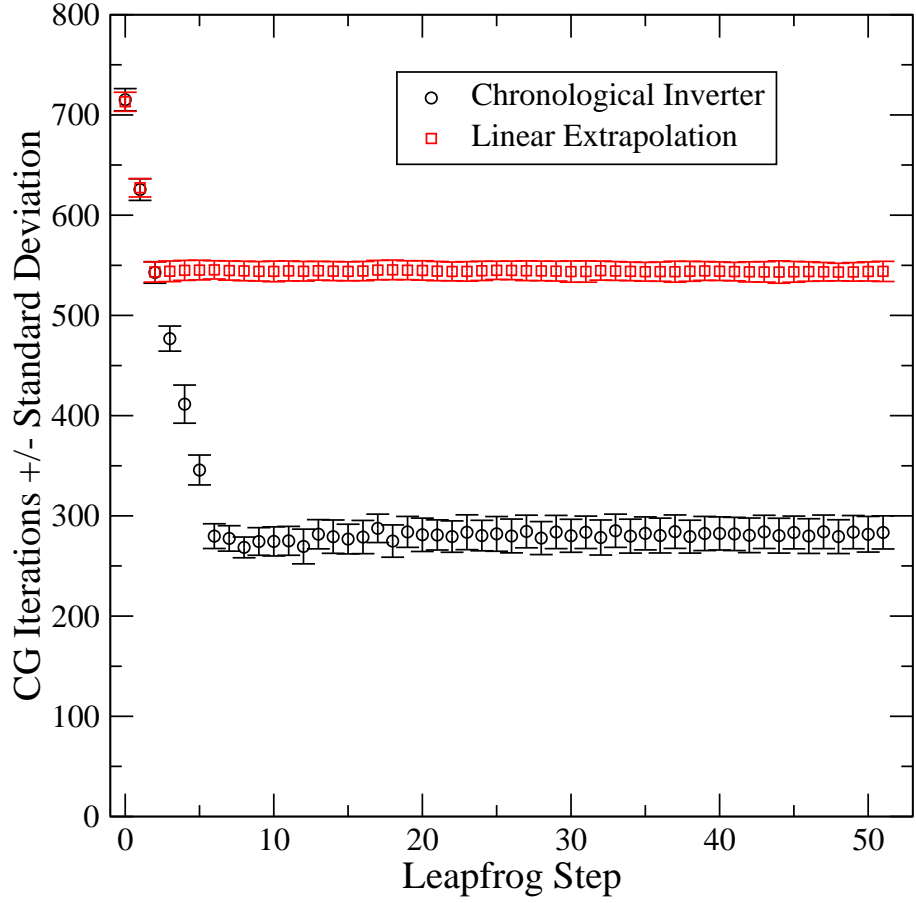}}
\caption{Conjugate gradient iteration count for the chronological inverter
using the previous 7 vectors compared with a linear extrapolation of the
previous two vectors for the $m_{\rm sea}=0.02$ ensemble.}
\label{fig:cgsprof}
\end{figure}

\begin{figure}[htbp]
\centerline {
\includegraphics[width=6in]{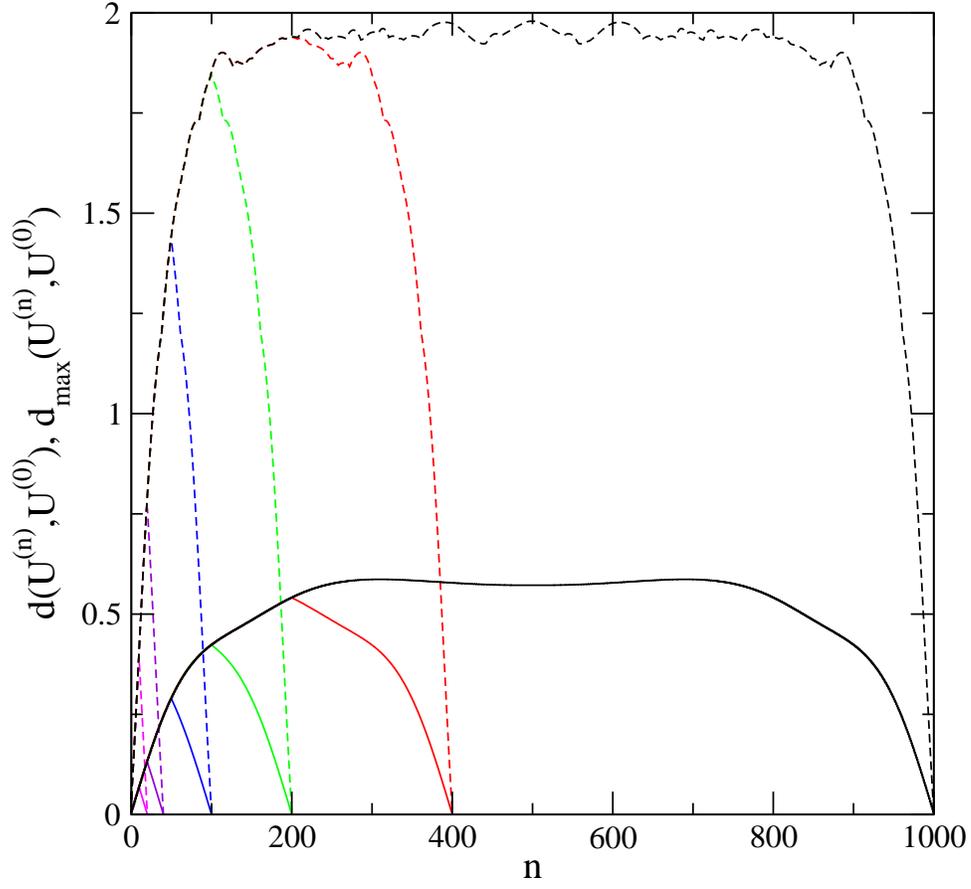}}
\caption{
$d(U_\mu^{(n)},U^{(I)}_\mu)$ and 
$d_{max}(U_\mu^{(n)},U^{(I)}_\mu)$ in a trajectory of
$m_{sea}=0.02$ on $16^3 \times 32$. 
The total step numbers are 10, 20, 50, 100, 200, and 500. 
The solid curves are $d(U_\mu^{(n)},U^{(I)}_\mu)$ while
the dashed are $d_{max}(U_\mu^{(n)},U^{(I)}_\mu)$. 
}
\label{fig:d_in_evol}
\end{figure}

\clearpage

\begin{figure}[htbp]
\centerline {
\includegraphics[width=6in]{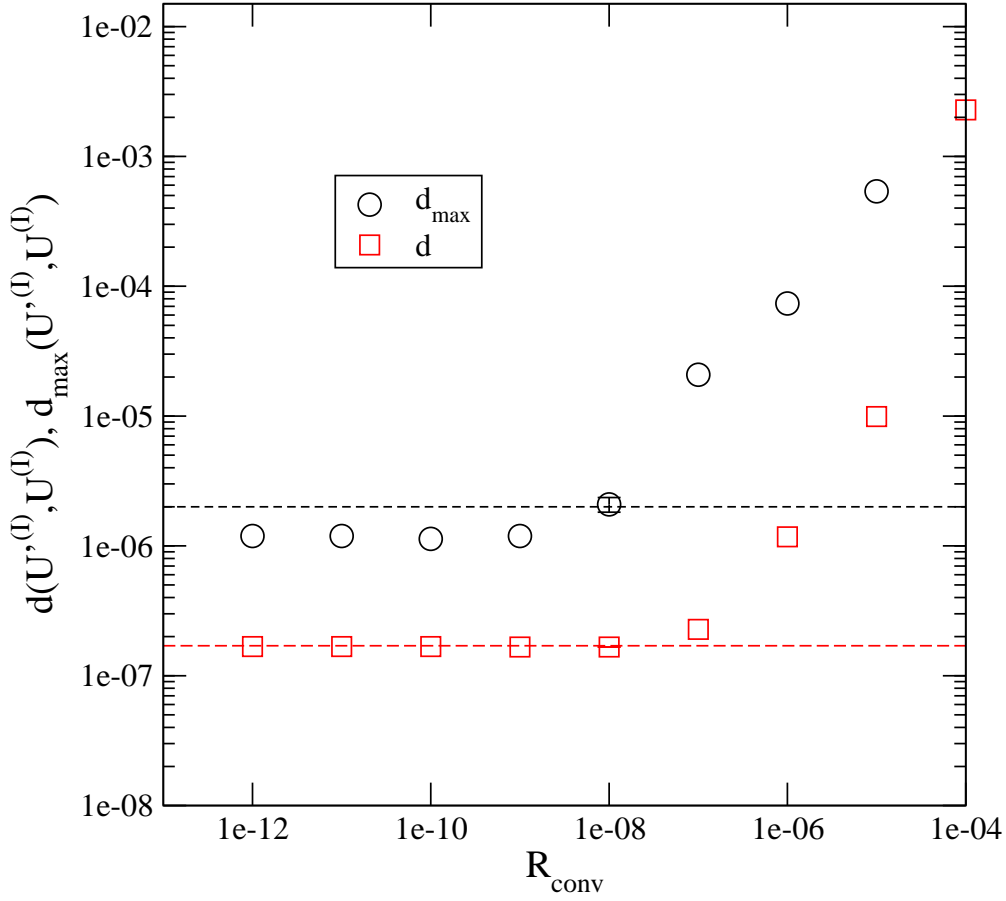}}
\caption{ The breaking of reversible dynamics measured by
$d(U^{'(I)}_\mu,U^{(I)}_\mu)$ and $d_{max}(U^{'(I)}_\mu,U^{(I)}_\mu)$ as a
function of the CG convergence criteria, $R_{conv}$.  The dotted lines are
observed upper bound of deviations due to the reunitarization process. The
error-bar at $R_{conv}$ = 1e-8 was obtained using five configurations.  }
\label{fig:rev_break_vs_res}
\end{figure}

\begin{figure}[htbp]
\centerline {
\includegraphics[width=6in]{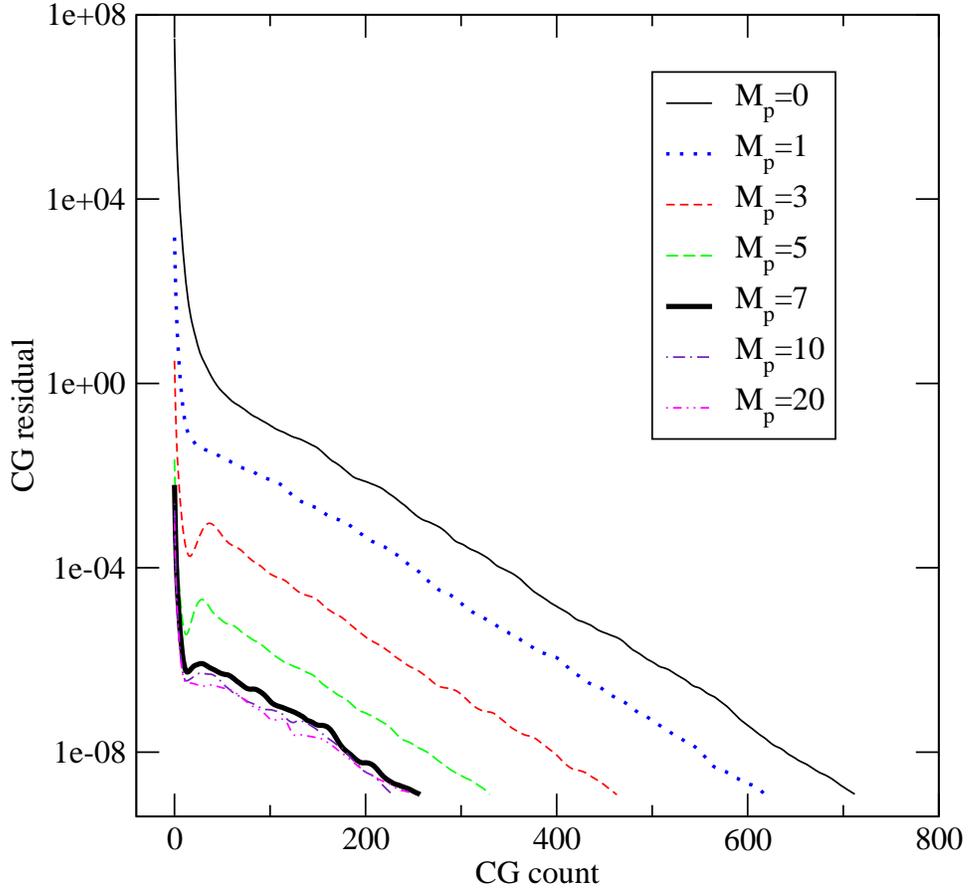}}
\caption{
The residues of CG, Eq. \ref{eq:resCG}, as
a function of the number of CG iteration are plotted
for various numbers of previous solutions, $N_p$, on a typical
configuration of the $m_{\rm sea}=0.02$ ensemble.
}
\label{fig:CG_vs_Nprecog}
\end{figure}

\begin{figure}[htbp]
\centerline {
\includegraphics[width=5in]{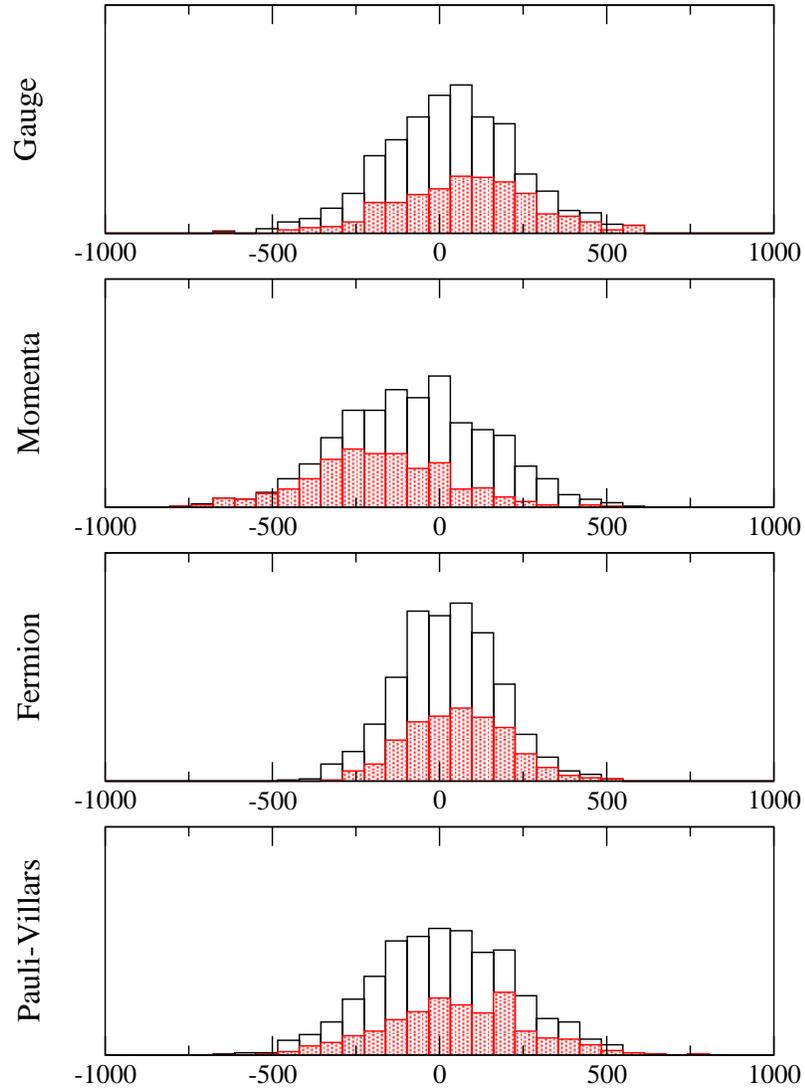}}
\caption{The individual contributions to the total change in the Hamiltonian
from the various components of the Hamiltonian for the large step-size, old
force term simulation described in Table~\protect\ref{imp:nforce_small}.  The
shaded bars represent the trajectories which failed the accept/reject step,
while the empty bars tally all trajectories.  }
\label{fig:hamold}
\end{figure}

\begin{figure}[htbp]
\centerline {
\includegraphics[width=5in]{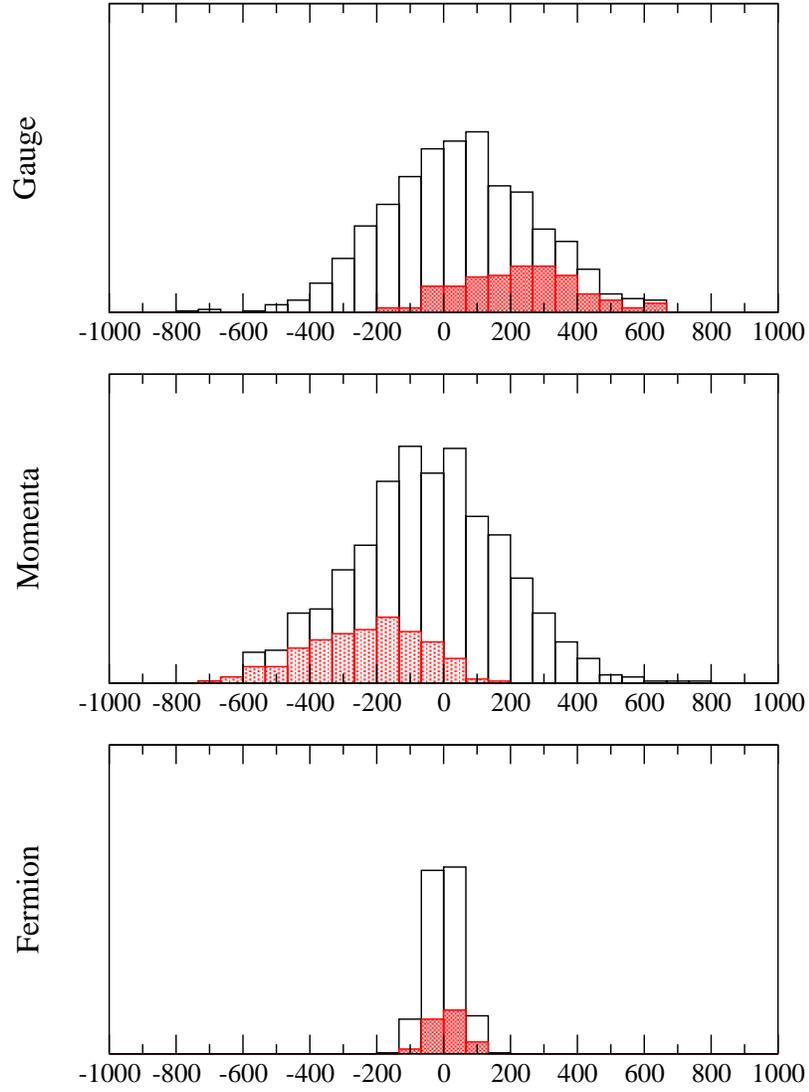}}
\caption{The individual contributions to the total change in the Hamiltonian
from the various components of the Hamiltonian for the large step-size, new
force term simulation described in Table \protect\ref{imp:nforce_small}. The
shaded bars represent the trajectories which failed the accept/reject step,
while the empty bars tally all trajectories. Note that the magnitude of the
differences is dominated by the gauge and momentum parts of the Hamiltonian,
and the noticeable trend that trajectories for which the change in the gauge
part of the Hamiltonian is positive are more likely to be rejected.}
\label{fig:hamnew}
\end{figure}

%% file: figures/thermalization/figures.tex
\begin{figure}[htbp]
\centerline {
\includegraphics[width=6in]{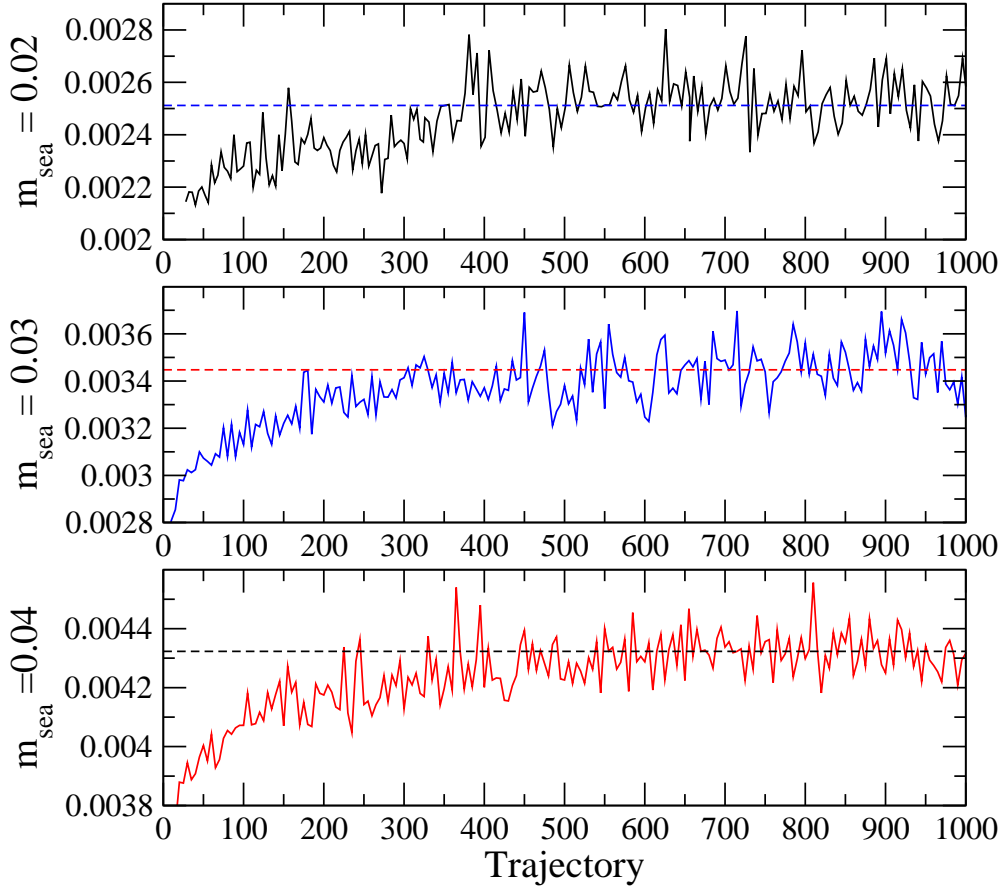}}
\caption{The history of $\overline{q} q$ ($m_{\rm val} = m_{\rm
dyn}$), up to trajectory 1000, for all evolutions. The average from
trajectory 3000 onwards is shown as a dashed horizontal line.}
\label{fig:pbptherm}
\end{figure}

\begin{figure}[htbp]
\centerline {
\includegraphics[width=6in]{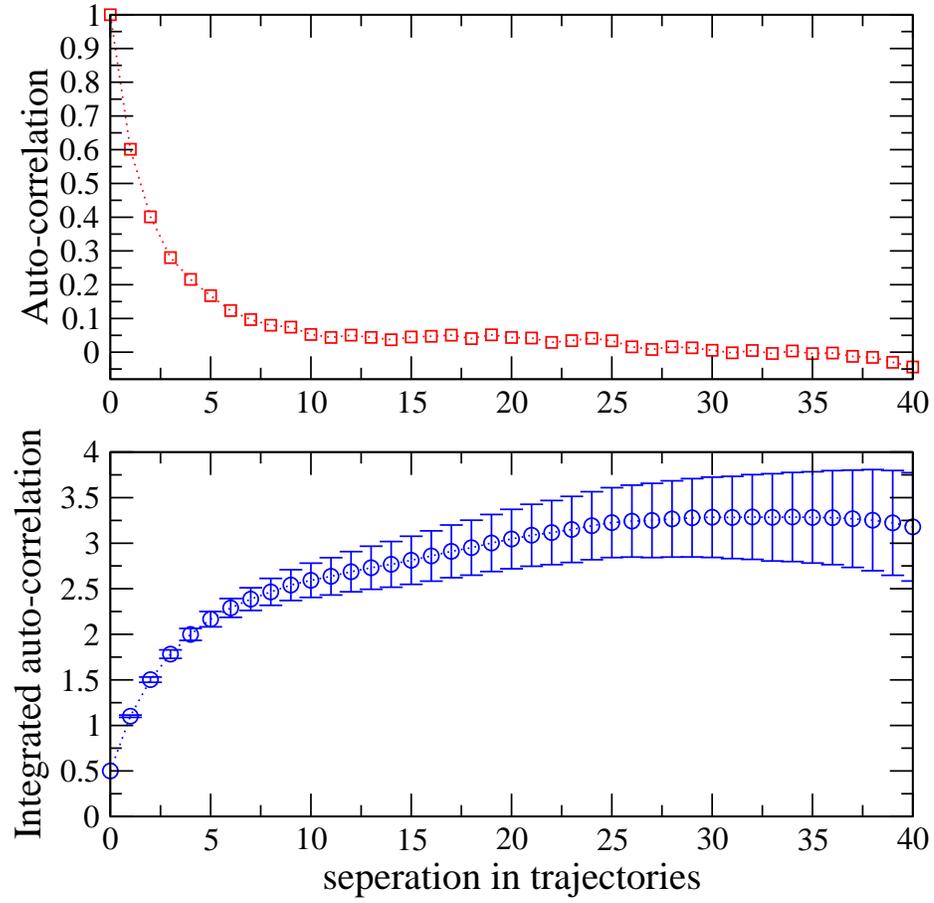}}
\caption{Plaquette auto-correlation function and integrated auto-correlation
  length for the $m_{\rm sea}=0.02$ ensemble.}
\label{fig:autoplaq}
\end{figure}

\begin{figure}[htbp]
\centerline {
\includegraphics[width=6in]{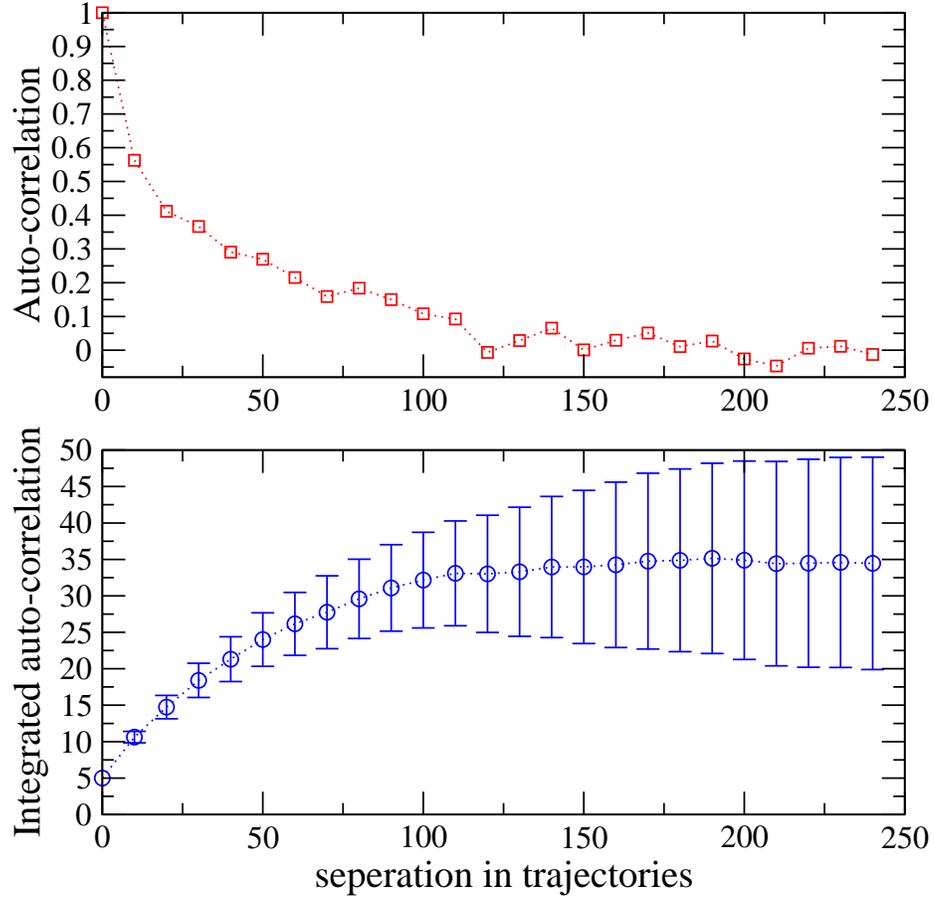}}
\caption{Auto-correlation function and integrated auto-correlation length
for timeslice 12 of the correlation function of the time component of the 
local axial vector current on the $m_{\rm sea}=0.02$ ensemble.}
\label{fig:autoax}
\end{figure}

\begin{figure}[htbp]
\centerline {
\includegraphics[width=6in]{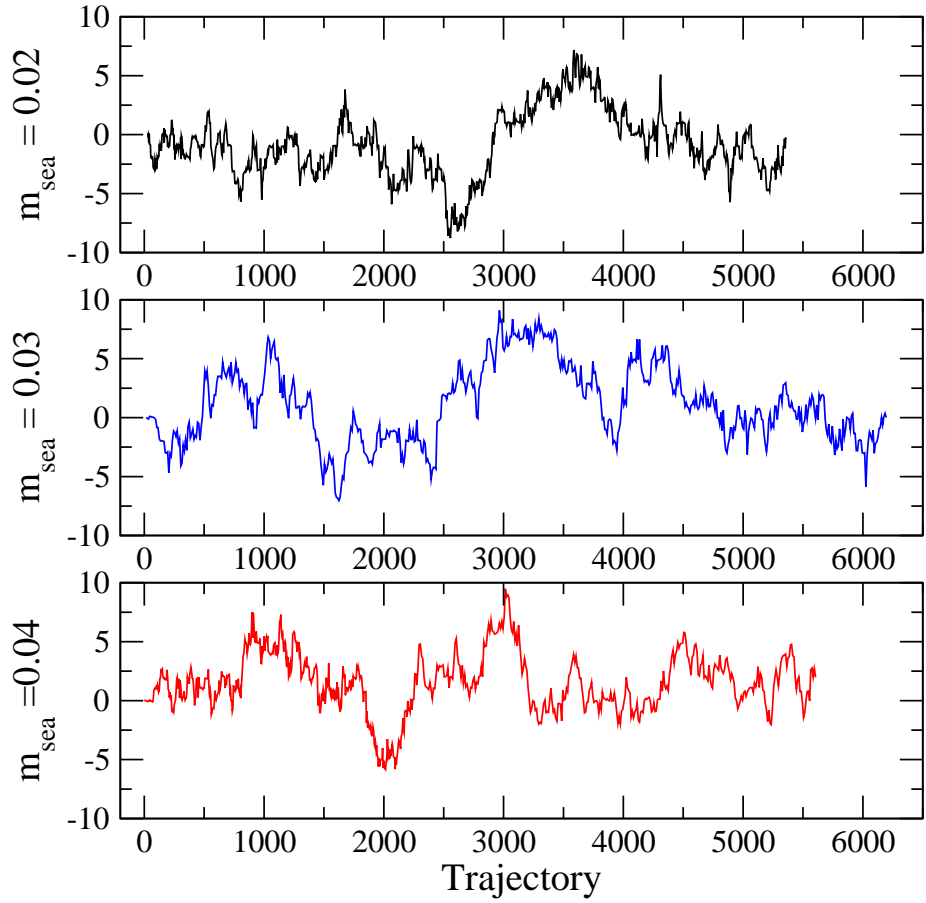}}
\caption{Topological charge history for all the ensembles. Note the
correlations of the scale of many hundreds of HMC trajectories.}
\label{fig:topo}
\end{figure}

%% file: figures/physical/figures.tex

\begin{figure}[htbp]
\centerline {
\includegraphics[width=6in]{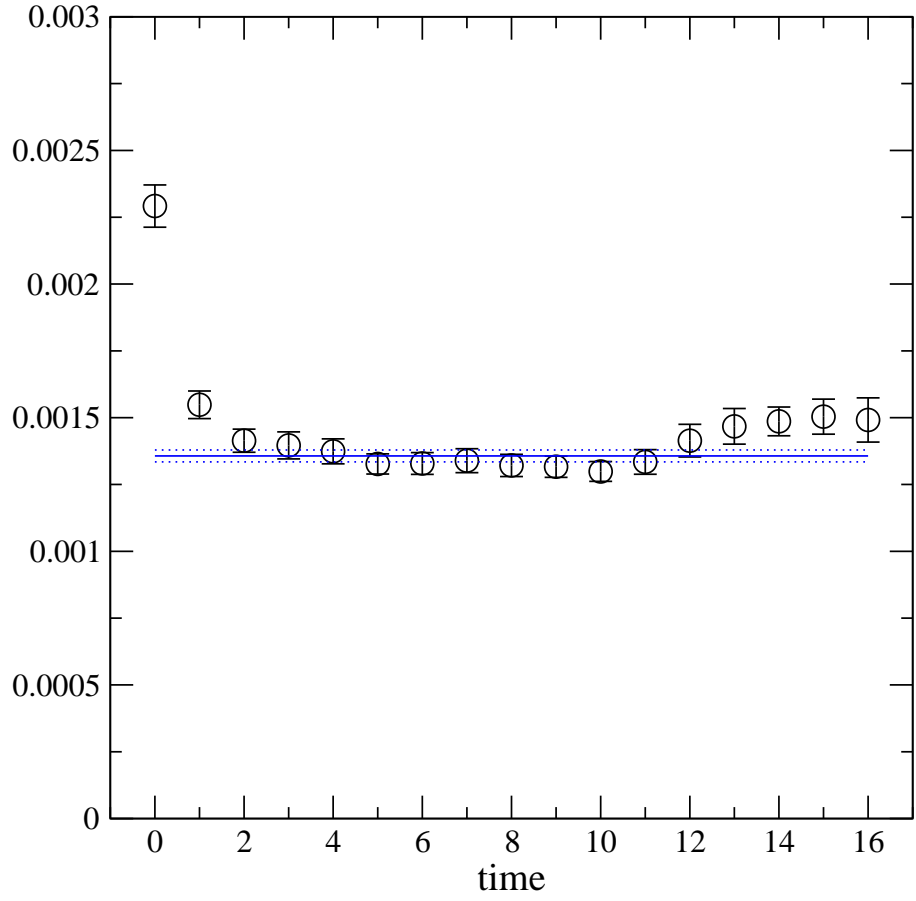}}
\caption{The residual mass plateau plot for $m_{sea}=m_{val}=0.02$. The
  error-weighted average between timeslices 4 and 16 is shown as a horizontal
  line.}
\label{fig:mres plateau}
\end{figure}\vspace{.2in}


\begin{figure}[htbp]
\centerline {
\includegraphics[width=6in]{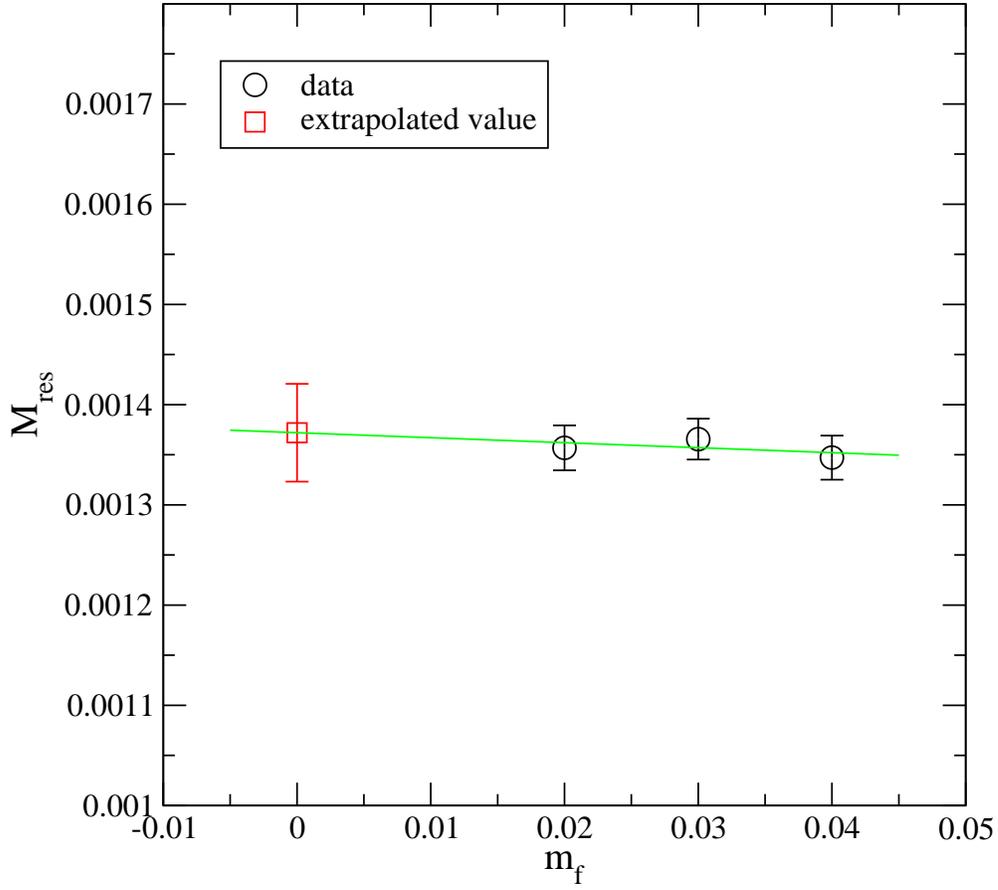}}
\caption{The residual mass extrapolated to $m_{sea}=m_{val}=0.0$.}
\label{fig:mres extr}
\end{figure}\vspace{.2in}


\begin{figure}[htbp]
\centerline {
\includegraphics[width=6in]{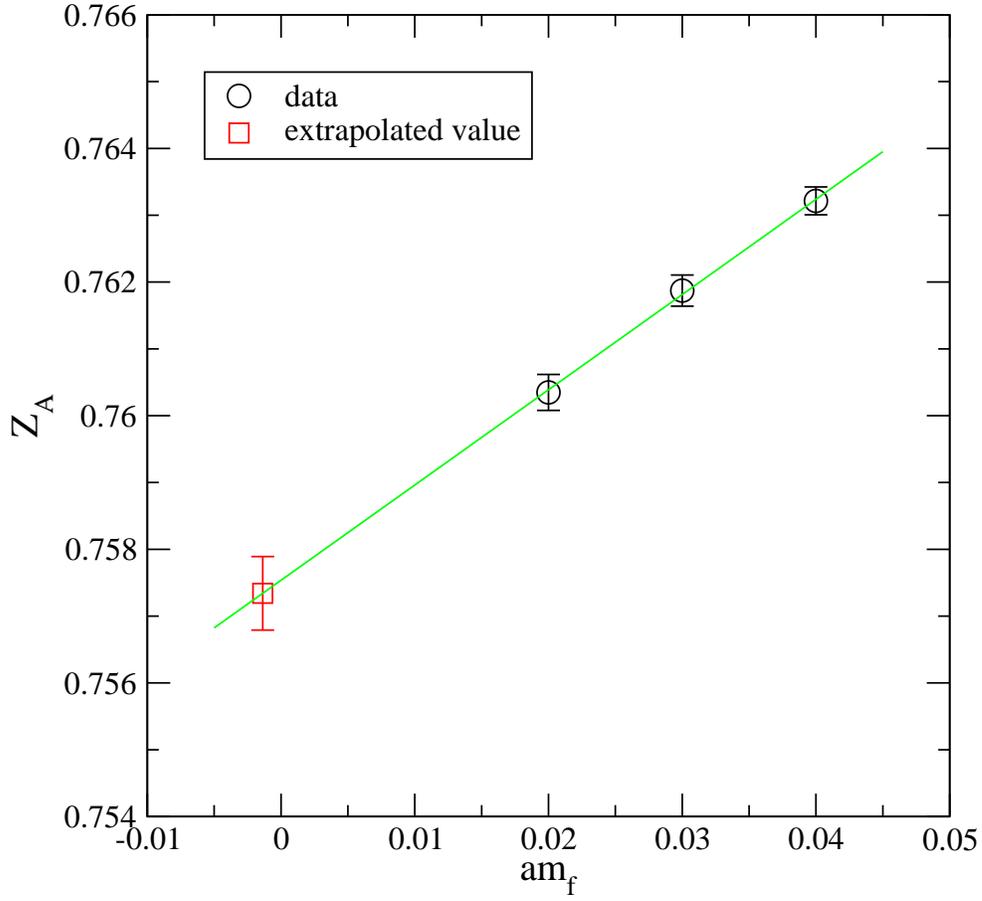}}
\caption{The renormalisation factor for the local, non-singlet, axial
current. This is defined in the chiral limit. Here we show the data at finite
mass for the fully dynamical points together with the results of a linear
extrapolation to the chiral limit. }
\label{fig:za}
\end{figure}\vspace{.2in}


\begin{figure}[htbp]
\centerline {
\includegraphics[width=6in]{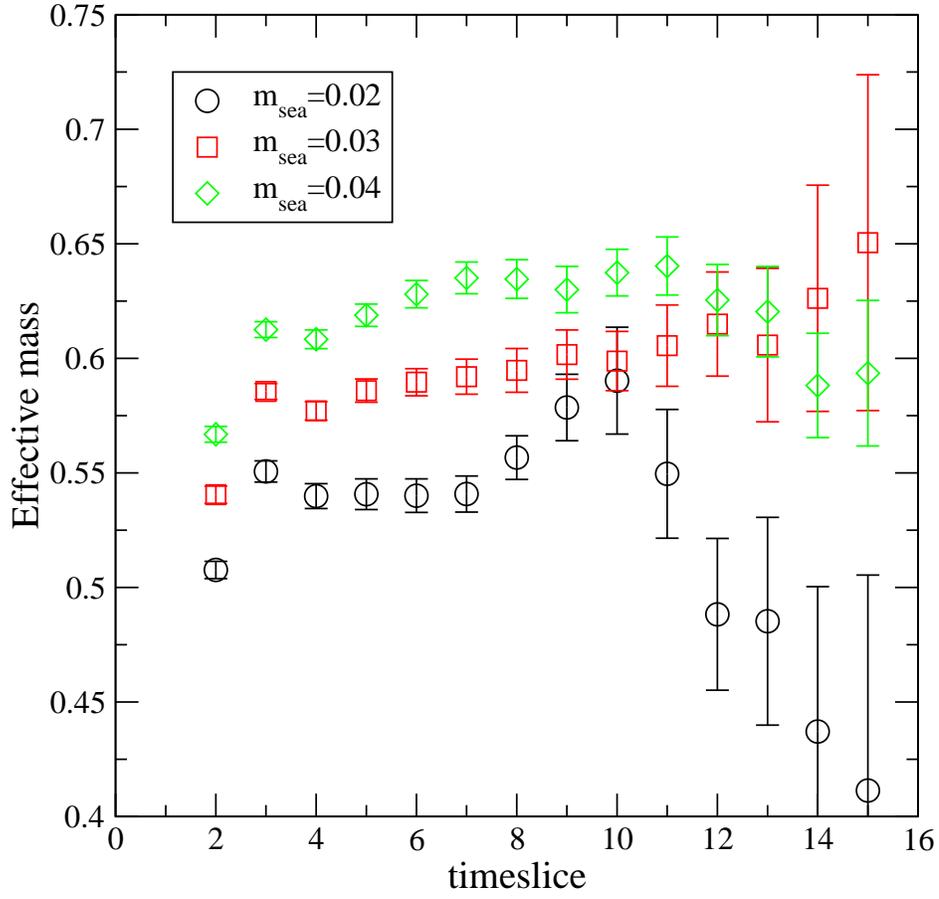}
}
\caption{The effective mass in the vector channel for $m_{sea}=m_{val}$.}
\label{fig:mvec-eff}
\end{figure}\vspace{.2in}


\begin{figure}[htbp]
\centerline {
\includegraphics[width=6in]{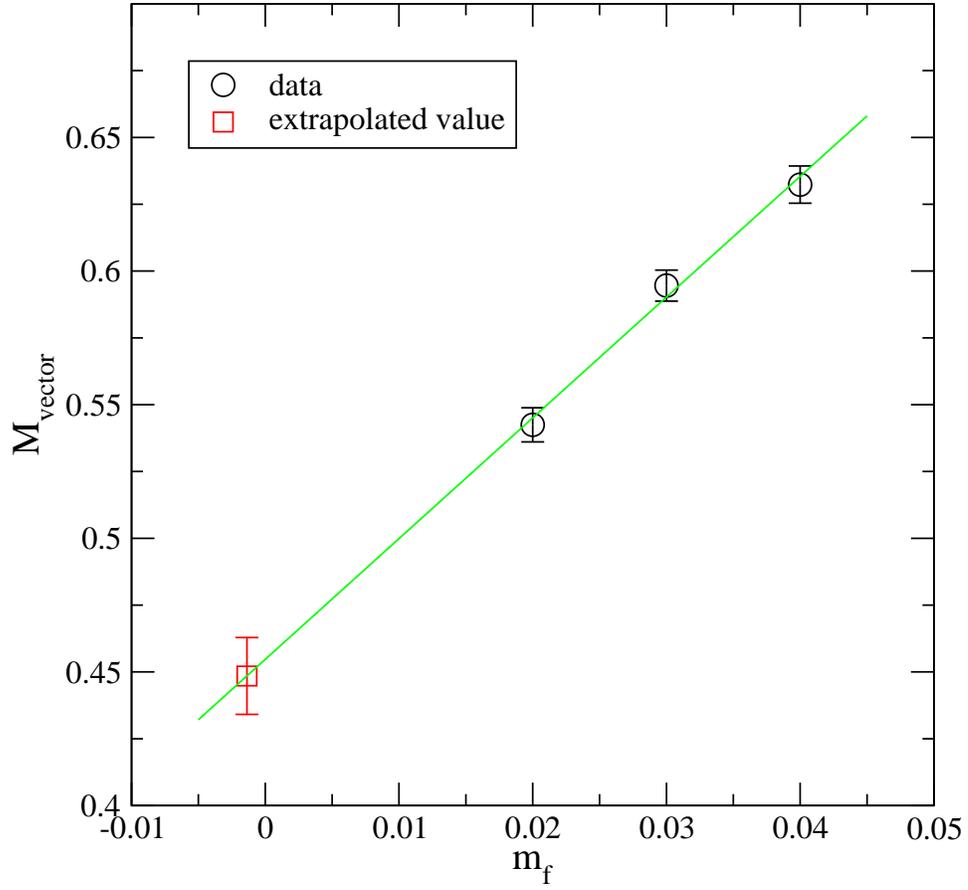}
}
\caption{The fitted vector mass in the vector channel for
$m_{sea}=m_{val}$. The solid line denotes a linear fit.}
\label{fig:mrho fit}
\end{figure}\vspace{.2in}

\begin{figure}[htbp]
\centerline {
\includegraphics[width=6in]{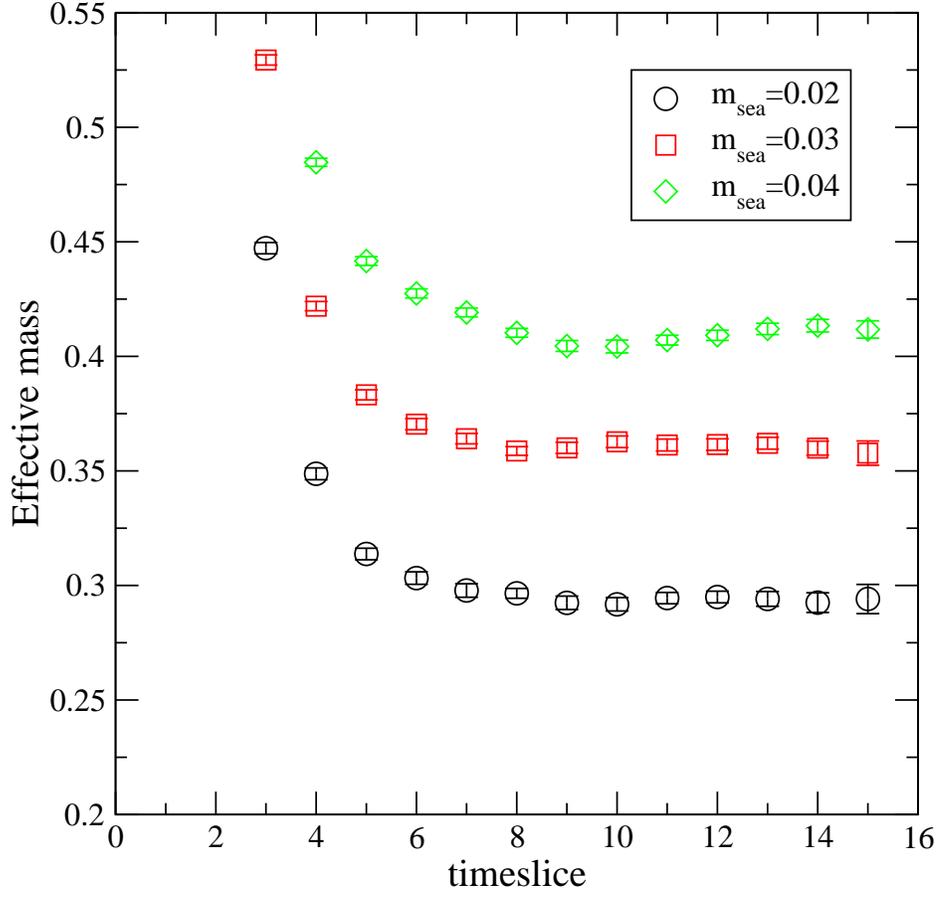}
}
\caption{The effective mass in the pseudo-scalar channel for
$m_{sea}=m_{val}=0.02$ (circles), 0.03 (squares), and 0.04 (triangles). Here
we use the point-point Kuramashi-source correlation function. Similar results
hold for $m_{sea}\neq m_{val}$.}
\label{fig:mpi-GAM-5-eff}
\end{figure}\vspace{.2in}

\begin{figure}[htbp]
\centerline {
\includegraphics[width=6in]{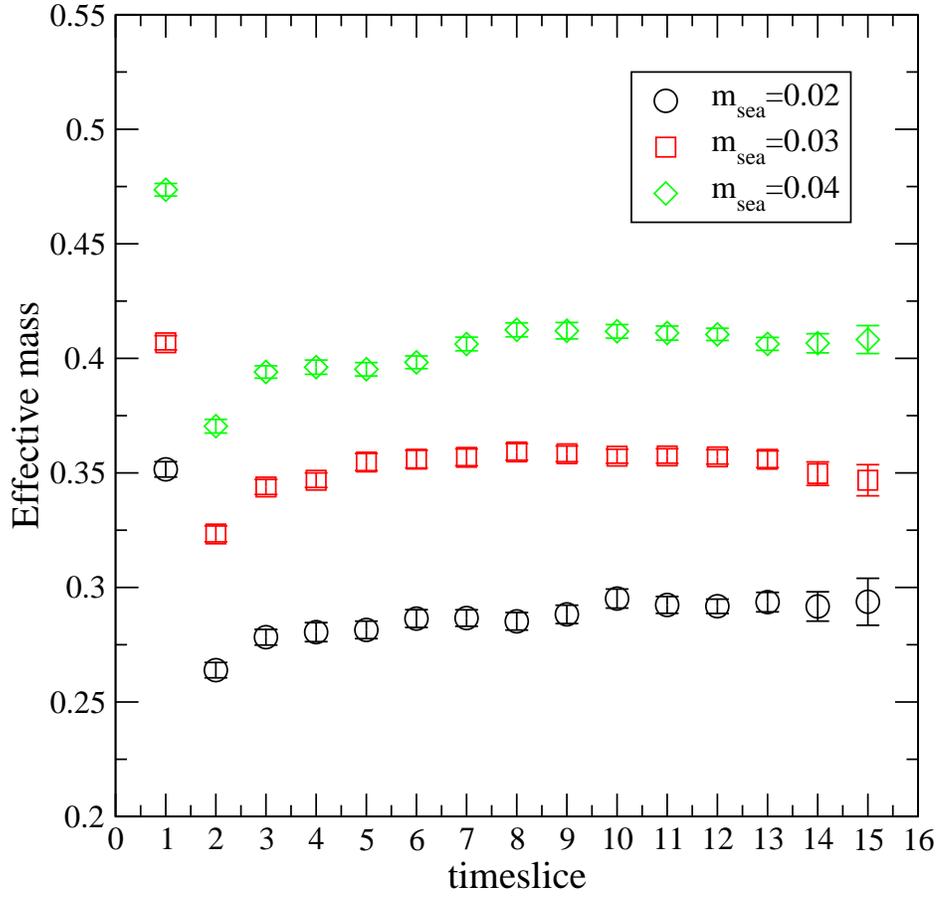}
}
\caption{The effective mass in the pseudo-scalar channel for
$m_{sea}=m_{val}=0.02$ (circles), 0.03 (squares), and 0.04 (triangles).  Here
we use a Coulomb-gauge-fixed wall sourcewith a point sink.  Similar results
hold for $m_{sea}\neq m_{val}$.}
\label{fig:mpi-dmes15-eff}
\end{figure}\vspace{.2in}

\begin{figure}[htbp]
\centerline {
\includegraphics[width=6in]{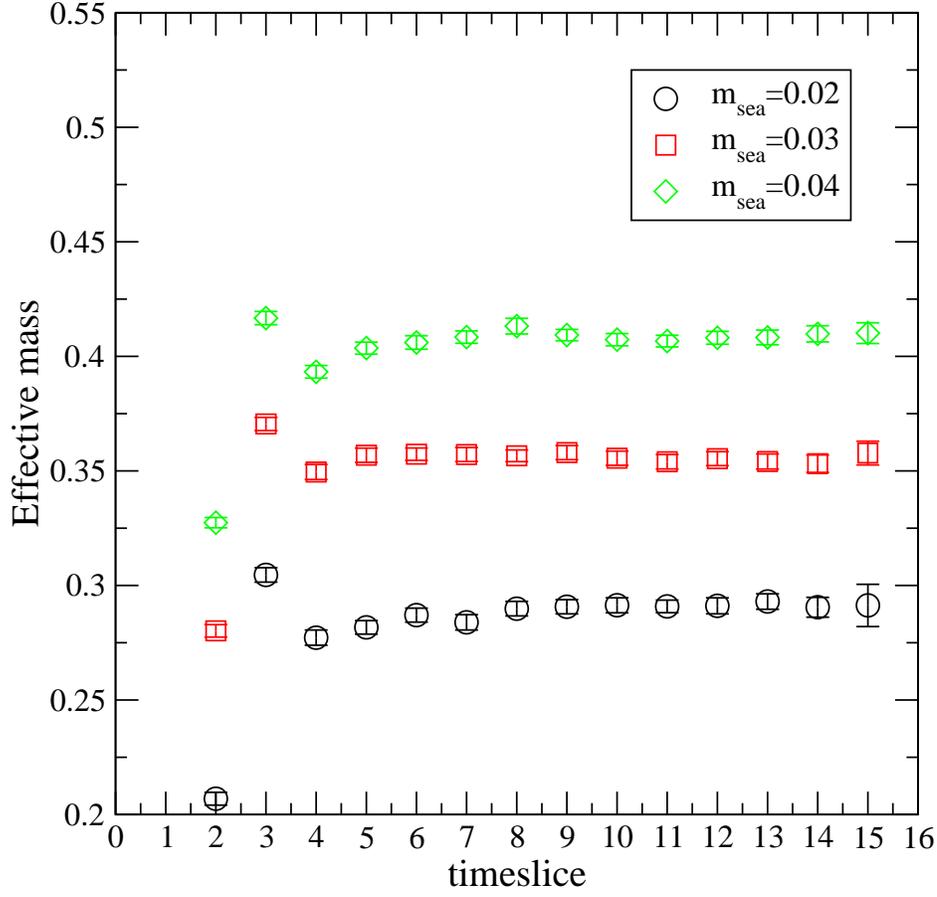}
}
\caption{The effective mass in the axial-vector channel for
$m_{sea}=m_{val}=0.02$ (circles), 0.03 (squares), and 0.04 (diamonds).  Here
we use a Coulomb-gauge-fixed wall sourcewith a point sink.  Similar results
hold for $m_{sea}\neq m_{val}$.}
\label{fig:mpi-dmes11-eff}
\end{figure}\vspace{.2in}

\begin{figure}[htbp]
\centerline {
\includegraphics[width=6in]{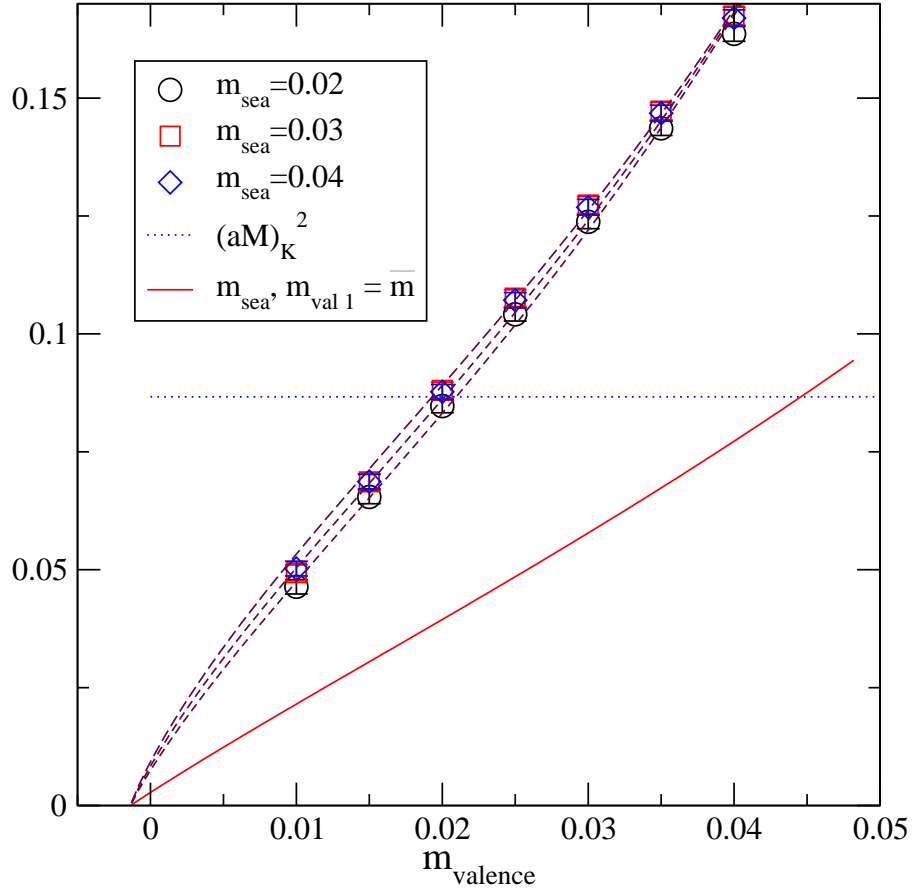}
}
\caption{The pseudo-scalar meson mass, squared, extracted from the wall-point
axial-vector correlation function. $m_{sea}=0.02$ (circles), 0.03 (squares),
and 0.04 (diamonds).  Dashed lines denote a next-to-leading order in chiral
perturbation theory fit. The dotted line marks the mass of the kaon. The solid
line shows the results of an extrapolation of the dynamical, and one of the
valence, quark masses to $\bar{m}$. In this case the x-axis represents the
remaining valence quark mass.}
\label{fig:nlo mpisq fit}
\end{figure}\vspace{.2in}

\begin{figure}[htbp]
\centerline {
\includegraphics[width=6in]{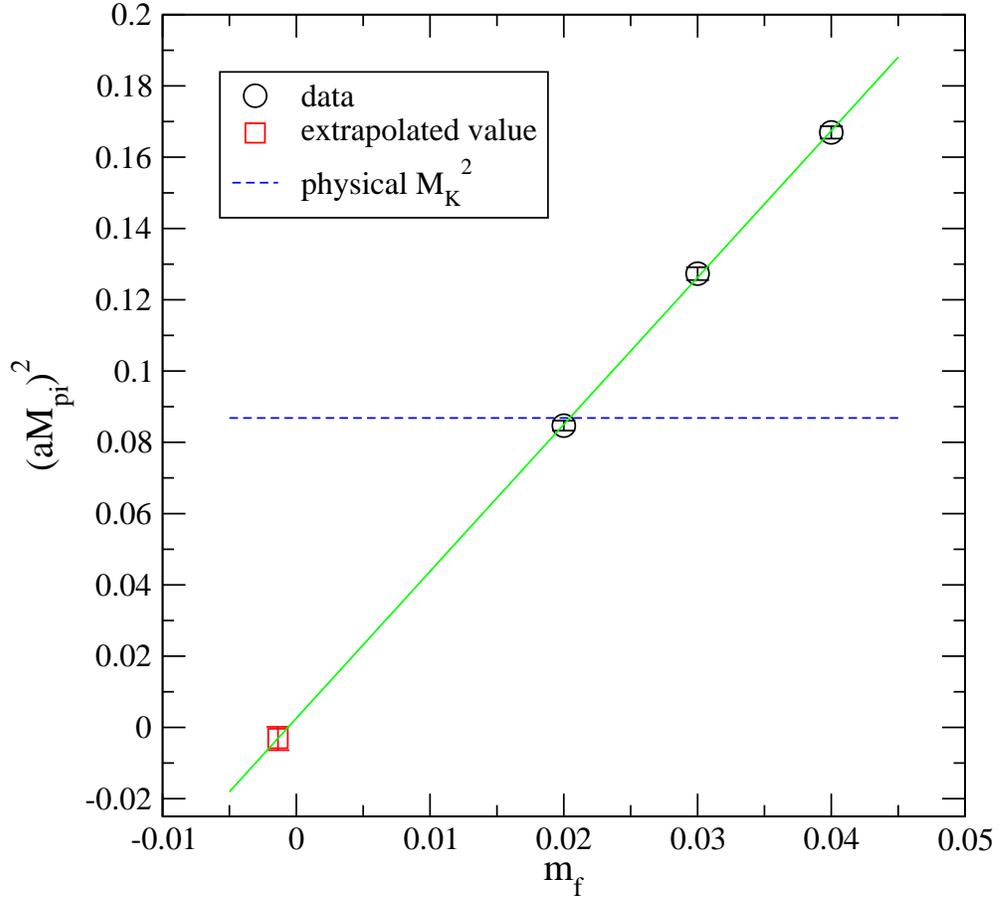}
}
\caption{The pseudo-scalar meson mass squared, extracted from the wall-point
axial-vector correlation function for the dynamical points ($m_f = m_{\rm sea}
= m_{\rm val}$), together with the
results of a LO chiral perturbation theory fit to the quark mass
dependence. The experimental value for the square of the kaon mass is shown as
a dashed line.}
\label{fig:lin mpisq fit}
\end{figure}\vspace{.2in}

\clearpage

\begin{figure}[htbp]
\centerline {
\includegraphics[width=6in]{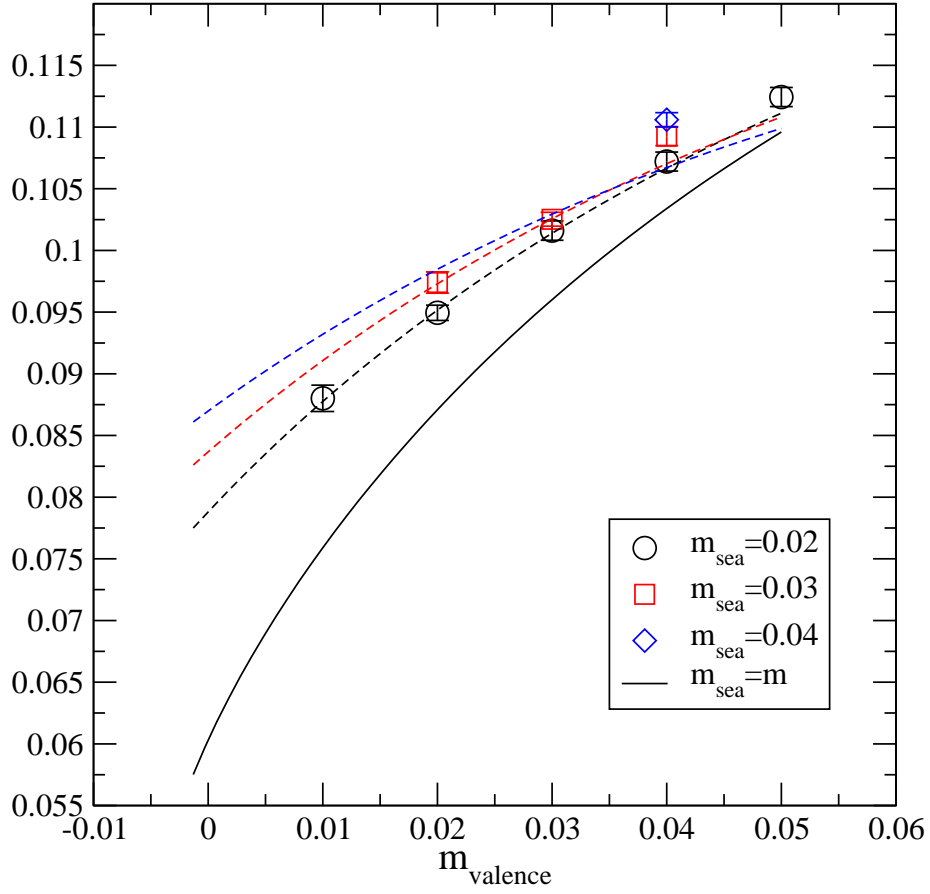}
}
\caption{The meson decay constant. $m_{sea}=0.02$ (circles), 0.03 (squares),
and 0.04 (diamonds).  Dashed lines denote a next-to-leading order in chiral
perturbation theory fit, with the solid line being the extrapolation to $m_{sea}=\overline{m}$.}
\label{fig:nlo decay fit}
\end{figure}\vspace{.2in}

\begin{figure}[htbp]
\centerline {
\includegraphics[width=6in]{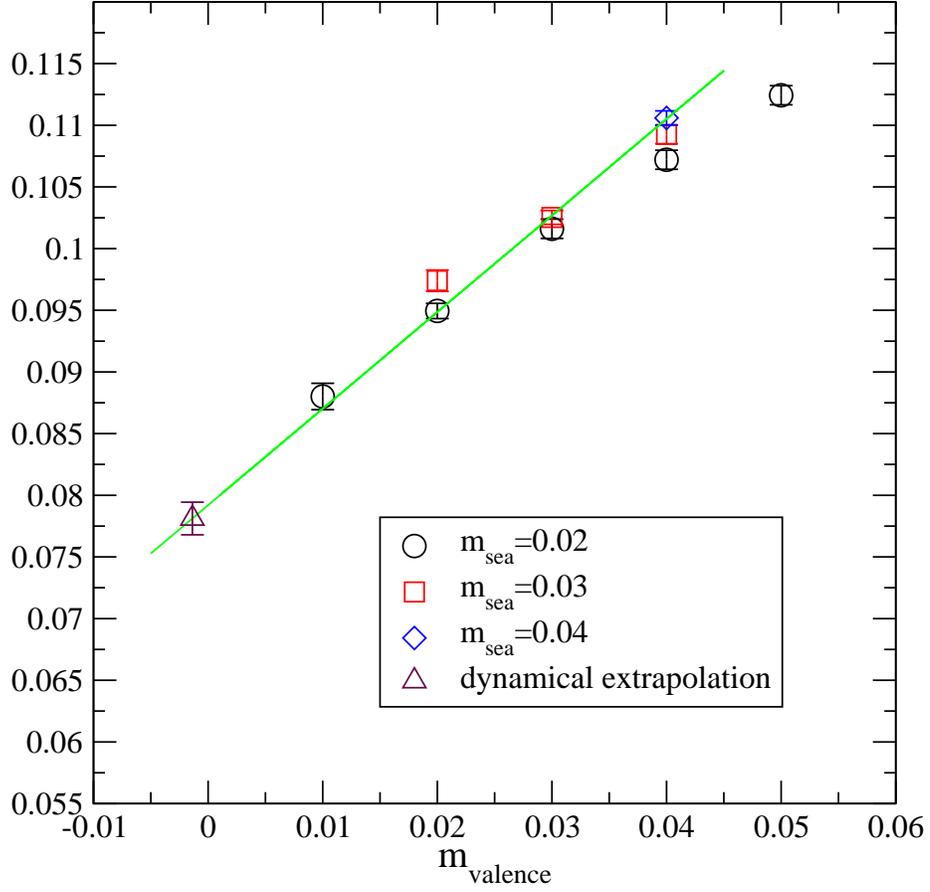}
}
\caption{The meson decay constant. $m_{sea}=0.02$ (circles), 0.03 (squares),
and 0.04 (diamonds), together with the results of a linear fit to the
dynamical points. The extrapolated value at $m_{sea}=-m_{\rm res}$ is also shown.}
\label{fig:lin decay fit}
\end{figure}\vspace{.2in}

\begin{figure}[htbp]
\centerline {
\includegraphics[width=6in]{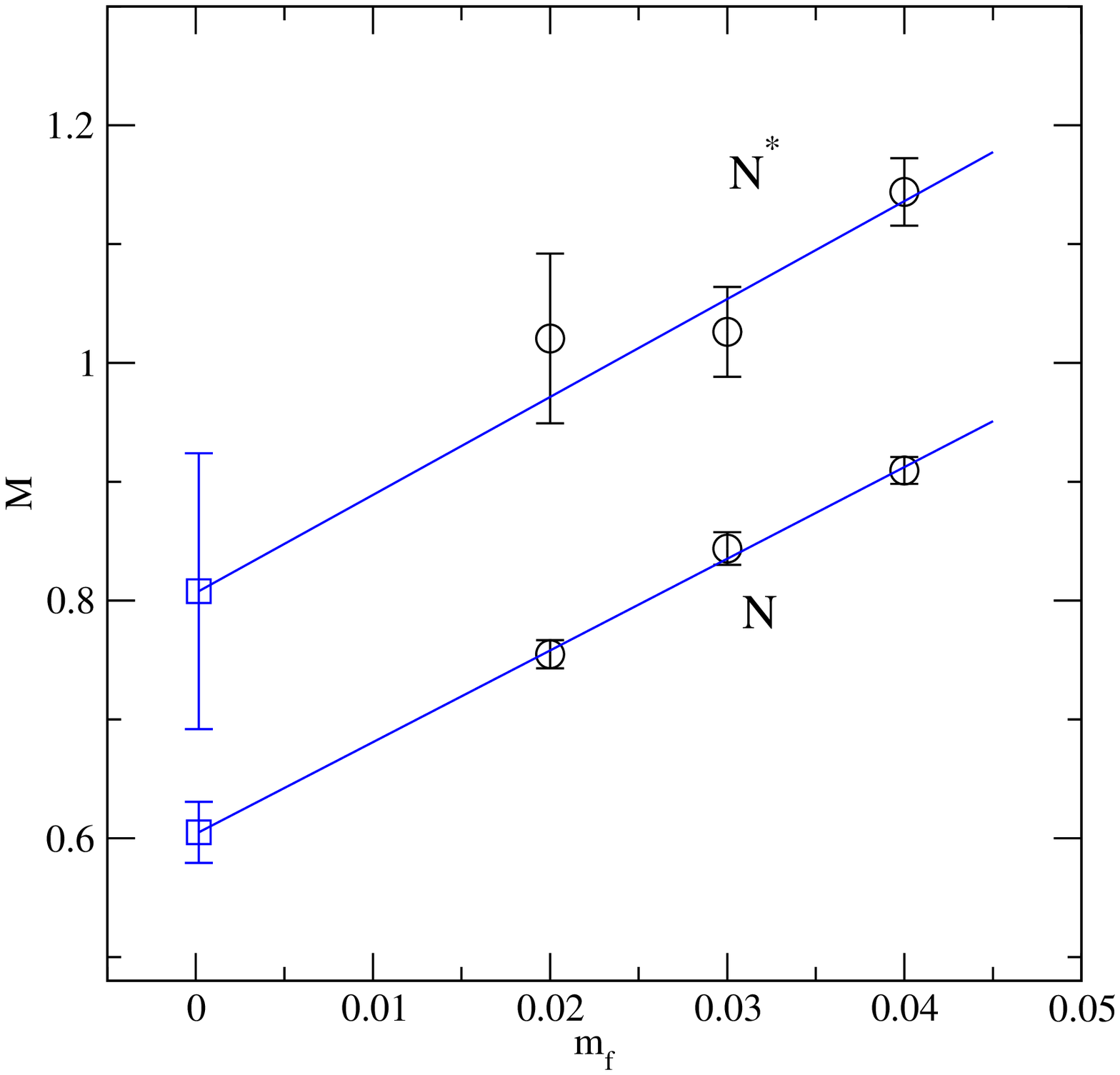}
}
\caption{The masses of nucleon and its parity partner for the dynamical points
($m_f=m_{\rm sea}=m_{\rm val}$).  The solid line is a simple linear fit.
Square symbols show the extrapolated results at $m_f=\bar{m}$.}
\label{fig:mnuc}
\end{figure}\vspace{.2in}
 
\begin{figure}[htbp]
\centerline {
\includegraphics[width=6in]{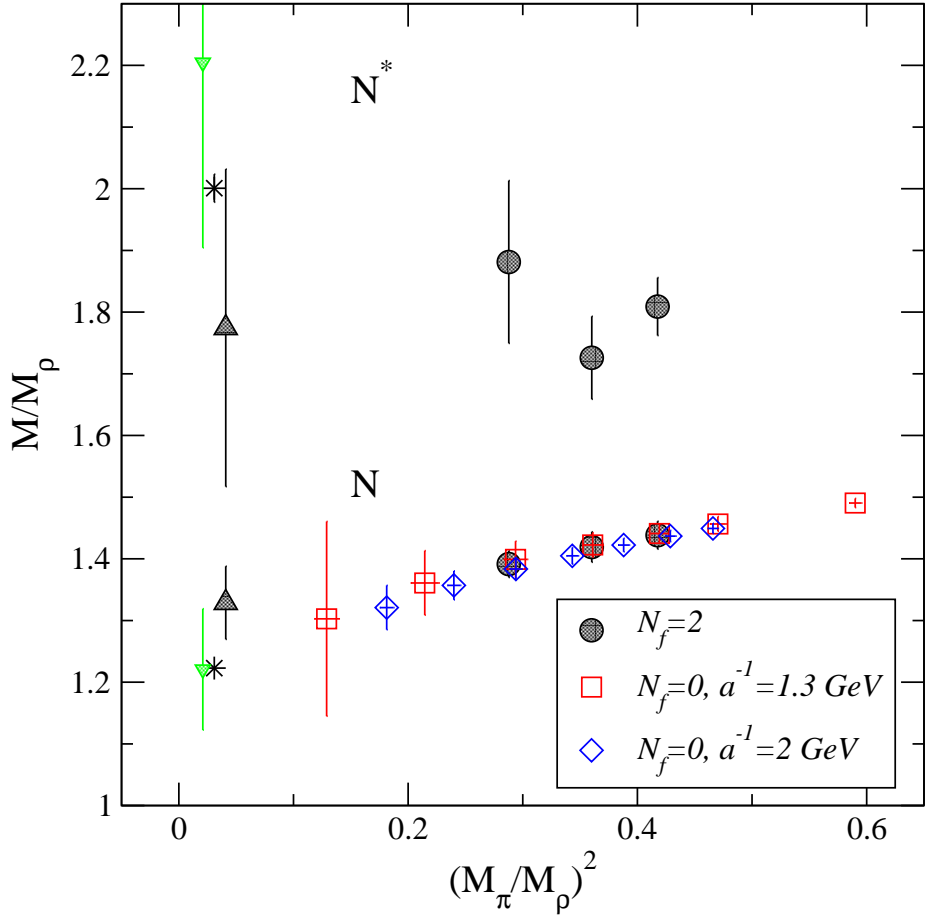}
}
\caption{APE plot of nucleon and its parity partner.
 Filled symbols show the dynamical DWF results, while 
 open symbols show the quenched DWF results \cite{Aoki:2002vt} with
 $a^{-1}\simeq 1.3$ GeV (squares) and $a^{-1}\simeq 2$ GeV (diamonds). 
 The upper and lower triangles are dynamical DWF results 
 at the physical light quark mass ($\bar{m}$) with diagonal
 and two stage chiral extrapolations from Table \ref{tab:mass ratio},
 Stars indicate experimental values, $N(939)$ and $N^{*}(1535)$.}
\label{fig:ape}
\end{figure}\vspace{.2in}

 
\begin{figure}[htbp]
\centerline {
\includegraphics[width=6in]{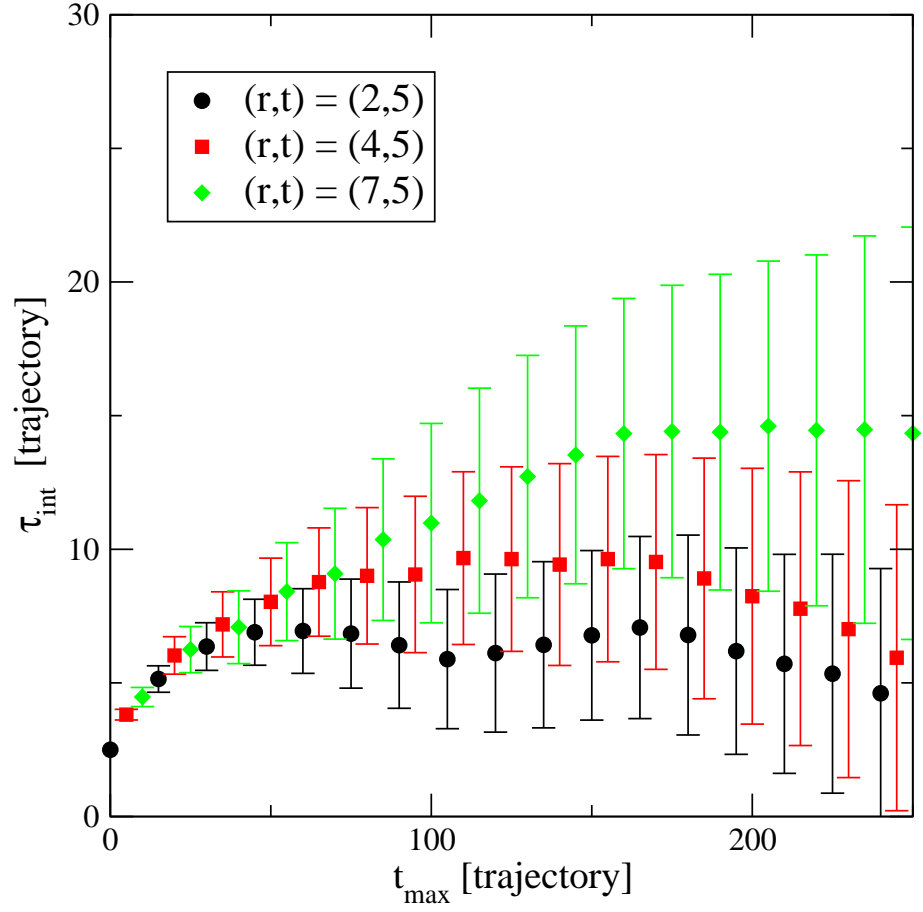}}
\caption{The integrated autocorrelation time, $\tau_{int}$,
of the Wilson loops, $\vev{W(r,t)}$, for $m_{sea}=0.02$.}
\label{fig:Wloop-tauint}
\end{figure}\vspace{.2in}


\begin{figure}[htbp]
\centerline {
\includegraphics[width=6in]{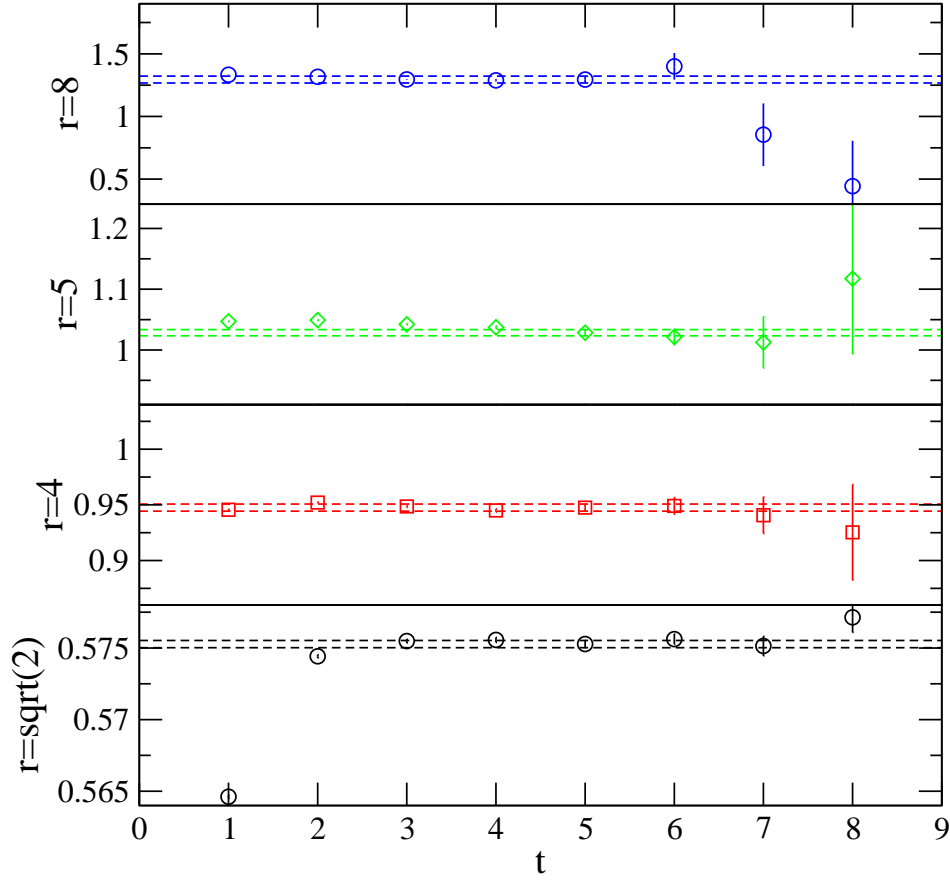}}
\caption{The static quark potential of $m_{sea}=0.02$ at $r$,extracted using
  Eq.~\ref{eq:potextract} at $t$ shown as the horiozntal axis in the
  graph.$r=\sqrt{2}, 4, 5,$ and 8.}
\label{fig:pot_vs_t}
\end{figure}\vspace{.2in}

\begin{figure}[htbp]
\centerline {
\includegraphics[width=6in]{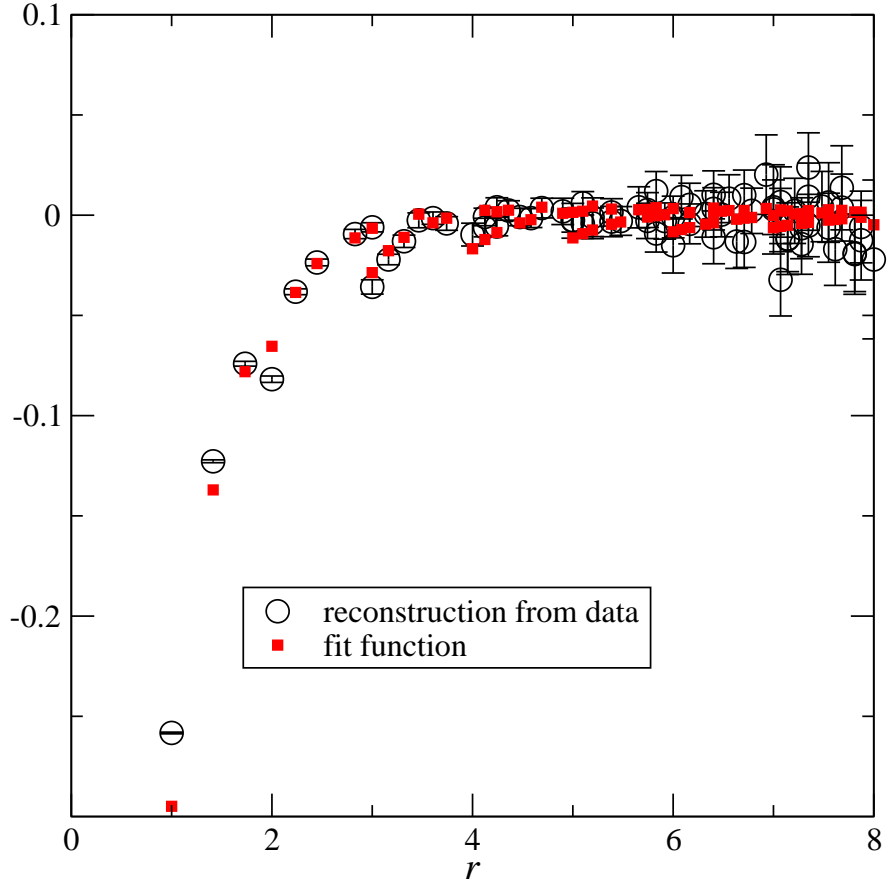}}
\caption{
$\delta V_L(\vec{r})$ and its corresponding reconstruction from
fit data, $[ V^{\rm (data)}(\vec{r})- V_{cont}(r) ] / l$.
Data extracted on $m_{sea}=0.02$ configuration 
at $t=5, r\in[\sqrt{3},8]$ is used in the fit.
}
\label{fig:cmp_lat_coulomb}
\end{figure}

\begin{figure}[htbp]
\centerline {
\includegraphics[width=6in]{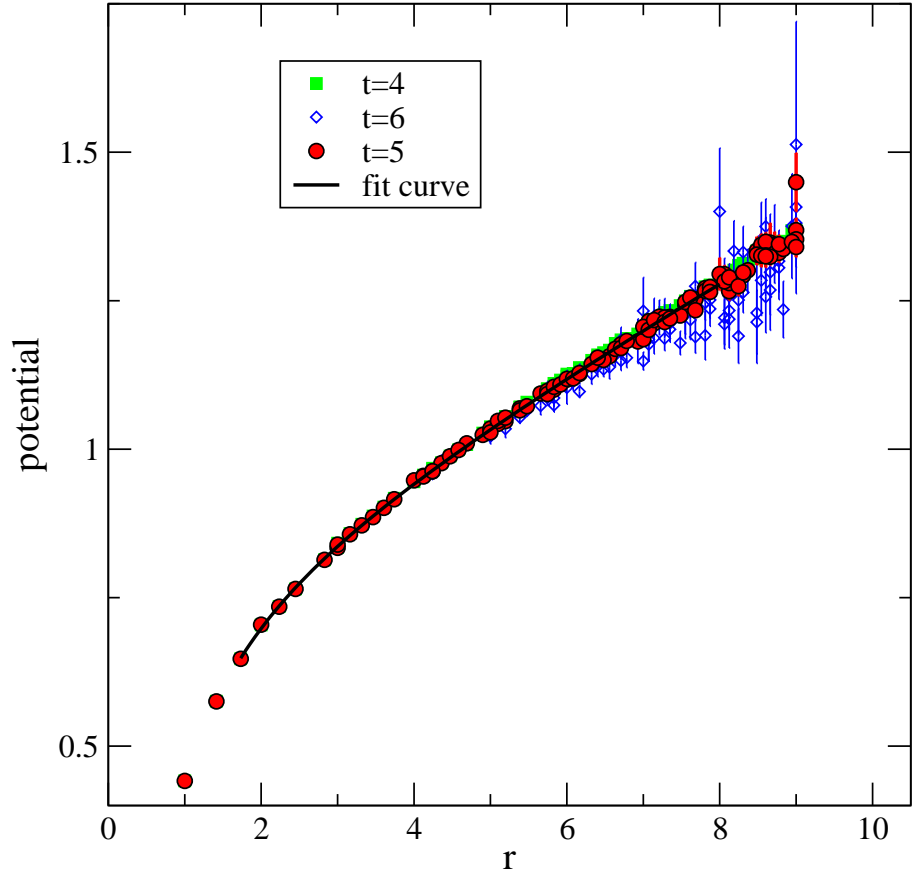}}
\caption{The static quark potential extracted from $t=4,5$ and 6 
for $m_{sea}=0.02$.
The black curve is the fit result to Eq.~\ref{eq:potfit} with $l=0$
using $t=5$ and $\sqrt{3} \leq r \leq 8$.}
\label{fig:pot}
\end{figure}
\vspace{.2in}
                                                                                
\begin{figure}[htbp]
\centerline {
\includegraphics[width=6in]{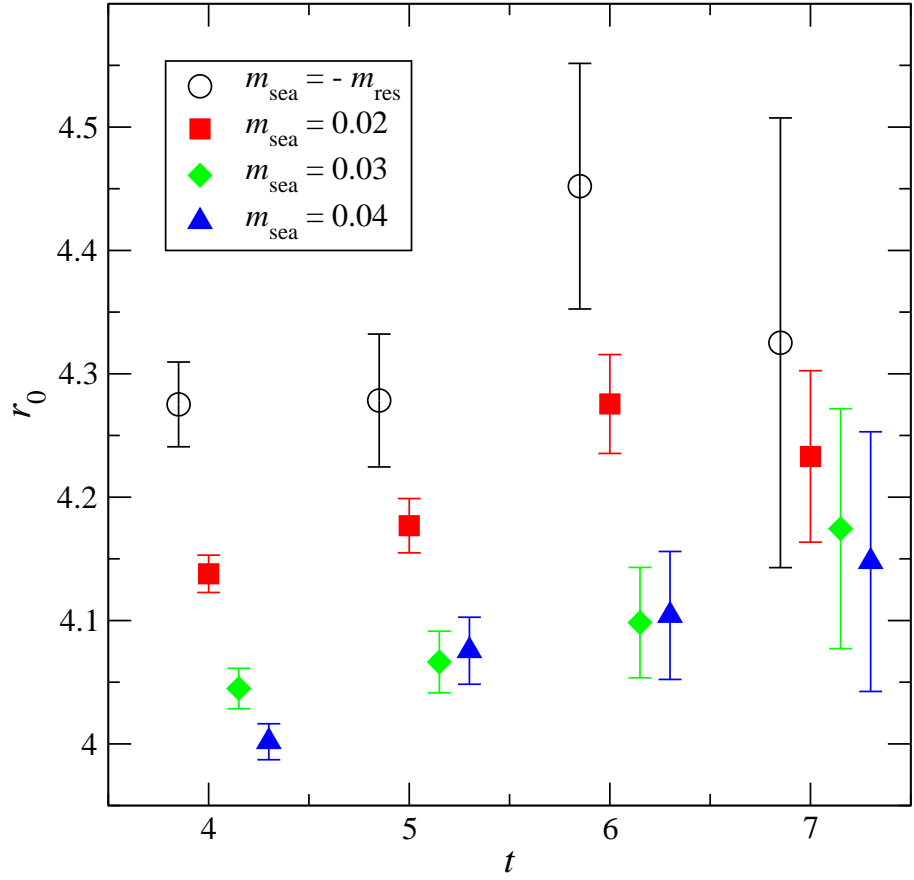}}
\caption{
$t$ dependence of the Sommer scale $r_0$ for $m_{sea}=0.02, 0.03, 0.04$ as well
as
the lineary extrapolated value at chiral point, $m_{sea}=-m_{res}$.
The fit formula with the lattice Colomb term correction
,Eq.~\ref{eq:potfit} with $l\neq0$, are used. }
\label{fig:r0_vs_tmin}
\end{figure}

\begin{figure}[htbp]
\centerline {
\includegraphics[width=6in]{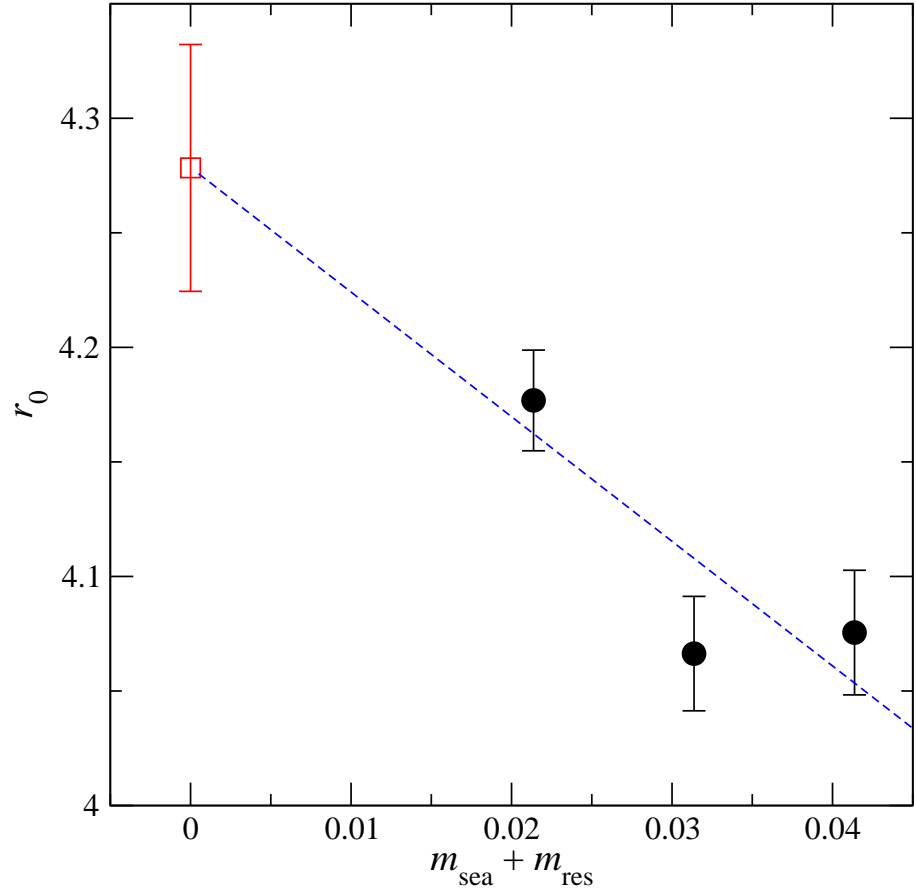}}
\caption{Sommer scale $r_0$ as a function of $m_{sea}$ and their chiral
extrapolation using  linear function. Error bars are statistical only. }
\label{fig:r0_vs_msea}
\end{figure}

\clearpage

\begin{figure}[htbp]
\centerline {
\includegraphics[width=6in]{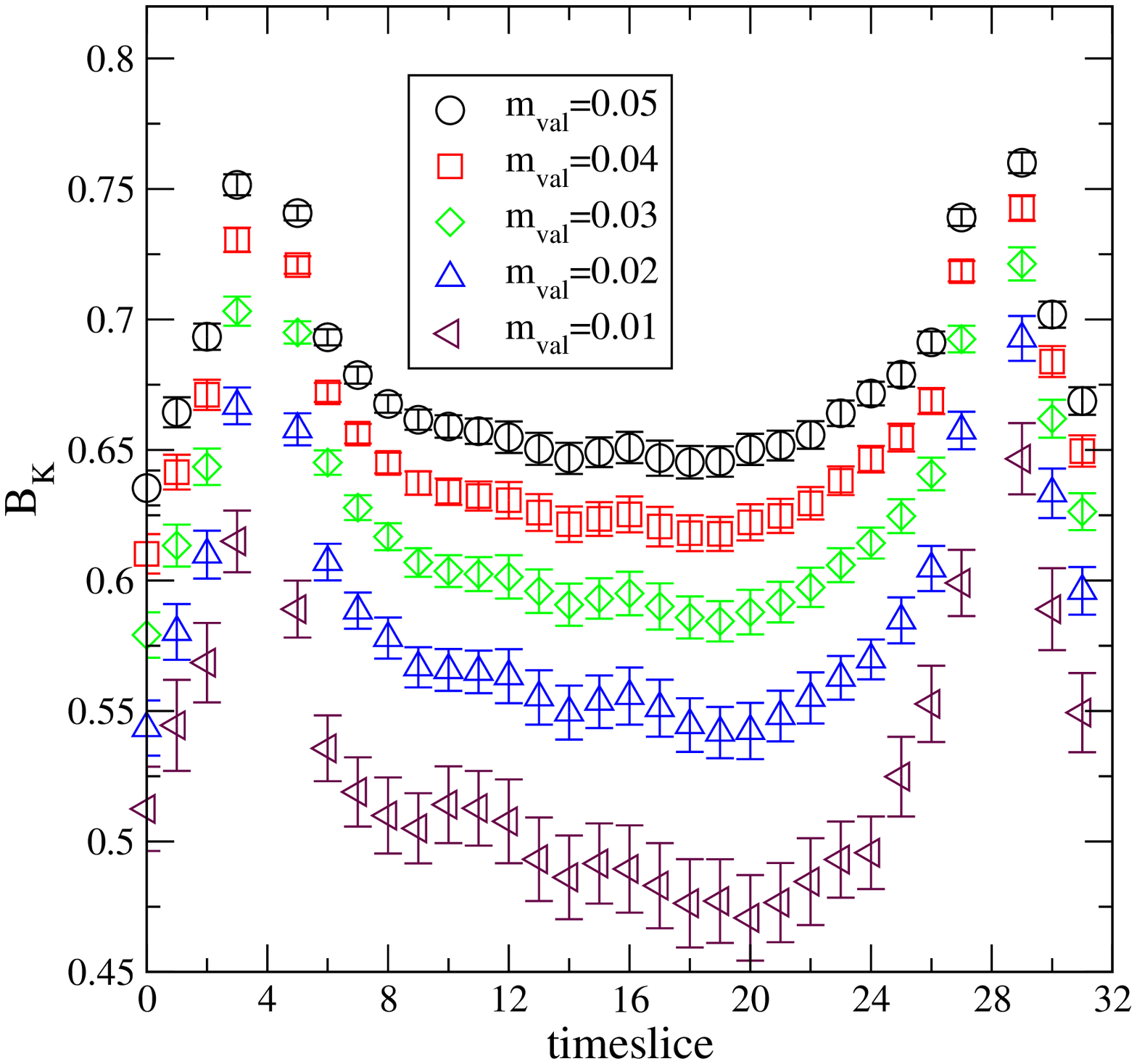}
}\caption{The pseudo-scalar B parameter for $m_{sea} = 
0.02$ and a range of degenerate valence quark masses
versus the time-slice of the operator insertion.}
\label{fig:bp plateaus 0.02}
\end{figure}\vspace{.2in}

\begin{figure}[htbp]
\centerline {
\includegraphics[width=6in]{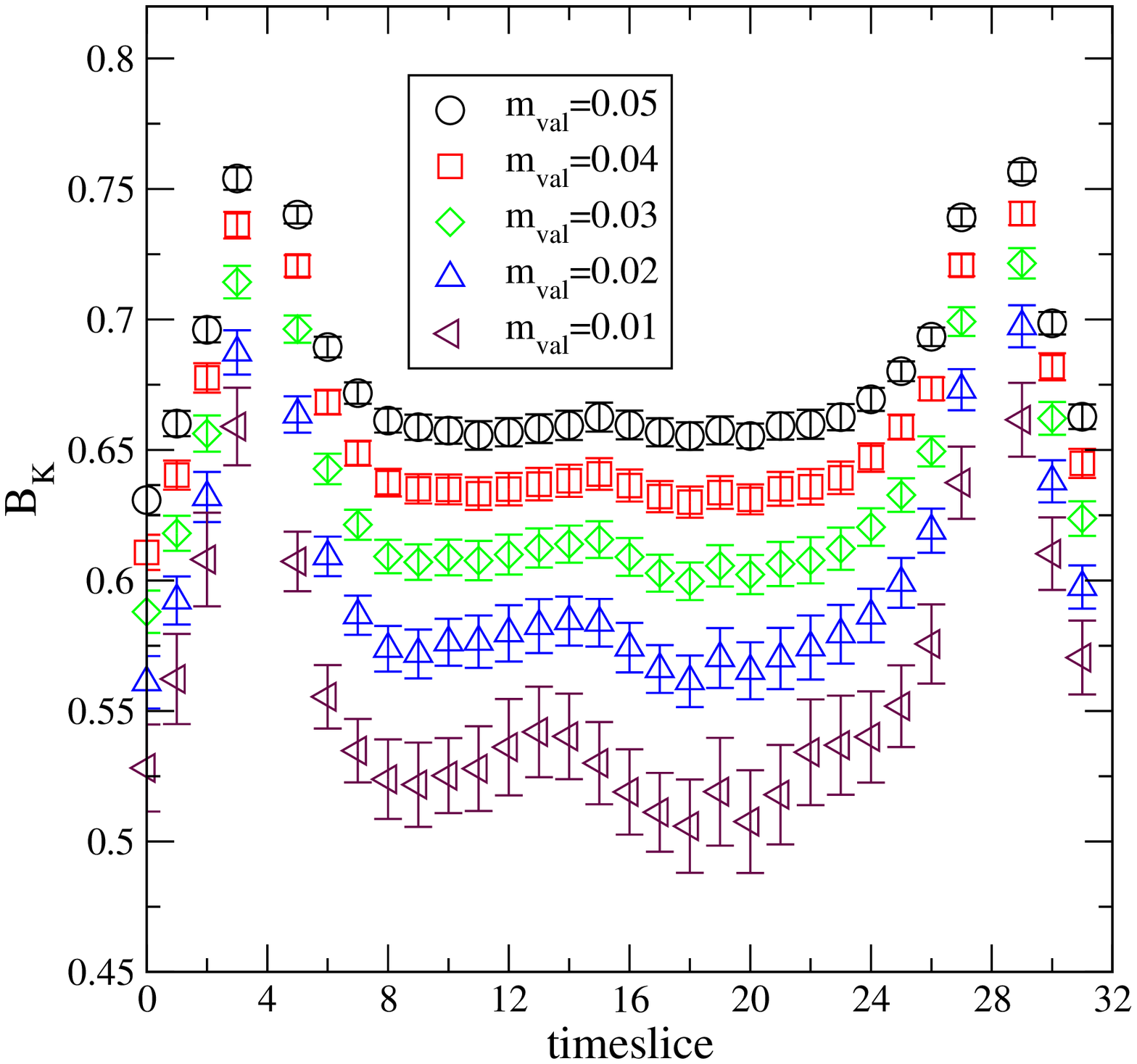}
}
\caption{The pseudo-scalar B parameter for $m_{sea} = 
0.03$ and a range of degenerate valence quark masses
versus the time-slice of the operator insertion.} 
\label{fig:bp plateaus 0.03}
\end{figure}\vspace{.2in}

\begin{figure}[htbp]
\centerline {
\includegraphics[width=6in]{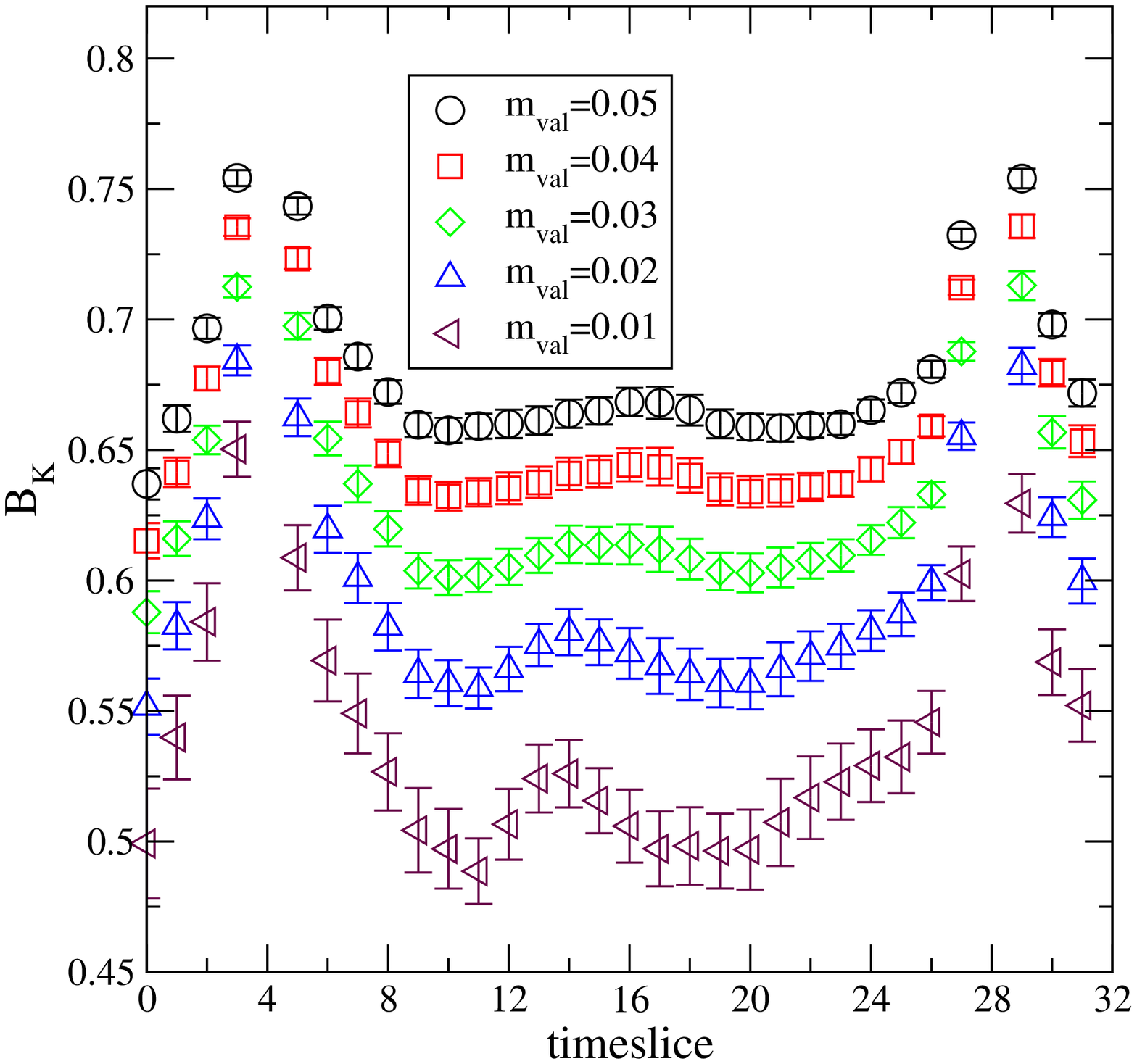}
}
\caption{The pseudo-scalar B parameter for $m_{sea} = 
0.04$ and a range of degenerate valence quark masses
versus the time-slice of the operator insertion.}
\label{fig:bp plateaus 0.04}
\end{figure}\vspace{.2in}

\begin{figure}[htbp]
\centerline {
\includegraphics[width=6in]{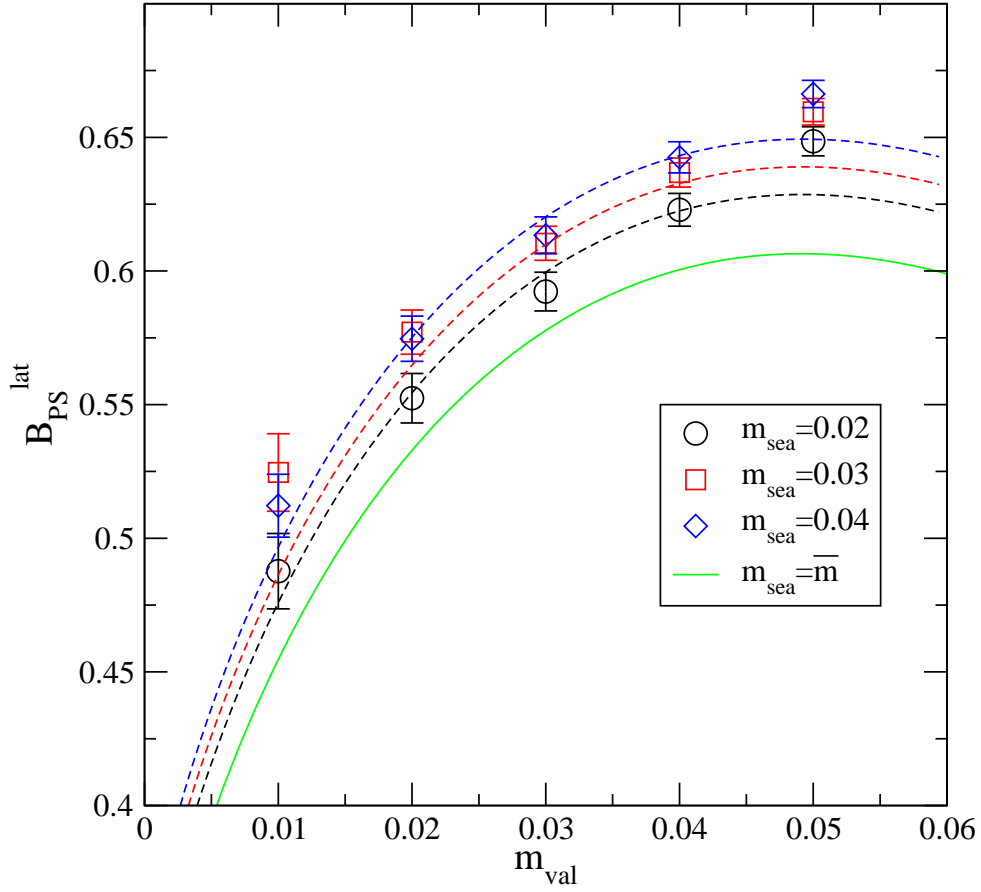}
}
\caption{{The pseudo-scalar B parameter, extracted from time-slices 14 to 17;
$m_{\rm sea}\neq m_{\rm val}$.  $m_{\rm sea}=0.02$ (circles), 0.03 (squares),
and 0.04 (diamonds).  The dashed lines are from a fit to Eq.~\ref{eq:bp-sea},
evaluated at $m_{\rm sea}=0.04$, 0.03, 0.02, and the solid line is an
extrapolation, in the dynamical mass, to $\bar{m}$.  The range of quark masses
used in this particular fit is $0.02 \le m_f \le 0.04$.}}
\label{fig:bp-sea}
\end{figure}\vspace{.2in}

\begin{figure}[htbp]
\centerline {
\includegraphics[width=6in]{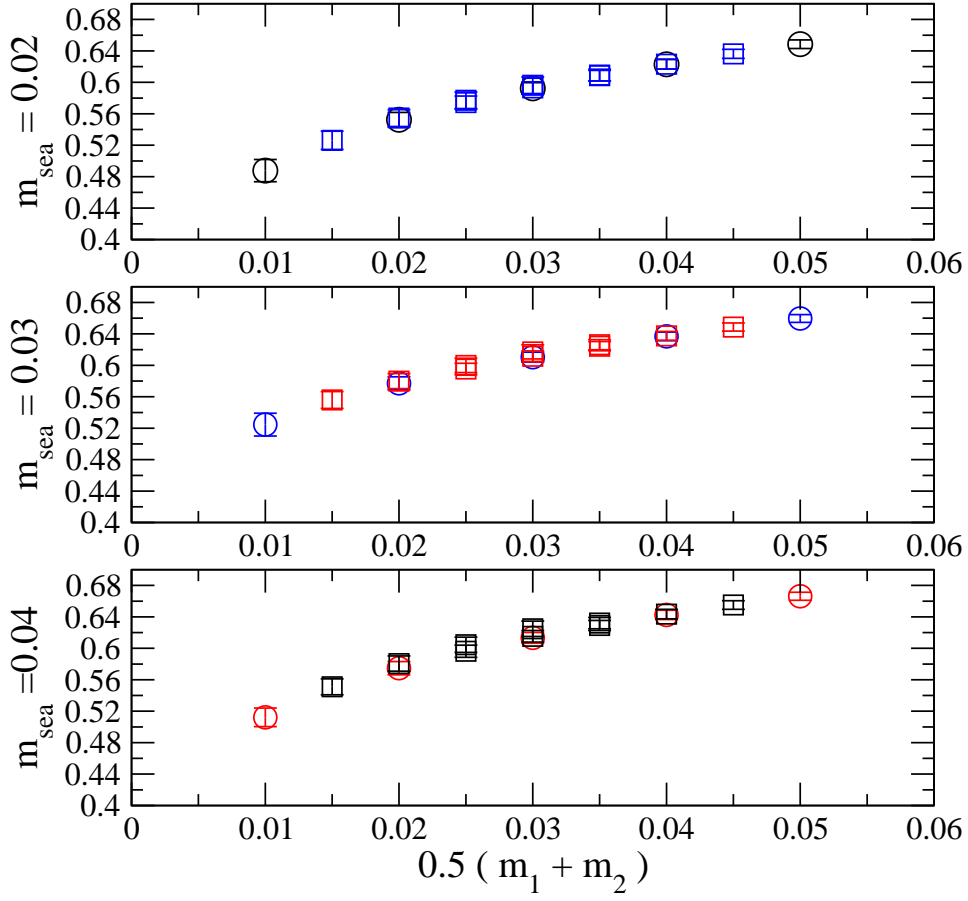}
}
\caption{The pseudo-scalar B parameter; including non-degenerate valence
  quarks, $m_{1}\neq m_{2}$ (squares), and degenerate valence quark
  masses (circles).}
\label{fig:bp nondegenerate}
\end{figure}\vspace{.2in}

%% file: figures/chiral/figures.tex
\begin{figure}[htbp]
\centerline {
\includegraphics[width=6in]{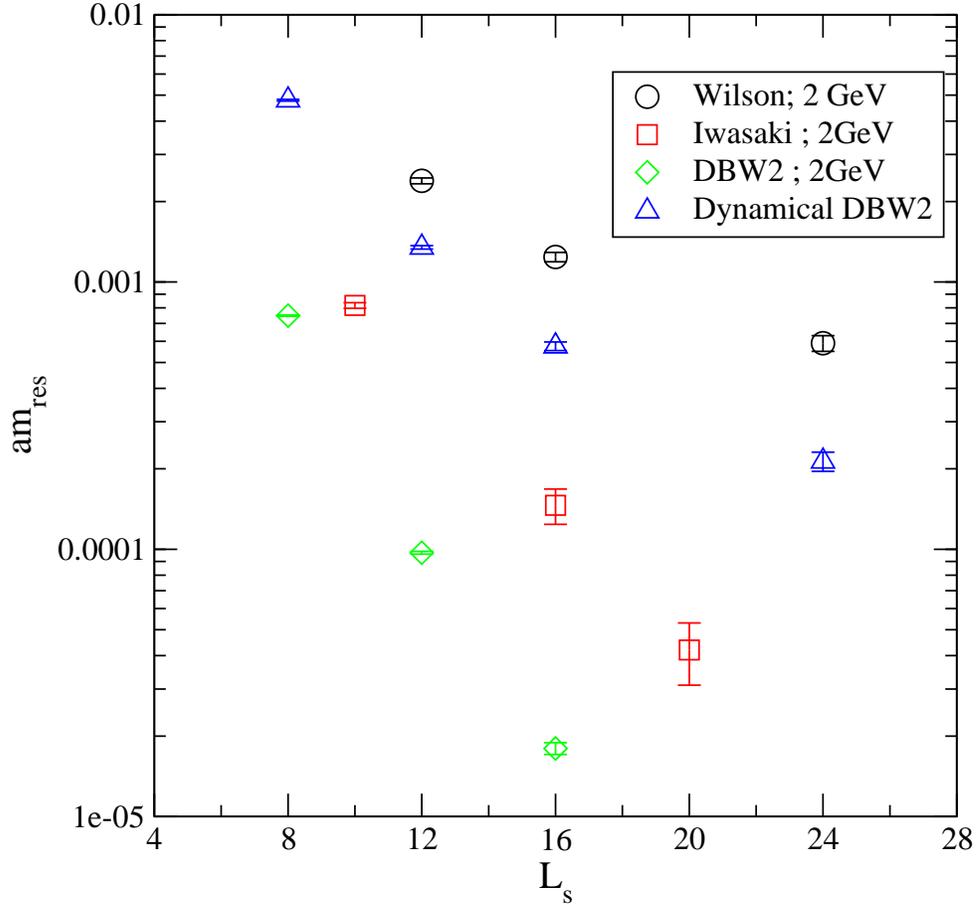}}
\caption{The residual mass versus $L_s$ for $m_{sea}=0.04$, compared with
that measured on three quenched ensembles: Wilson $\beta=6.0$, Iwasaki
$\beta=2.6$ and DBW2 $\beta=1.04$ ; all of which correspond to $a^{-1} \approx
2 {\rm GeV}$.}
\label{lsmres}
\end{figure}

\begin{figure}[htbp]
\centerline{\includegraphics[width=6in]{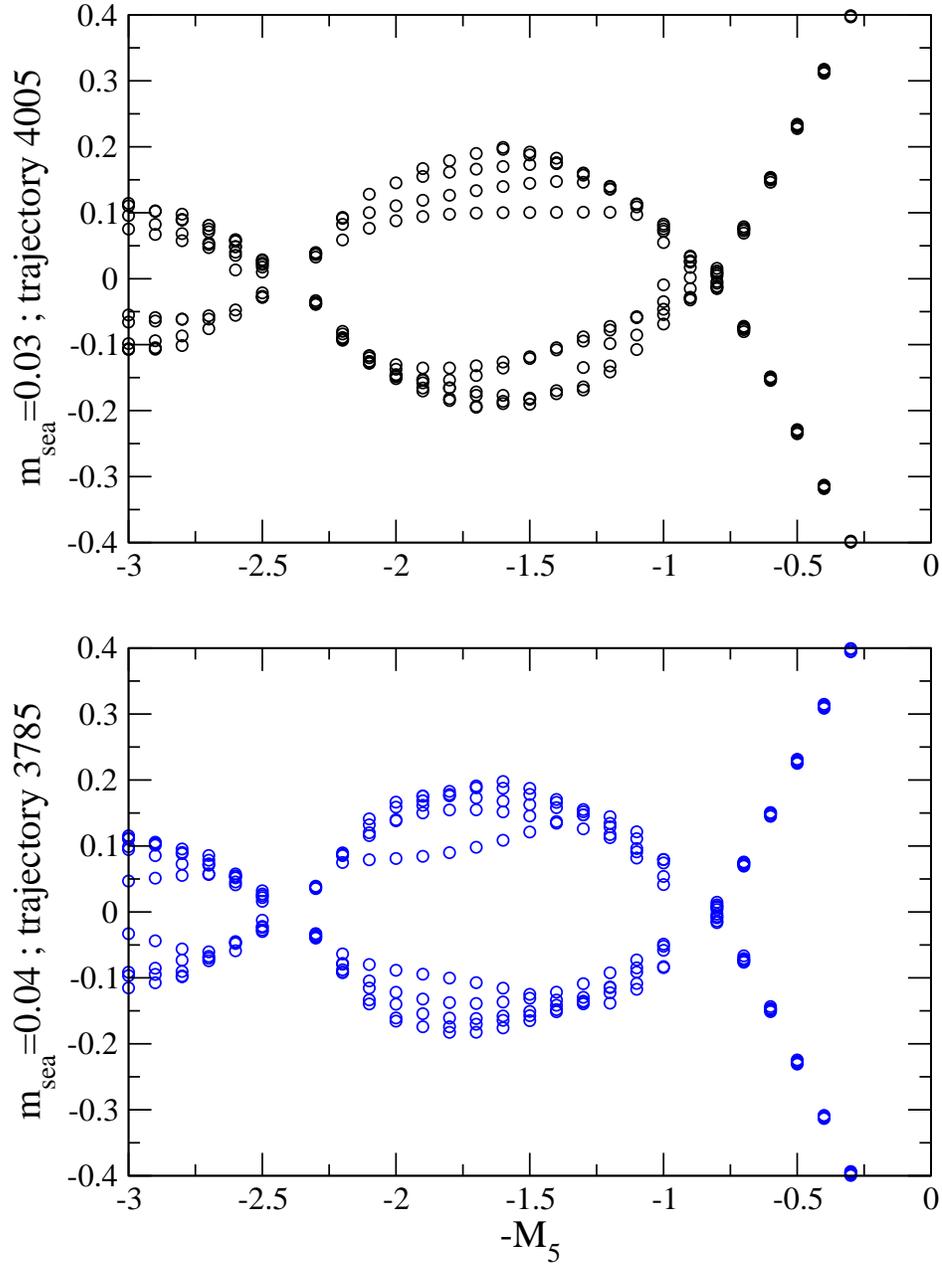}}
\caption{Two typical spectral flows: one from the $m_{sea}=0.03$ evolution and one
from the $m_{sea}=0.04$ evolution.}
\label{spectral}
\end{figure}

\begin{figure}[htbp]
\centerline {
\includegraphics[width=6in]{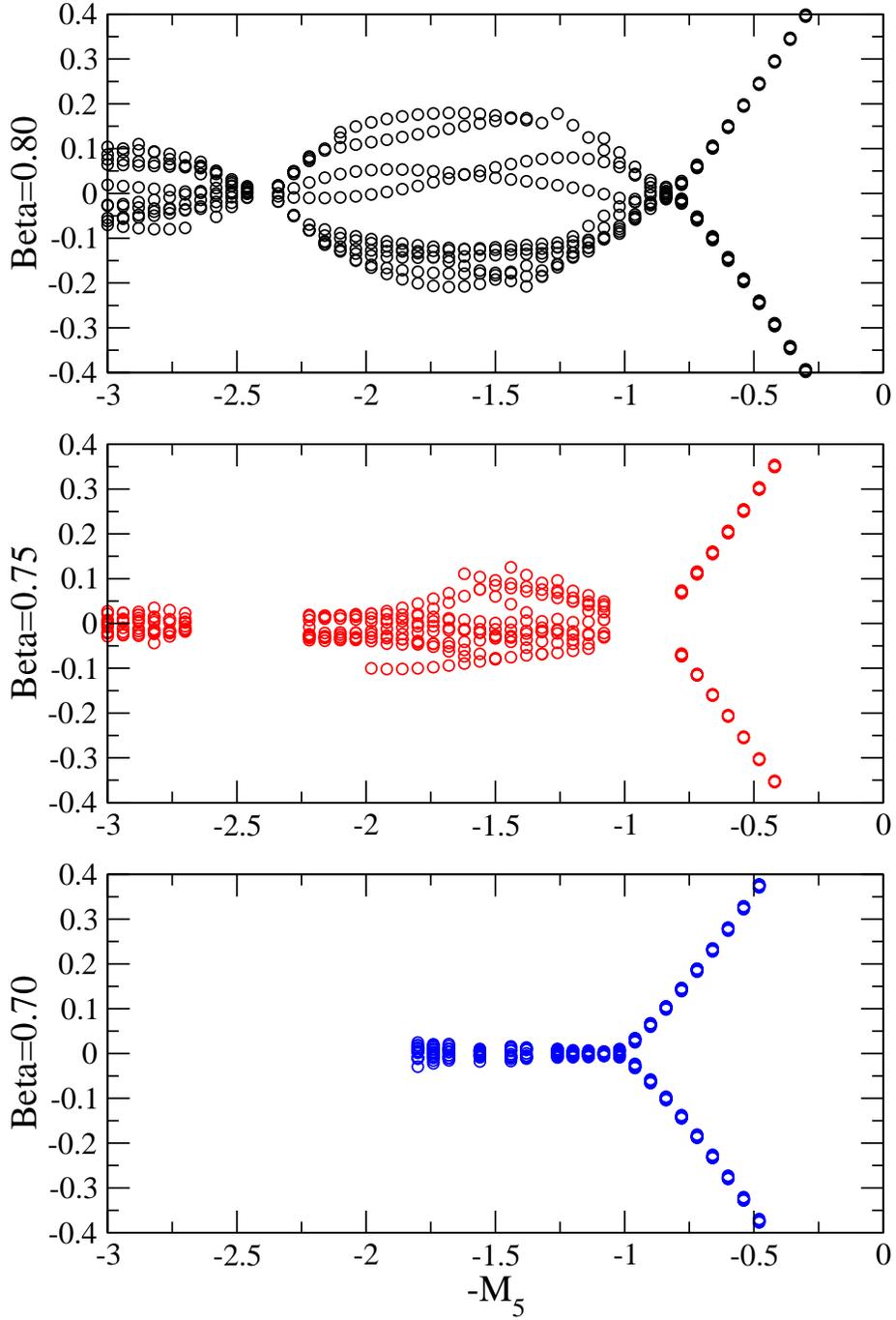}}
\caption{A comparison of three spectral flows: all generated with the DBW2
  action, using $\beta=0.8$, $\beta=0.75$ and $\beta=0.7$ respectively.}
\label{spectralbeta}
\end{figure}